\documentclass[preprint,12pt]{elsarticle}
\usepackage[T1]{fontenc}
\usepackage[utf8]{inputenc}
\usepackage{lmodern}
\usepackage{graphicx}
\usepackage{amsmath}
\usepackage{xcolor}
\usepackage{float}
\usepackage{hyperref}
\usepackage{braket}
\usepackage{amssymb,amsmath,amsthm, array}
\usepackage{mdframed}
\usepackage{systeme,mathtools}
\usepackage{yfonts}
\usepackage{comment}

\def  \bnabla   {\mbox{\boldmath$\nabla $}}

\DeclareUnicodeCharacter{0301}{-}

\journal{Physica D: Nonlinear Phenomena}

\begin{document}
\renewcommand{\vec}[1]{\mathbf{#1}}
\newcommand{\ii}{\mathrm{i}}
\def\ya#1{{\color{orange}{#1}}}

\title{Entanglement properties of photon-magnon  crystal from nonlinear perspective}

\author[PRz]{M. Wanic}
\affiliation[PRz]{organization={Department of Physics and Medical Engineering, Rzeszow University of Technology},
             city={Rzeszow},
             postcode={35-959 Rzeszow},
             country={Poland}}
\author[PRz]{C. Jasiukiewicz}

\author[Georgia]{Z. Toklikishvili}
\affiliation[Georgia]{organization={Faculty of Exact and Natural Sciences, Tbilisi State University},
             addressline={Chavchavadze av.3},
             city={Tbilisi},
             postcode={0128 Tbilisi},
             country={Georgia}}
\author[Germany]{V. Jandieri}
\affiliation[Germany]{organization={General and Theoretical Electrical Engineering (ATE), Faculty of Engineering, University of Duisburg-Essen and CENIDE - Center for Nanointegration Duisburg-Essen},
             city={Duisburg-Essen},
             postcode={D-47048 Duisburg},
             country={Germany}}
\author[PRz]{M. Trybus}

\author[Lublin]{E. Jartych}
\affiliation[Lublin]{organization={Department of Electronics and Information Technology, Faculty of Electrical Engineering and Computer Science, Lublin University of Technology},
             city={Lublin},
             postcode={20-618 Lublin},
             country={Poland}}

\author[India]{S. K. Mishra}
\affiliation[India]{organization={Department of Physics, Indian Institute of Technology (Banaras Hindu University)},
             city={Varanasi},
             postcode={Varanasi - 221005},
             country={India}}

\author[PRz]{L. Chotorlishvili}

\date{\today}
\begin{abstract}
Quantifying the entanglement between two continuous bosonic modes, such as magnons and photons, is not trivial. The state-of-the-art for today is the logarithmic negativity, calculated through the quantum Langevin equations subjected to thermal noise. However, due to its complexity, this method requires further approximation. Namely, after the linearization procedure, quantum operators are replaced by their semiclassical expectation values calculated near the steady state. However,  the phase space of a generic nonlinear system contains topologically different regions, and the steady state may correspond to the different types of fixed points, such as Saddle Points, Stable or unstable Spirals, and Nodes. Through the conventional linearization procedure, one obtains equations for the photon and magnon number operators, but the character of the fixed point is unexplored. In the present work, we propose a new procedure. Namely, we derived the complete set of nonlinear equations, which includes equations for the magnon and photon number operators and phases. We show that not only number operators but also phases are important for exploring the character of the fixed point, and the character of the fixed point influences the magnon-photon entanglement. We showed that methods of the qualitative theory of nonlinear differential equations are also relevant for photon-magnon entanglement problems. Our main finding is that entanglement is not defined in the  Saddle Point region. On the other hand, the maximum of the entanglement corresponds to the region near the border between the Stable node and Stable spiral regions. Our approach is quite general. However, we did calculations for a particular system: photon-magnon crystal based on the yttrium iron garnet (YIG) film with the periodic air holes drilled in the film. Our interest focuses on magnons with a particular wavelength and frequency corresponding to the magnon condensate. Those magnons couple strongly with the photons of similar frequency. We discuss in detail the interaction between magnons and photons originating from the magneto-electric coupling and the effective Dzyaloshinskii-Moriya interaction. We show that this interaction is responsible for the robust photon-magnon entanglement in the system.  
\end{abstract}

\maketitle

\section{Introduction}
\label{sec:Introduction}

Entangled photon-magnon systems is the research topic formed during the last decade where fundamental quantum physics and technological applications met and have been developing together, mutually enriching each other with new ideas and novel technological solutions \cite{PhysRevLett.111.127003, PhysRevLett.113.083603, PhysRevLett.113.156401, PhysRevB.91.104410, PhysRevResearch.3.013277,pantazopoulos2019high,PhysRevB.96.104425,PhysRevB.111.075415}. Quantum computation and functionality tasks on the nanoscale level raised specific requirements for the systems. Typically, platforms designed for quantum information protocols are binary, constituent of two different subsystems, and incorporate different parameters. While nanomechanical systems joint mainly mechanical and spin excitations \cite{bachtold2022mesoscopic,aspelmeyer2014cavity,singh2022hybrid}, constituent parts of optomagnon systems are photons and magnons. The study of entangled photon-magnon systems follows the spirit of nanomechanics and nanomechanical systems, merging quantum optics and solid-state physics. Photon-phonon interaction in photonic crystals plays a crucial role in the entanglement properties \cite{PhysRevA.102.022816,manzoni2017designing,goban2014atom,hung2013trapped,douglas2015quantum}. Interest in photon-magnon entanglement concerns the enhanced interactions between magnons and photons. Previous studies showed that entanglement between two continuous bosonic modes requires interactions beyond the bilinear form \cite{PhysRevLett.121.203601, PhysRevB.107.115126}. The paradigmatic photonic-magnonic setup is an optical cavity and a yttrium iron garnet (YIG) sphere. Magnetocrystalline anisotropy in the YIG sphere leads to the nonlinear magnon interaction and  Magnon Kerr Eﬀect (MKE) \cite{PhysRevB.106.054425, PhysRevA.104.033708, PhysRevA.106.012419, PhysRevApplied.12.034001, PhysRevB.107.064417, PhysRevLett.129.123601}. The interaction term $K_0\hat m^+\hat m\hat m^+\hat m$, (where $\hat m^+$,   $\hat m$  are magnon creation and annihilation operators) is usually small, characterized by the interaction constant $K_0=\mu_0K_{0an}\gamma_e/M^2V_m$, with $\mu_0$,  being the magnetic permeability, $K_{0an}$ the anisotropy constant, $\gamma_e$ the gyromagnetic ratio  $M$ the saturation magnetization, and $V_m$ is the volume of the YIG sphere. The smaller the volume of the YIG sphere, the stronger the MKE is.

In the recent work \cite{amazioug2023enhancement} was proposed a coherent feedback loop scheme for enhancing the magnon–photon–phonon entanglement in cavity magnomechanics. It was shown that the steady state and dynamical state of the system form a tripartite entanglement state. The feedback control of quantum correlations in the context of magnon squeezing was studied in \cite{amazioug2023feedback}. The robust magnon-photon entanglement in the systems with  Dzyaloshinskii-Moriya (DM) interaction can be exploited in quantum sensors \cite{adani2024critical,PhysRevB.111.024315}.
\\
In the present work, we propose the model of photon-magnon crystal and study magnon-photon entanglement based on the magnon condensation effect. We consider a YIG film with a periodic array of holes in the YIG film To quantify entanglement between two continuous bosonic modes \cite{PhysRevLett.84.2726, PhysRevLett.84.2722, PhysRevLett.96.050503} (photonic and magnonic), we exploit measures of the entanglement such as logarithmic negativity  \cite{adesso2007entanglement, PhysRevLett.95.090503}. Instead of the YIG sphere and weak Kerr effect, we propose to exploit the magnon condensation effect and intraband magnon-magnon interaction originated from the perpendicular magnetocrystalline anisotropy (PMA) and bismuth doping on the YIG film. The magnon-magnon interband interaction term in the magnon condensate mimics the effective Kerr effect. We propose to control of the effective Kerr nonlinearity through the doping of the YIG film by bismuth impurities. Utilizing the multiferroic properties of YIG \cite{PhysRevLett.106.247203}  and magnetoelectric effect (ME) \cite{PhysRevB.80.140412,trybus2024dielectric}, we derive photon-magnon coupling in terms of the effective DM interaction \cite{kolesnikov2024energy,kolesnikov2023influence,kolesnikov2022influence,wang2020optical}. The effective DM interaction and magnon condensation effect allow us to control nonlinearity in the system through two parameters: external electric field and concentration of the doped bismuth impurities. We note that the DM interaction term is a product of three operators (see section 3). Therefore, the DM interaction term has a bilinear operator contribution in a quantum Langevin equation. After semi-classical approximation, the DM term leads to nonlinear effects, which are in the scope of interest of our work. \vspace{0.2cm}\\
While the magnon spectrum of YIG film is broad, we are interested in the magnon mode corresponding to the magnon condensate. It is a specific mode characterized by the deep minimum in the dispersion curve. This deep minimum stabilizes the magnon condensate. To have a strong effect, we consider resonant or near resonant photons with frequencies close to condensate magnons. In large detuning cases, photons and magnons are disentangled. Thus in the focus of our discussion are those selected modes.
The study of magnon-photon entanglement at finite temperatures requires the method of the quantum Langevin equation \cite{PhysRevLett.121.203601}. Due to the complexity of this approach, a certain approximation is needed.
Namely, general quantum operator $\hat Q$ is replaced by $\hat Q=\langle\hat Q\rangle+\delta\hat Q$,
where $\delta\hat Q$ is the deviation of operators from its steady state expectation value $\langle\hat Q\rangle$. In particular, magnon creation and annihilation operators are replaced by canonical momentum and coordinate operators \cite{PhysRevLett.121.203601}: $\hat q_m=\left(\hat m^\dag+\hat m\right)/\sqrt{2}$,
$\hat p_m=i\left(\hat m^\dag-\hat m\right)/\sqrt{2}$, $\hat q_a=\left(\hat a^\dag+\hat a\right)/\sqrt{2}$,
$\hat p_a=i\left(\hat a^\dag-\hat a\right)/\sqrt{2}$, where $\hat m^\dag$  $\hat m$, $\hat a^\dag$  $\hat a$ are magnon (photon) creation (annihilation) operators. 
The canonical momentum and coordinate operators are linearized in the vicinity of the steady state \cite{PhysRevLett.121.203601}. 
The expectation values of the magnon $\langle\hat m^\dag\hat m\rangle=|\beta|^2$ and photon $\langle\hat a^\dag\hat a\rangle=|\alpha|^2$ number operators and phases $\phi$, $\varphi$, $\langle\hat m\rangle=\beta e^{i\varphi}$, $\langle\hat a\rangle=\alpha e^{i\phi}$ usually are determined in the vicinity of the steady state via the semiclassical approximation \cite{PhysRevB.107.115126}. However, the information about the character of the steady state is not explored in detail. In essence, the steady state is a fixed point of the dynamical system, and fixed points of the nonlinear dynamical system have different topological characteristics and properties. Therefore, in the present work, we propose an alternative approach. At first, we derive a nonlinear set of equations for the variables $\phi$, $\varphi$, $\alpha$,  $\beta$ and explore the phase diagram of the fixed points. We show that the phases $\phi$, $\varphi$, which are not paid enough attention in the linear approach, play a crucial role in nonlinear cases, leading to the instability and transitions between the different dynamical regimes and influence magnon-photon entanglement. We apply methods of the dynamical system's theory and find bifurcation points and different dynamical regions in the system's phase space. We show that separatrix lines separate those regions, and magnon-photon entanglement differs in the topologically distinct regions of the phase space. The manuscript is organized as follows: Section \textbf{2} describes the magnetic subsystem. In section \textbf{3}, we derive coupling term between magnons and photons based on the ME effect. In section \textbf{4}, we study photon modes.  In section \textbf{5}, we consider the system's semiclassical approach and nonlinear dynamical aspect. In section \textbf{6}, we analyze magnon-photon entanglement in the system and conclude the work.

\section{Magnetic subsystem}
\label{sec:Model}

\begin{figure}[ht!]
\centering
\includegraphics[width=\columnwidth]{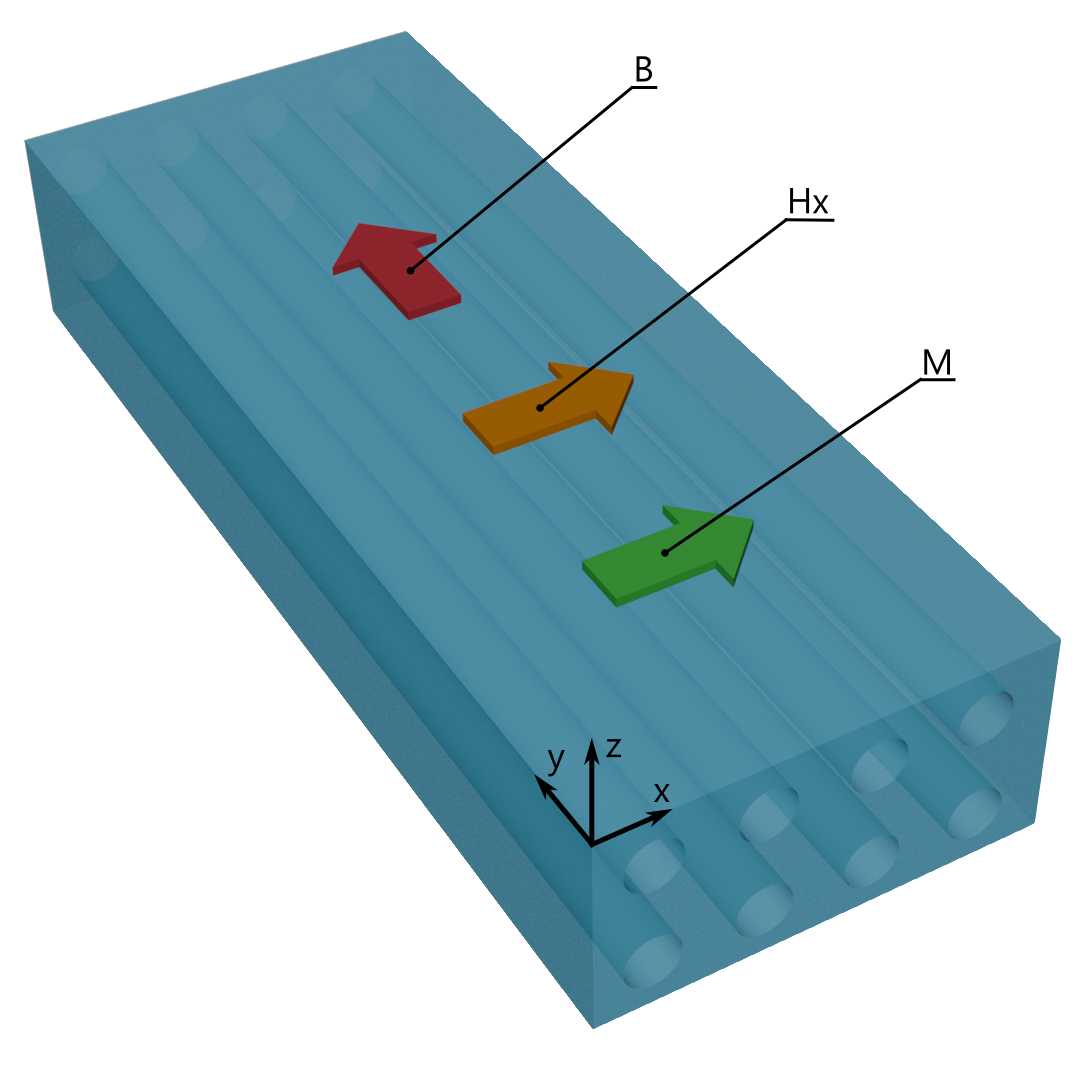}
\caption{ The photon-magnon crystal is based on the YIG sample, with the periodic chain composed of the air holes having a radius $a=0.46h$ ($h$ is the period of the structure). The perpendicular magnetocrystalline anisotropy (PMA) oriented in the $\textbf{z}$ direction $H_{an}=-K_{an}\sum\limits_n(\textbf{S}_n\textbf{z})^2$, and constant magnetic field $\textbf{H}=\textbf{x}H_x$ applied along the $\textbf{x}$ axis lead to the ground state magnetization along the $\textbf{x}$ axis.}
\label{fig:schematic}
\end{figure}

The effect of magnon condensation was discovered about twenty years ago \cite{demokritov2006bose}. Afterward, magnon condensation was observed in YIG \cite{schneider2020bose}. The interaction between magnons is weak at low densities of the condensate but becomes significant at higher magnon densities. Magnon condensate is stable when magnons repel each other. The stability of the magnon condensate is granted by dipole-dipole interaction between magnetic moments and is associated with the enhancement of the band depth \cite{mohseni2020bose}. We consider the YIG sample in the $XOY$ plane with the characteristic geometry $L_y\gg L_{x,z}$. Equilibrium magnetization is oriented along $\textbf{x}$ axis, parallel to the applied constant magnetic field  $H_0\textbf{x}$ as shown in Fig.\ref{fig:schematic}. 
The  Hamiltonian of the system magnetic subsystem reads: 
\begin{eqnarray}\label{new1}
\hat H_M=\hat H_{ex}+H_z+\hat H_{dip}+\hat H_{an}.
\end{eqnarray}
Here $H_z$ is the Zeeman interaction term, PMA is oriented in the $\textbf{z}$ direction $H_{an}=-K_{an}\sum\limits_i(\textbf{S}_i\textbf{z})^2$, $\hat H_{dip}$ is the dipole-dipole interaction term, $\textbf{S}_i$ is the vector of
spin operator at site $i$. The Hamiltonian Eq.(\ref{new1}) was studied in details in the recent work \cite{PhysRevResearch.6.L012011}. Using the Holstein-Primakoff spin-boson transformation authors obtained the effective Hamiltonian. Here we present part of this effective Hamiltonian corresponding to the intraband processes:
\begin{eqnarray}\label{magnetic Hamiltonian}
&&\hat H_{m}=\sum\limits_q\omega_q \hat m^\dag_q\hat m_q+ K\sum\limits_q\hat m^\dag_q\hat m_q\hat m^\dag_q\hat m_q.
\end{eqnarray}
Here, $\omega_q=\sqrt{\eta_1^2-\eta_2^2}-K$, with $\eta_1=Jq^2+\gamma_e(H_{0x}+2\pi M_sf_q)-K_{an}S$ and $\eta_2=2\pi M_sf_s-K_{an}S$ is the magnon frequency, which is in the order of several GHz, the amplitude of the effective spin per unit cell in YIG  $S=14$, $J$ is the exchange constant, $M_s$ saturation magnetization, $f_q$ form factor, $\gamma_e$ is gyromagnetic ratio, $\hat m^\dag$, $\hat m$ are magnon creation annihilation operators,  $q$ is the magnon wave vector. We are not interested in the whole magnon spectrum but particular magnons from the magnon condensate. Therefore we chose a particular magnon mode with wave vector $q\approx25\mu m^{-1}$ corresponding to the concave shape of the magnon band (a deep stable minimum) in the magnon condensate \cite{PhysRevResearch.6.L012011}. For the applied magnetic field $H_0=1$kOe and $K_{an}=0.8\mu eV$, the frequency of magnons in the condensate is equal to $\omega_c=2$GHz. We set dimensionless parameters by introducing the dimensionless time $t=\tau\omega_c$. The dimensionless Hamiltonian is given by $\hat H/\hbar\omega_c$. The second term in Eq.(\ref{magnetic Hamiltonian}) describes the intraband magnon-magnon interaction. The interaction constant $K$ in Eq.(\ref{magnetic Hamiltonian}) has the form \cite{PhysRevResearch.6.L012011}: $K=-\frac{\gamma_e\pi M_s}{SN}\left[(\alpha_1+\alpha_3)f_Q-2\alpha_2(1-f_{2Q})\right]-\frac{JQ^2}{2SN}(\alpha_1-4\alpha_2)+\frac{K_{an}}{2N}(\alpha_1+\alpha_3)$. Here $N$ is the total number of spin sites,  $Q$ is the wave vector of the condensate magnons, $\alpha_{1,2}$ are Bogoliubov transformation coefficients. The last term is the PMA $H_{an}=-K_{an}\sum\limits_i(\textbf{S}_i\textbf{z})^2$. The value of $K_{an}$ can be controlled by bismuth doping. For more details about magnon spectrum and intraband magnon-magnon interaction term we refer to the work \cite{PhysRevResearch.6.L012011}. We note that PMA and, consequently, nonlinearity constant $K$ can be controlled not only through bismuth doping but also using straintronics methods, i. e., by applying mechanical stress. For more details we refer to the review paper \cite{bukharaev2018straintronics} and references therein. 

\section{Magnon-photon coupling}
\label{sec:Magnon-photon coupling}

YIG is a material characterized by magneto-electric (ME) coupling. This fact is proved experimentally \cite{seki2012observation} and by density functional calculations \cite{PhysRevLett.106.247203}. The ME term for YIG has the form \cite{seki2012observation}:
\begin{eqnarray}\label{Vignale1}
H_{DM}=\textbf{P}\textbf{E}.
\end{eqnarray}
Here $E$ is the cavity electric field and 
\begin{eqnarray}\label{Vignale2}
\textbf{P}=-J\frac{ea_0}{E_{SO}}\sum\limits_i\textbf{e}_{i,i+1}\times\left(\textbf{S}_i\times \textbf{S}_{i+1}\right),
\end{eqnarray}
is the ferroelectric polarization, with $a_0$ being the distance the magnetic ions, which in dimensionless units we set to $a_0=1$, $J$ is the exchange interaction constant,  $\textbf{e}_{i,i+1}$ is the unit vector connecting adjacent spins $\textbf{S}_i$ and $\textbf{S}_{i+1}$, $E_{SO}$ is the spin orbit constant, $e$ is the electron charge. To shorten notations in what follows we introduce the magneto-electric coupling constant $g_{ME}=-J\frac{eq}{E_{SO}}$. The essence of our photon-magnon crystal is the coupling between photons and magnons through the effective DM term \cite{PhysRevB.96.054440,wang2020optical,khomeriki2016positive,PhysRevB.91.041408,PhysRevLett.125.227201}. Taking into account ME effect in YIG  ,
we couple quantized cavity field $\textbf{E}=\ii\mathbf{U}\sum_k\left(\hat{a}_k - \hat{a}^{\dag}_k\right)$ where $\hat a^\dag_k$, $\hat a_k$ are the photon creation and annihilation operators corresponding to the particular $k$ mode. The cavity mode function $\textbf{U}(\textbf{r})$ is the solution of the Helmholtz equation $\triangle U_{x,y,z}+k^{2}U_{x,y,z}=0$. In the experiment one drills the cylindrical holes in the YIG film along $\textbf{y}$ axis as shown in Fig.\ref{fig:schematic}. Due to the Coulomb gauge $\bnabla\cdot\textbf{U}=0$ and elongated along the $\textbf{y}$ axis structure of the photon-magnon crystal, i.e., $L_y\gg L_{x,z}$, $k_{x,y,z}=\pi/L_{x,y,z}$, $k_yy\approx 0$ the leading mode is $U_y=U_0\sin(k_xx)\sin(k_zz)$. Taking into account that only $U_y$ mode survives, in the scalar product Eq.(\ref{Vignale1})  $\textbf{P}\textbf{E}=g_{ME}E_y\sum\limits_i[\textbf{e}_{i,i+1}\times\left(\textbf{S}_i\times \textbf{S}_{i+1}\right)]_y$. According to the definition of the polarization vector Eq.(\ref{Vignale2}) this implies two possibilities in the choice of the vector components: $\textbf{e}_{i,i+1}^x\left(\textbf{S}_i\times \textbf{S}_{i+1}\right)_z$  or $\textbf{e}_{i,i+1}^z\left(\textbf{S}_i\times \textbf{S}_{i+1}\right)_x$. Taking into account that magnetization direction in our case is $\textbf{x}$ axis, relevant in our case is $\textbf{x}$ component of chirality \cite{PhysRevB.88.184404}: $\textbf{P}\textbf{E}=g_{ME}E_y\sum\limits_i\left(\textbf{S}_i\times \textbf{S}_{i+1}\right)_x$.
Therefore the ME Hamiltonian $H_{DM}$ simplifies and takes the form:
\begin{eqnarray}\label{Vignale3}
H_{DM}=\frac{D_0}{2}\sum\limits_{i,k}\left(\textbf{S}_i\times \textbf{S}_{i+1}\right)_z\left(\hat{a}_k - \hat{a}^{\dag}_k\right).
\end{eqnarray}
Here we introduced the effective DMI constant $D_0=U_yg_{ME}$. Because quantization axis is along $\textbf{x}$ we perform the standard rotation \cite{tiablikov2013methods}: 
$S^j_i=e_j(1/2-\hat m^\dag_i\hat m_i)+A_j\hat m_i+A_j^*\hat m^\dag_i$, where 
$e_j$, $j=x,y,z$ is the $j$ component of the unit vector $\textbf{e}$ directed along the quantization axis, and coefficients have the form: $A_x=e^{-i\Theta}(1-e_z)/4-e^{i\Theta}(1+e_z)/4$, $A_y=ie^{i\Theta}(1+e_z)/4+e^{-i\Theta}(1-e_z)/4$, $A_z=\sqrt{1-e_z^2}/2$, $\tan\Theta=e_y/e_x$. After performing the Fourier transform into the momentum space we deduce:
\begin{eqnarray}\label{Vignale3}
\hat H_{DM}=D_0\sum\limits_{q,k}\sin(qa_0)\hat m^\dag_q\hat m_q\left(\hat{a}_k-\hat{a}^{\dag}_k\right),
\end{eqnarray}
where in what follows we set $a_0=1$. 

\section{Photonic cavity}
\label{sec:photonic cavity}

\begin{figure}[!t]
\centerline{\includegraphics[width=\columnwidth]{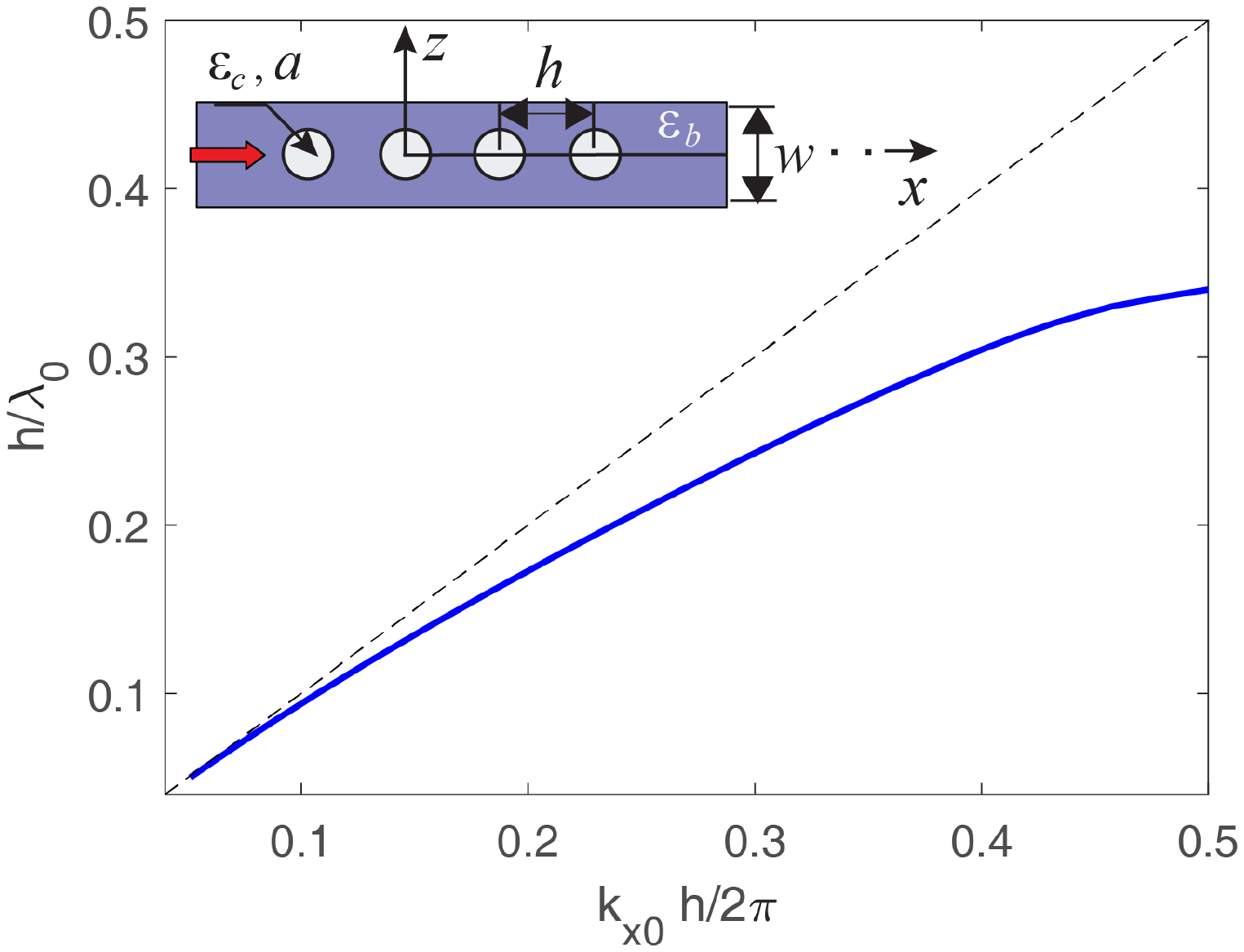}}
\caption{Brillouin diagram of the fundamental TM mode of the periodic chain composed of the air-holes having a radius $a=0.46h$ ($h$ is period of the structure) and situated in the background medium with the relative dielectric permittivity $\epsilon_b=5$. The thickness of the film is $w=h$, and $\lambda_0$ is the wavelength in the free space. The dashed line defines the light line. To control the frequency of the photons, we tune the wavelength at the fixed period of the structure $h$. We are interested in the GHz frequency range and the desired frequency can be easily determined from the dispersion diagram that is shown in non-dimensional parameters. For $h=10$mm, photon frequency is the order of $\omega_f=5$GHz. }
\label{fig2}
\end{figure}
The geometry of the problem is shown as inset in Fig. 2. The structure composed of periodically distributed air-holes with circular section and radius $a$ and relative dielectric permittivity $\epsilon_c$ ($\epsilon_c=1$) is geometrically invariant along the $y$-axis and infinite and periodic with a period $h$ along the $x$-axis. The surrounding medium is material with a relative dielectric permittivity $\epsilon_b$ and permeability $\mu_0$. The structure is located in the free space. Applying the addition theorem for the cylindrical functions and using the boundary condition on the surfaces of the circular rod at the global origin, a simple dispersion equation for the eigenmodes is obtained in the following form:
\begin{align}
\mathrm{det}[\bf{I}-{\mathbf{T}}(\mathit{k}_\textit{c}\it{a},\mathit{k}_\textit{b}\it{a})    
\bf{L}(\mathit{k}_\textit{b}\mathit{h},
\mathit{k}_\text{$x$0}\mathit{h})]=\text{0},
\end{align}
where $\bf{I}$ is the unit matrix. The transition matrix ${\mathbf{T}}(\mathit{k}_\textit{c}\it{a},\mathit{k}_\textit{b}\it{a})$ (T-matrix), which describes the nature of this particular scatterer, can be written in a compact form in terms of a diagonal matrix. It is given for the TM wave ($E_y, H_x, H_z$) and the TE wave ($H_y, E_x, E_z$) in the closed form:
\begin{align}
 T_m^{\rm TE}=\frac{-\sqrt{\epsilon_b}J_{m}\tilde{J}_{m}'
+\sqrt{\epsilon_c} {J_{m}'\tilde{J}_{m}}}
{\sqrt{\epsilon_b}H^{(1)}_{m}\tilde{J}_{m}'
-\sqrt{\epsilon_c}{{H^{(1)}_{m}}'\tilde{J}_{m}}}
\end{align}
\begin{align}
 T_m^{\rm TM}=\frac{-\sqrt{\epsilon_b}\tilde{J}_{m}J_{m}'
+\sqrt{\epsilon_c}{J_{m}\tilde{J}_{m}'}}
{\sqrt{\epsilon_b}{H^{(1)}_{m}}'\tilde{J}_{m}
-\sqrt{\epsilon_c}{H^{(1)}_{m}\tilde{J}_{m}'}}
\end{align}
Here $J_m(k_b a)$, $\tilde{J}_m(k_c a)$, and $H^{(1)}_m(k_b a)$ are Bessel and Hankel functions of the $m$-th order, respectively, the prime notation denotes their derivatives concerning the arguments,  $k_b=\omega\sqrt{\mu_0\epsilon_0\epsilon_b}$ and 
$k_c=\omega\sqrt{\mu_0\epsilon_0\epsilon_c}$ are the wavenumbers in the background medium and air-hole respectively, $\omega$ is the angular frequency. Note that the Fresnel matrices are the diagonal matrices implemented in the analysis on the upper ($w=h$) and lower ($w=0$) boundaries of the structure (see inset in Fig.\ref{fig2}).
The modal field in a periodic chain can be expressed in terms of an infinite series of space harmonics, each having a complex propagation wavenumber $k_{xn}=\beta_n+i\alpha$, with $\beta_n = k_{x0} + 2\pi n/h$ and $n=0,\pm 1,\pm 2,\cdot\cdot\cdot$. The $n$-th space-harmonic along the transverse $z$-axis behaves as $e^{jk_{zn}|z|}$, where
$k_{zn}=\sqrt{{k^2_0}-{k^2_{xn}}}$ is the relevant wavenumber. The formalism presented in the paper allows for the appropriate choice of the proper or improper spectral determination (positive or negative sign for $\text{Im}(k_{yn})$, respectively) for each space harmonic \cite{jandieri20191,jandieri2020modal}. \vspace{0.3cm}\\
\textbf{Lattice-Sums.}
Lattice Sums that appear in the dispersion equation for the eigenmodes play a crucial role in our formalism. It is expressed through the infinite series of the Hankel functions. It does not depend on the polarization of the incident field, the location of the observation points, and the material properties of the scatterers. Hence, it represents a very efficient technique for solving periodic structures' electromagnetic scattering problems. The $m$-th component of the Lattice Sums is given in the following form \cite{jandieri20191,jandieri2020modal,yasumoto1999efficient}:
\begin{eqnarray}
&&L_m=\sum_{n=1}^\infty H^{(1)}_{m}(k_bnh)[\exp(ink_{x0}h)+\nonumber\\
&&(-1)^{m}\exp(-ink_{x0}h)].
\end{eqnarray}
The series converges very slowly for the real $k_{x0}$ and diverges for the complex $k_{x0}$. In order to speed up the calculation, applying the Ewald transformation, the Lattice Sums can be calculated as a sum of the spectral and the spatial series \cite{jandieri20191,jandieri2020modal}. The details of derivation are omitted in the work, we give here the final expressions for each term of the Lattice Sums: 
\begin{eqnarray}
&&L_m^{spectral}=\frac{1}{k_{zn}}\times\nonumber\\
&&\sum_{n=-\infty}^\infty \left( \frac{k_{xn}}{k_b} \right)^{m}\sum_{q=0}^{[m/2]}(-1)^q \binom{m}{2q}\ 
 \left( \frac{k_{zn}}{k_{xn}} \right)^{2q}\ C_{q,n},
\end{eqnarray}
with
\begin{eqnarray}
&&C_{q,n}=\frac{1}{k_{zn}}\times\nonumber\\
&&\text{erfc}\left( -i\frac{hk_{zn}}{2E_{spl}} \right)\-
\frac{e^\frac{h^2{k^2_{zn}}}{4E_{spl}^2}}{k_{zn}}
\sum_{s=1}^q\frac{\left(-i\frac{hk_{zn}}{2E_{spl}} \right)^{1-2s}}{\Gamma(1.5-s)},
\end{eqnarray}
and
\begin{eqnarray}
 L_m^{spatial}=\delta_{m0}
 \left[-1-\frac{i}{\pi}Ei{\left(\frac{k_b^2h^2}{4E_{spl}^2} \right)}  \right]+
\frac{2^{m+1}}{i\pi}\sum_{n=1}^\infty [\exp(ink_{x0}h) \\ 
+(-1)^m exp(-ink_{x0}h)]
\left( \frac{n}{k_0h} \right)^m\  
\int_{E_{spl}}^{\infty}\,d\eta \
\frac{\exp\left (-n^2\eta^2+\frac{h^2k_b^2}{4\eta^2} \right)}{\eta^{-2m+1}}.
\end{eqnarray}
Here $\text{erfc(...)}$ is the error function. 
In the numerical results that follow, the Ewald splitting parameter $E_{spl}$=$\sqrt{\pi}$. The proposed formalism gives Gaussian convergence even in the case of complex and leaky waves \cite{jandieri20191,jandieri2020modal,yasumoto1999efficient}. Yasumoto presented another efficient method based on a simple trapezoidal integration formulae \cite{yasumoto1999efficient}. Although the authors mention that their method works only for the real propagation constant $k_{x0}$, the identity theorem in the complex analysis allows us to calculate the Lattice Sums for the complex $k_{x0}$. However, the main advantage of the formalism in this work is that it allows for the appropriate choice of the spectral determination for each space harmonic (choice of a sign for $k_{zn}$), which is not the case in \cite{yasumoto1999efficient}. Although for the mode analysis, we calculate only the bound modes ($k_{x0}$ is a real value), the determination of each space-harmonic is essential for a flexible design of Fabry-Perot cavities, EBG-based resonators, open waveguides or plasmonic guiding devices with intrinsic losses.

\section{The total Hamiltonian}
\label{sec:total hamiltonian}

The total Hamiltonian of the system consists of the Hamiltonian magnetic subsystem $\hat H_m$ Eq.(\ref{magnetic Hamiltonian}), Hamiltonian of the photonic subsystem $\hat H_f=\sum_k\omega(k)\hat a_k^\dag\hat a_k$ and interaction Hamiltonian between photons and magnons $\hat H_{DM}$ Eq.(\ref{Vignale3}). While the magnon spectrum is broad, we are interested only in the magnons from the magnon condensate. As was shown in the recent work
\cite{PhysRevResearch.6.L012011} those magnons have a particular wave vector $q_c\approx25\mu m^{-1}$
which corresponds to the deep stable minimum in the magnon spectrum. For the applied magnetic field $H_0=1$kOe and $K_{an}=0.8\mu eV$, the frequency of magnons in the condensate is equal to $\omega_c\approx2$GHz. Dispersion of the cavity photons was studied in the section \ref{sec:photonic cavity}. We are interested only in the resonant photons, which are strongly coupled with the magnons from the magnon condensate and have similar frequency $\omega_f(k_c)\approx \omega_c(q_c)$. The resonant photons, together with the magnons from the magnon condensate, form the photonic-magnonic crystal. The Hamiltonian of the system of interest takes the form:
\begin{eqnarray}\label{total Hamiltonian}
&&\hat H_{tot}=\omega_f\hat a^\dag\hat a+\omega \hat m^\dag\hat m+ K\hat m^\dag\hat m\hat m^\dag\hat m+\nonumber\\
&&\ii B\left(\hat m^\dag e^{-\ii\omega_0t}-\hat m e^{\ii\omega_0t}\right)+ \ii D\hat m^+\hat m\left(\hat a-\hat a^\dag\right).
\end{eqnarray}
Here $B$, $\omega_0$ are the amplitude and frequency of the time-dependent magnetic field applied along $y$ axis, $D=D_0\sin(q_c)$. All constants in Eq.(\ref{total Hamiltonian}) are dimensionless and are related to the dimensional constants via 
$K\rightarrow K'/\hbar\omega_c$, 
$D\rightarrow D'/\hbar\omega_c$,
$\omega_f\rightarrow \omega_f'/\omega_c$, 
$\omega\rightarrow  \omega'/\omega_c$, 
$B\rightarrow \gamma_eB'/\omega_c$.

\section{Nonlinear dynamics}
\label{sec:MNonlinear dynamics}
Nonlinearity can substantially influence the dynamic properties of of the system
\cite{chotorlishvili2010quantum,PhysRevE.70.026219,PhysRevE.71.056211}. Following \cite{PhysRevLett.121.203601, PhysRevB.107.115126}, we derive the set of quantum
Langevin equations (QLEs) describing the system:
\begin{eqnarray}\label{equations of motion}
&&\frac{d\hat a(t)}{dt}=-i\left[\hat a(t),\hat H_{tot}\right]-\gamma_f\hat a(t)+\sqrt{2\gamma_f}\hat a_{in}(t),\\
&&\frac{d\hat m(t)}{dt}=-i\left[\hat m(t),\hat H_{tot}\right]-\gamma_m\hat m(t)+\sqrt{2\gamma_m}\hat m_{in}(t).\nonumber
\end{eqnarray}

\begin{figure}[]
\centerline{\includegraphics[width=\columnwidth]{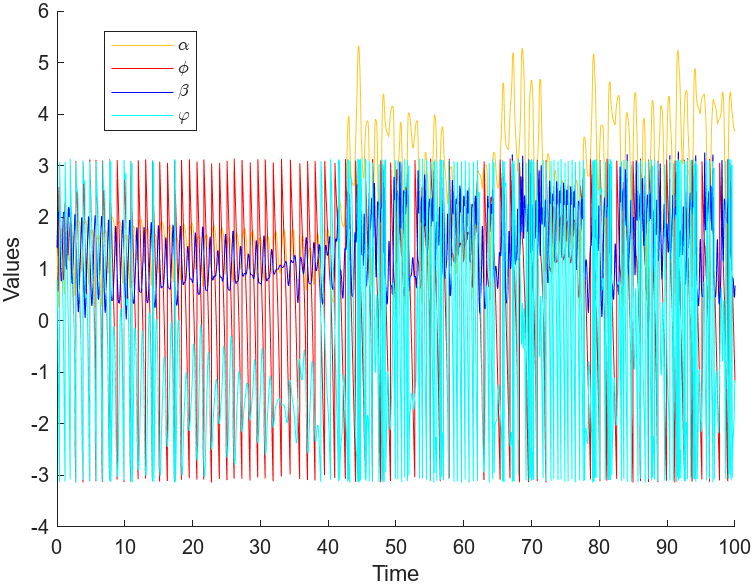}}
\caption{Dependence of expectation values of the mean magnon and photon operators  $\beta$ (blue), $\alpha$ (orange), and phases $\varphi$ (red), $\phi$ (cyan) on time. The result is obtained via the numerical integration of Eq.\ref{semi-classical phase}. The values of the parameters read: $D = 2.0, B = 3.5, K = 1.6, \omega = 1.5, \omega_0 = 1.5,\, \omega_f = 5.0, \gamma_f = \gamma_m = \gamma_0 = 0.02$. As we can see, mean magnon and photon numbers change irregularly over time, which is a signature of complex nonlinear dynamics.}
\label{numeric_solve_eq17}
\end{figure}

\begin{figure}[]
\centerline{\includegraphics[width=\columnwidth]{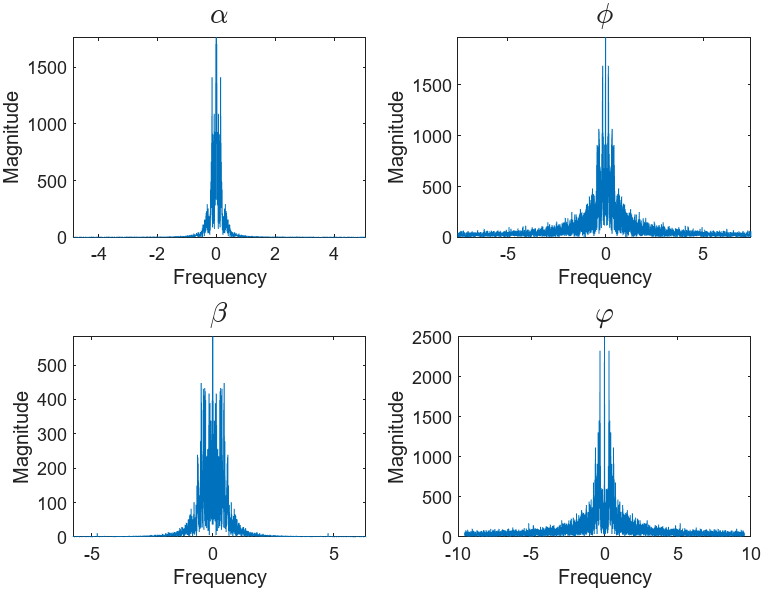}}
\caption{Plots of the fast Fourier transform for $\alpha(t), \phi(t), \beta(t)$ and $\varphi(t)$ 
calculated based on Eq.\ref{A.23}. The values of the parameters read: $D = 2.0, B = 3.5, K = 1.6, \omega = 1.5, \omega_0 = 1.5 \omega_f = 5.0, \gamma_f = \gamma_m = \gamma_0 = 0.02$. }
\label{2numeric_solve_eq17}
\end{figure}

Here, $\gamma_f$ and $\gamma_m$ are damping constants for cavity photons and magnons, respectively. For a high fineness photon-magnon crystal, damping constants are small. Here 
$a_{in}(t)$, $m_{in}(t)$ are input noise bosonic operators for the photonic and magnonic subsystems respectively, characterized by the correlation functions $\langle a_{in}^\dag(t)a_{in}(t')\rangle=N_a(\omega_f)\delta(t-t')$, and $\langle m_{in}^\dag(t)m_{in}(t')\rangle=N_m(\omega)\delta(t-t')$, where $N(\omega)=\left[\exp[\hbar\omega/k_BT]-1\right]^{-1} $ is the Bose–Einstein distribution function.
In Eq.(\ref{equations of motion}) we implement semi-classical and rotating wave approximations through the transformations $ O=\langle\hat O\rangle$, $(\hat a-\hat a^+)\hat m=(a-a^*)m$, $\langle\hat m^+\hat m\rangle=|m|^2$, $\hat m\rightarrow \hat m e^{-i\omega_0t}$, $\hat 
m_{in}\rightarrow m_{in}e^{-i\omega_0t}$ and $a\equiv |a|e^{i\phi}=\alpha e^{i\phi}$, $m\equiv |m| e^{i\varphi}=\beta e^{i\varphi}$ and deduce:
\begin{eqnarray}\label{semi-classical phase}
&&\dot{\alpha}=-D\beta^2\cos\phi-\gamma_f\alpha,\nonumber\\
&&\dot{\phi}=-\omega_f+\frac{D\beta^2}{\alpha}\sin\phi,\nonumber\\
&&\dot{\beta}=-\gamma_m\beta+B\cos\varphi,\nonumber\\
&&\dot{\varphi}=\omega-\omega_0+2D\alpha\sin\phi-2K\beta^2-\frac{B\sin\varphi}{\beta}. 
\end{eqnarray}

\begin{figure}[]
\centerline{\includegraphics[width=\columnwidth]{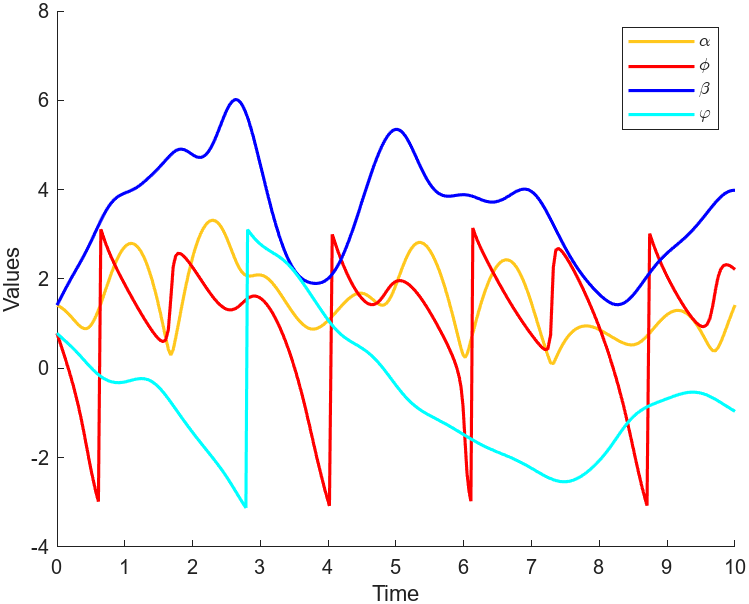}}
\caption{Dependence of expectation values of the mean magnon and photon operators  $\beta$ (blue), $\alpha$ (orange), and phases $\phi$ (red), $\varphi$ (cyan) on time plotted for a shorter time interval. The result is obtained via the numerical integration of Eq.\ref{semi-classical phase}. The values of the parameters read: $D = 0.4, B = 3.5, K = 0.06, \omega = 1.5, \omega_0 = 1.5 \omega_f = 5.0, \gamma_f = \gamma_m = \gamma_0 = 0.1$. As we can see, mean magnon and photon numbers change irregularly over time, and that is a signature of complex nonlinear dynamics. The phases change in time abruptly.}
\label{3numeric_solve_eq17}
\end{figure}

In what follows, our main concern is the nonlinear set of equations Eq.(\ref{semi-classical phase}). 
We note that there is correspondence between variables in Eq.(\ref{semi-classical phase}) and momentum coordinate variables used in the standard approach  \cite{PhysRevLett.121.203601, PhysRevB.107.115126} ($q_m=\sqrt{2}\beta\cos\varphi$, $p_m=\sqrt{2}\beta\sin\varphi$), ($q_a=-\frac{\sqrt{2}D\gamma_f}{\omega_f^2+\gamma_f^2}\beta^2$, $p_a=-\frac{\sqrt{2}D\omega_f}{\omega_f^2+\gamma_f^2}\beta^2$). When studying entanglement properties, most of the authors linearize Eq.(\ref{equations of motion}), see e.g., \cite{PhysRevLett.121.203601, PhysRevB.107.115126} and references therein. In the first step, one implements the following ansatz to the magnon and photon operators: $\langle\hat m^+\hat m\rangle=\langle\hat m^+\rangle\langle\hat m\rangle$, $\hat m=\langle\hat m\rangle+\delta\hat m$, $\hat m^+=\langle\hat m^+\rangle+\delta\hat m^+$, $\hat a=\langle\hat a\rangle+\delta\hat a$, $\hat a^+=\langle\hat a^+\rangle+\delta\hat a^+$. The second step is the transition to the canonical variables: $\hat x_m=(\hat m^++ \hat m)/2$,  $\hat p_m=i(\hat m^+- \hat m)/2$, $\hat x_a=(\hat a^++ \hat a)/2$,  $\hat p_a=i(\hat a^+- \hat a)/2$. The expectation values $\langle\hat m\rangle$, $\langle\hat a\rangle$ are replaced by their semi-classical mean values, and the system of equations  Eq.(\ref{equations of motion}) is linearized in terms of $\delta\hat x_m$, $\delta\hat x_p$, $\delta\hat x_a$, $\delta\hat p_a$ in the vicinity of the steady state. The steady state after linearization defines the mean magnon and photon numbers. However, the steady state, in essence, corresponds to the fixed point of the nonlinear dynamical system, and fixed points can be of different types, such as Saddle Points, Stable or unstable Spirals, and Nodes. The same we expect for Eq.(\ref{semi-classical phase}). Typically, nonlinear systems have different types of fixed points, and the different regions in the phase space are divided by those fixed points. In what follows, we explore the phase space of the fixed points and show that magnon-photon entanglement depends on the region of the phase space, i.e., on the character of the fixed point. We proceed with nonlinear system Eq.(\ref{semi-classical phase}) and implement methods of the theory of Dynamical system \cite{katok1995introduction}. Due to its nonlinearity, we expect the general solution of Eq.(\ref{semi-classical phase}) to be irregular, far beyond the steady state. In Fig.\ref{numeric_solve_eq17}, we see such irregular behavior. The finite broadening of the Fourier transform signal shown in Fig.\ref{2numeric_solve_eq17} confirms the presence of chaos in the system. 
In Fig.\ref{3numeric_solve_eq17}, we see the anharmonic character of the dynamics and abrupt changes in the phases. The description of the entanglement of bosonic modes is based on the covariance matrix (either time-dependent or steady) obtained through the linearization of equations of motion in the vicinity of the steady state. However, we note that fixed points of the nonlinear dynamical system can be unstable, e.g., unstable spiral characterized by positive large Lyapunov exponents of the phase trajectories. Linearization of dynamics near such unstable fixed points is an invalid procedure since it leads to the exponential growth of initial infinitesimal errors. The long-term time behavior of the system becomes unpredictable since any two-phase trajectories starting from almost similar initial conditions diverge in time. Therefore, before deriving the covariance matrix, we need to explore the phase space of the system and explore the character of the fixed points in detail. The fixed point of the dynamical system Eq.(\ref{semi-classical phase}) is given by:
\begin{eqnarray}\label{the stationary case}
&&\alpha^2=\frac{D\beta^4}{\omega_f^2+\gamma_f^2},\nonumber\\
&&\phi=-\arctan(\omega_f/\gamma_f),\nonumber\\
&&\Delta_m\left[ \gamma^2_m+(\omega_0-\omega+\Delta_m)^2\right] =\left(\frac{2D^2\omega_f}{\omega_f^2+\gamma_f^2}-2K\right)B^2,\nonumber\\
&& \tan\varphi=\frac{\omega_0-\omega+2D\alpha\sin\phi-2K\beta^2}{\gamma_m}.
\end{eqnarray}
Here $\Delta_m=\beta^2\delta_K$, $\delta_K=\left(\frac{2D^2\omega_f}{\omega_f^2+\gamma_f^2}-2K\right)$.
Eq.(\ref{the stationary case}) is the essence of a nonlinear set of equations with respect to the variables: $\alpha(D, B, K)$, $\beta(D, B, K)$,  $\phi(D, B, K)$, $\varphi(D, B, K)$. The amplitude-frequency characteristics of the nonlinear resonance follow the singular points where the derivative with respect to the detuning $\delta\omega=\omega_0-\omega$ turns to be zero \cite{rabinovich2012oscillations}
\begin{eqnarray}\label{where the derivative}
\frac{d}{d\delta\omega}\left\lbrace\Delta_m\left[ \gamma^2_m+(\Delta_m+\delta\omega)^2\right]\right\rbrace =0.
\end{eqnarray}
From Eq.(\ref{where the derivative}), we derive the equation of the amplitude-frequency characteristics. 
\begin{eqnarray}
\frac{d\beta^2}{d\delta\omega}=-\frac{2\beta (\delta_k\beta^2+\delta\omega)}{3(\delta_k\beta^2)^2+4\delta_k\beta^2\delta\omega+\gamma_m^2+\delta\omega^2}.
\end{eqnarray}
The singular points correspond to the case $3(\delta_k\beta^2)^2+4\delta_k\beta^2\delta\omega+\gamma_m^2+\delta\omega^2=0$. i.e., frequencies:
\begin{eqnarray}\label{frequencies}
\delta\omega_{1,2}=-2\delta_k\beta^2\mp\sqrt{(\delta_k\beta^2)^2-\gamma_m^2}.
\end{eqnarray} 
It is easy to see that in the limit of small $\delta_k\ll 1$ in the fixed point
\begin{eqnarray}\label{weak1}
\beta^2_{01}=\frac{B^2}{\delta\omega^2+\gamma_m^2}.
\end{eqnarray}
Then the set of parameters [$\alpha_{01}^2=D\beta_{01}^4/(\omega_f^2+\gamma^2_f)$, $\phi=-\pi/2$, $\varphi_{01}=\arctan((\delta\omega-2D\alpha_{01}-2K\beta_{01}^2)/\gamma_m)$], we easily find from Eq.(\ref{the stationary case}).
In the nonlinear case our interest concerns two driving frequencies of the external magnetic field 
\begin{eqnarray}\label{frequencies magnetic field}
\omega_0^{(1,2)}=\omega-2\delta_k\beta^2\mp\sqrt{(\delta_k\beta^2)^2-\gamma_m^2}.
\end{eqnarray}
In the nonlinear case, the set of the amplitude-frequency characteristic equations 
\begin{eqnarray}\label{Amplitude-frequency characteristic equation }
&&\delta\omega_{1,2}=-2\delta_k\beta^2\mp\sqrt{(\delta_k\beta^2)^2-\gamma_m^2},\nonumber\\
&&\beta^2\left[\gamma_m^2+\left(\beta^2\delta_k+\delta\omega\right)\right]=B^2, 
\end{eqnarray}
has two different solutions $\beta_{1,2}^{\pm}(\delta\omega)$ for each $\delta\omega=\omega_0^{(1,2)}-\omega$ as it is shown in the $\beta(\delta\omega)$ plot below.
\begin{figure}[]
\centerline{\includegraphics[width=\columnwidth]{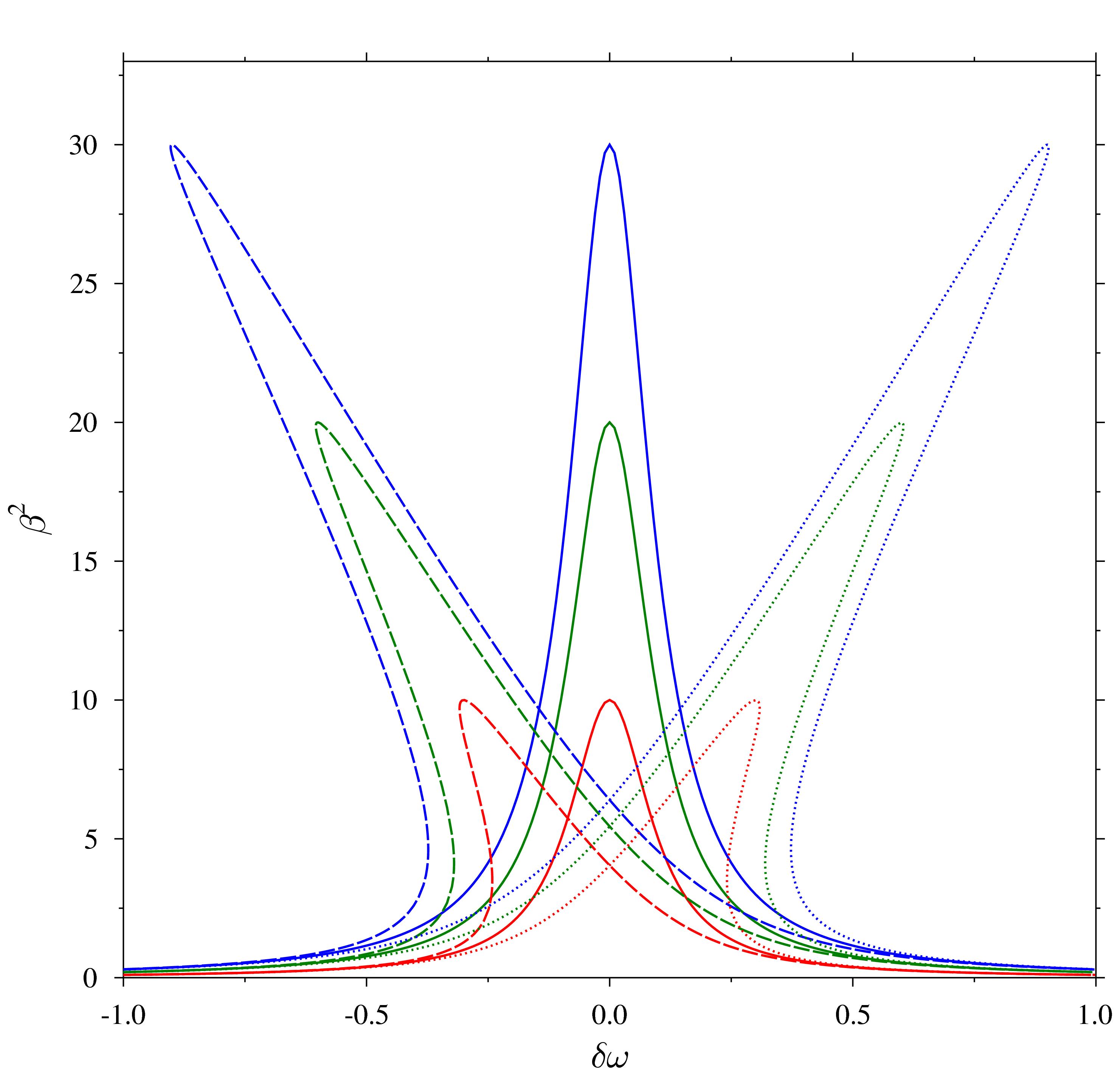}}
\caption{Dependence of the steady-state mean magnon number $n_m=\beta^2$ on the detuning $\delta\omega=\omega_0-\omega$ for the different values of the dimensionless magnetic field $B\equiv \gamma_e B/\omega$ and $\delta_K=\left(\frac{2D^2\omega_f}{\omega_f^2+\gamma_f^2}-2K\right)$. In particular, solid lines correspond to $\delta_k=0$, while dashed lines to $\delta_k=0.03$ and the dot lines to $\delta_k=-0.03$. The red color corresponds to the dimensionless magnetic field $B\equiv \gamma_e B/\hbar\omega$,  $B^2=0.1$, the green color to $B^2=0.2$, and the blue color to $B^2=0.3$.}
\label{fig3}
\end{figure}
The dependence of the steady-state mean magnon number on the detuning $n_m=\beta^2(\delta\omega)$ is plotted in Fig.\ref{fig3}. For 
\begin{eqnarray}
\delta_K=\left(\frac{2D^2\omega_f}{\omega_f^2+\gamma_f^2}-2K\right)=0, 
\end{eqnarray}
we have a linear regime in the system. All three solid lines show that the steady-state mean magnon number depends on the magnetic field $B$ and increases with $B$. Condition $\delta_k=0$ defines the specific values of the external electric field $E=D/g_{\text{ME}}$ when the dynamic in the system is linear. For nonzero values of $\delta_k$, dynamic in the system is non-linear, and for every single value of detuning $\delta\omega$, we have two specific values of the steady-state  mean magnon numbers
$[\beta_{1,2}^{\pm}(\delta\omega)]^2=n_{m}^{\pm}(1,2)=\langle\hat m^+\hat m\rangle^{(\pm)}_{1,2}$. Indexes $1,2$ here correspond to the negative and positive detuning cases. The jumps between magnon numbers $\Delta\beta^2=[\beta_{1,2}^{+}(\delta\omega)]^2-[\beta_{1,2}^{-}(\delta\omega)]^2=\Delta n_m(1,2)$ depend on the values of detuning. Besides, we clearly see the mirror symmetry $\delta_k\rightarrow-\delta_k$ between dashed and dotted lines. In what follows, we explore the magnon-photon entanglement in the vicinity of both bifurcation points $\beta_{1,2}^{\pm}(\delta\omega)$. We note that the fixed point value is related to the mean magnon number in the system $[\beta_{1,2}^{\pm}(\delta\omega)]^2=n_{m}^{\pm}(1,2)=\langle\hat m^+\hat m\rangle^{(\pm)}_{1,2}$. Therefore, bifurcation jumps $\Delta\beta^2=[\beta_{1,2}^{+}(\delta\omega)]^2-[\beta_{1,2}^{-}(\delta\omega)]^2=\Delta n_m(1,2)$ define the mean magnon number jumps in the system near the bifurcation points. Bifurcations in the system and jumps in the steady-state mean magnon numbers exert a certain impact on the magnon-photon entanglement. Our next aim is to explore steady state mean magnon number as a function of the magnetic field. 
\begin{figure}[]
\centerline{\includegraphics[width=\columnwidth]{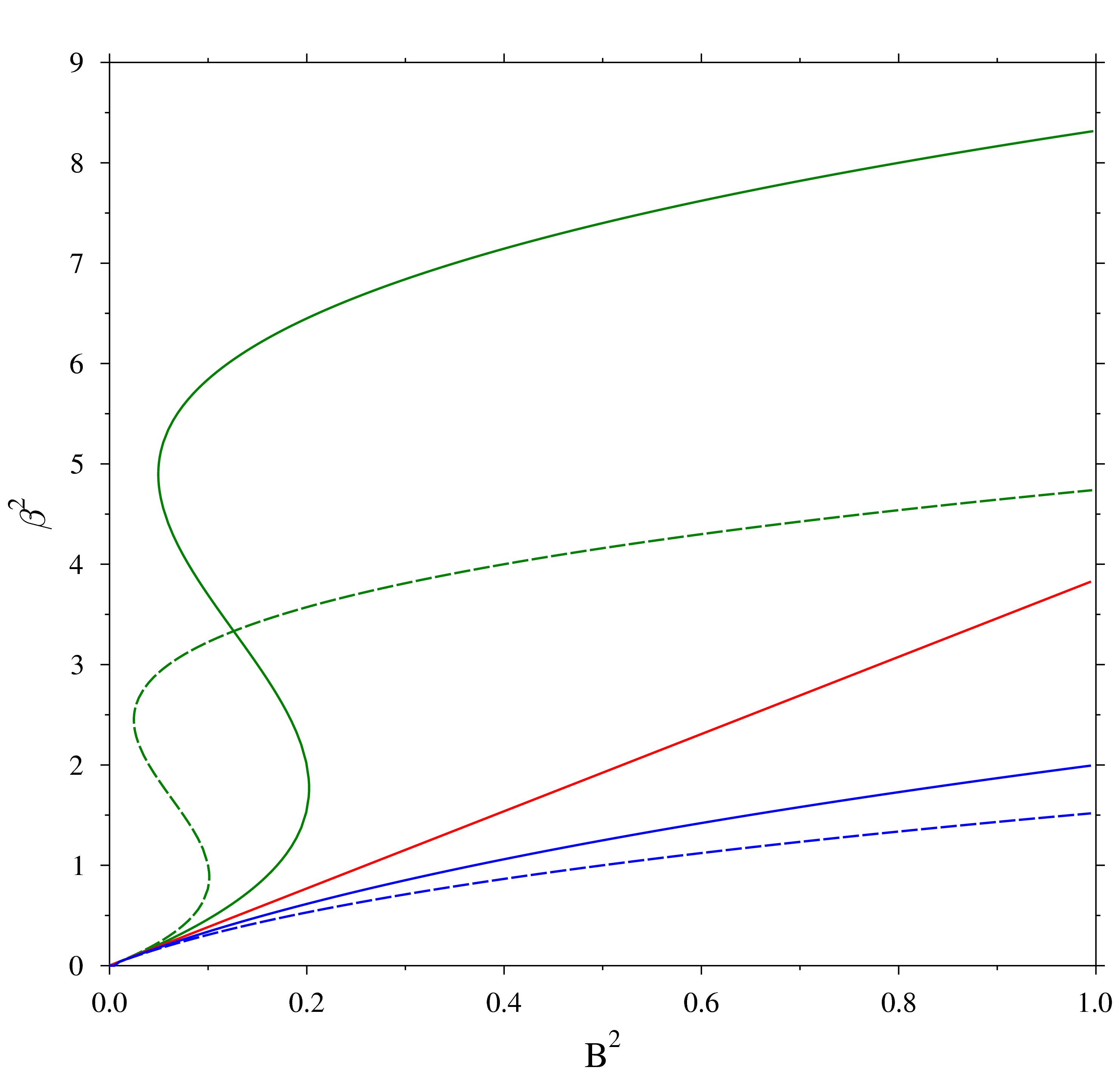}}
\caption{Dependence of the steady-state mean magnon number $n_m=\beta^2$ on the external dimensionless magnetic field $B\equiv \gamma_e B/\omega$ for the fixed detuning $\delta\omega=\omega_0-\omega=0.5$ and the different values of $\delta_K=\left(\frac{2D^2\omega_f}{\omega_f^2+\gamma_f^2}-2K\right)$. In particular, the red solid lines corresponds to $\delta_k=0$, while the green solid and green dashed lines to $\delta_k=-0.1$ and $\delta_k=-0.2$ respectively. The blue solid line corresponds to $\delta_k=0.1$ and the blue dashed line to $\delta_k=0.2$. Because of the $n_m(B^2,-\delta_k,\delta\omega)=n_m(B^2,\delta_K,-\delta\omega)$, the plot for the negative detuning $\delta\omega=-0.5$ is the same upon the exchange of green and blue colors.}
\label{fig4}
\end{figure}
\begin{figure}[]
\centerline{\includegraphics[width=\columnwidth]{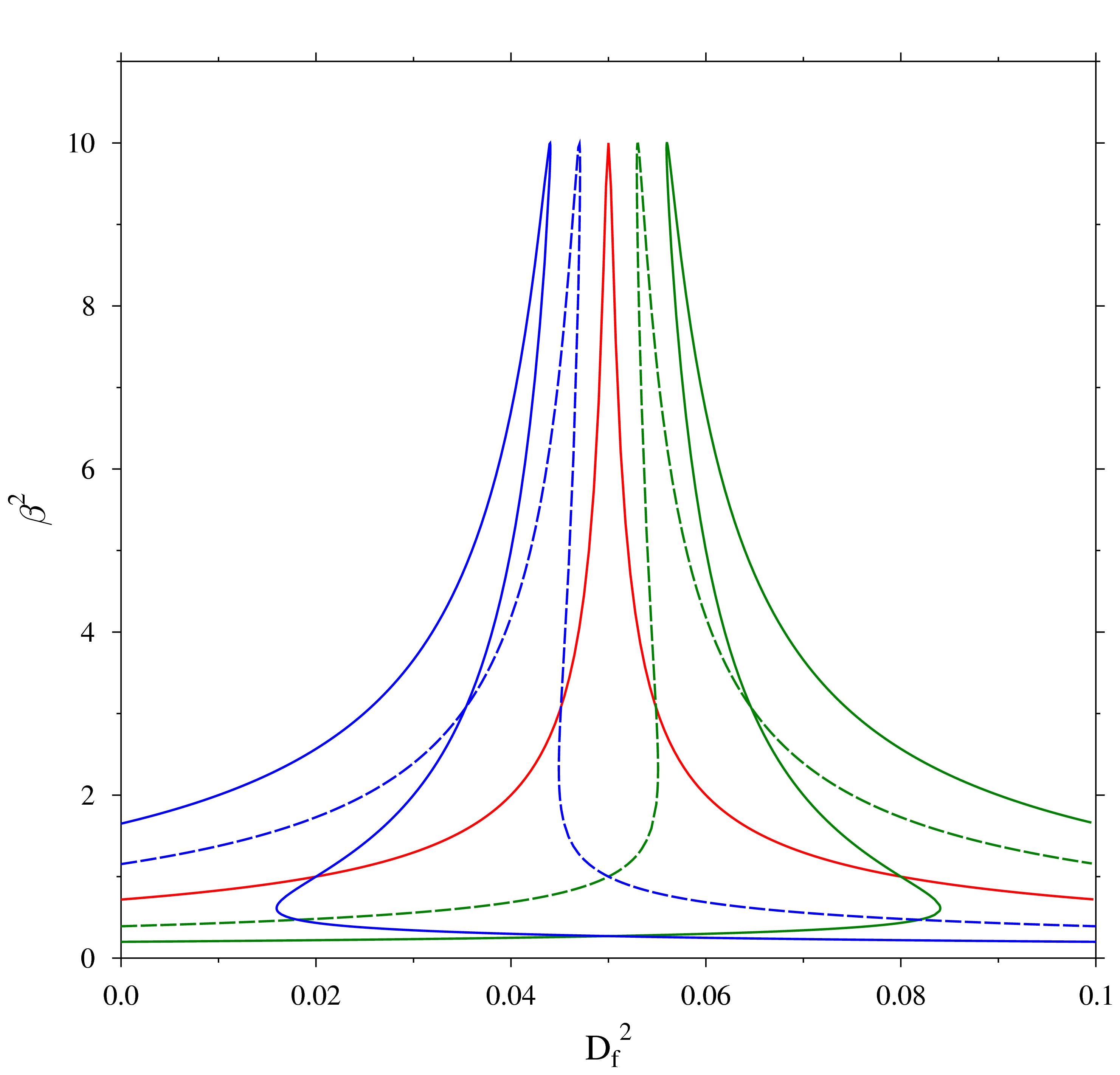}}
\caption{Dependence of the steady-state mean magnon number $n_m=\beta^2$ on the external electric field $E=D_f/g_{\text{ME}}$, $D_f=D/\sqrt{\omega_f^2+\gamma^2}$ for the constant magnetic field $B^2=0.1$ and the different values of the detuning: $\delta\omega=0$ (red solid), $\delta\omega=-0.6$ (green solid), $\delta\omega=-0.3$ (green dashed), $\delta\omega=0.6$ (blue solid). }
\label{fig5}
\end{figure}
Taking into account Eq.(\ref{the stationary case}) 
we deduce the closed polynomial equation for the equilibrium magnon number
\begin{eqnarray}\label{closed equation for the equilibrium magnon number}
&&P(\beta)=\delta_k^2\beta^6+2\delta\omega\delta_k\beta^4+\nonumber\\
&&(\gamma_m^2+\delta\omega^2)\beta^2-B^2=0.
\end{eqnarray}
Analyses of the roots of $P(\beta)$ and derivatives $\partial_{x}P(x)=0$, $\partial_{xx}P(x)=0$, $x\equiv \beta^2$ show the magnon number instability interval as a function of the magnetic field $B$. 
Namely, the equation
\begin{eqnarray}\label{czeslaw1}
\frac{dP(\beta^2)}{d\beta^2}=3\delta_k^2\beta^4+4\delta\omega\delta_k\beta^2+(\gamma_m^2+\delta\omega^2)=0,
\end{eqnarray}
for $\delta\omega^2\geqslant3\gamma^2_m$ has two roots
\begin{eqnarray}\label{czeslaw2}
&&\beta_{1,2}^2=-\frac{2\delta\omega}{3\delta_k}\pm\frac{1}{3\delta_k}\sqrt{\delta\omega^2-3\gamma_m^2}.
\end{eqnarray}
The equations 
\begin{eqnarray}\label{czeslaw3}
&&P\left(\beta_{1,2}^2\right)=-\frac{2}{27\delta_k}\delta\omega(\delta\omega^2+9\gamma_m^2)\pm\nonumber\\
&&\frac{2}{27\delta_k}\left(\delta\omega^2-3\gamma_m^2\right)^{3/2}-B^2=0,
\end{eqnarray}
define constraints between parameters in the vicinity of the fixed point:
\begin{eqnarray}\label{czeslaw4}
B^2_{1,2}=\frac{2\left(-\delta\omega(\delta\omega^2+9\gamma_m^2)\pm(\delta\omega^2-3\gamma_m^2)^{3/2}\right)}{27\delta_k}.
\end{eqnarray}
After solving Eq.(\ref{czeslaw4}) with respect to the detuning $\delta\omega$ we obtain an equation for the boundary of the instability region in the parametric space $\left(\delta\omega, D/\sqrt{\omega_f^2+\gamma_f^2}\right)$ as it follows:
\begin{eqnarray}\label{czeslaw5}
&&\Sigma(\delta\omega)=\delta\omega^4+\frac{B^2\delta_k}{\gamma_m^2}\delta\omega^3+2\gamma_m^2\delta\omega^2+9B^2\delta_k\delta\omega+\nonumber\\
&&\frac{27}{4\gamma_m^2}(B^2\delta_k)^2+\gamma_m^4.
\end{eqnarray}
Eq.(\ref{czeslaw5}) has two real roots for $\frac{2D^2\omega_f}{\omega_f^2+\gamma_f^2}-2K\leqslant \frac{8\sqrt{3}\gamma_m^3}{9B^2}$ and complex roots otherwise. The jump in the mean magnon number calculated along the borderline of the instability region reads:
\begin{eqnarray}\label{czeslaw6}
\Delta n_m=\Delta\beta^2=\frac{\sqrt{\delta\omega^2-3\gamma_m^2}}{2(D^2\omega_f/(\omega_f^2+\gamma_f^2)-K)}.
\end{eqnarray}
The steady-state mean magnon number as a function of the external magnetic field $n_m(B)=\beta^2(B)$ is plotted in Fig.\ref{fig4}. The red line corresponds to the border of linear regime $\delta_k=0$ meaning the particular value of the external electric field $E=\sqrt{k(\omega_f^2+\gamma_f^2)/(\omega_fg_{\text{ME}})}$. The green lines correspond to the nonlinear regime. We clearly see the region where two or more values of magnon number $\beta^2$ correspond to the single value of the magnetic field. We note that the equation Eq.(\ref{closed equation for the equilibrium magnon number}) possesses certain symmetry with respect to the simultaneous transformations $\delta\omega\rightarrow-\delta\omega$ and $\delta_k\rightarrow-\delta_k$. Therefore, linear and nonlinear regimes plotted in Fig.\ref{fig4} by blue and green lines for the positive $\delta\omega=0.5$, mutually exchange for the negative detuning $\delta\omega=-0.5$ case. 
In Fig.\ref{fig5} we plot mean magnon number as a function of the applied external electric field. As we see from Fig.\ref{fig5} by proper tuning of the electric field magnon number can be increased drastically. 

\begin{figure*}[ht]
    \centering
    \begin{minipage}{0.48\textwidth}
        \begin{flushleft}
            a)
        \end{flushleft}
        \vspace{-1em}
        \centering
        \includegraphics[width=\textwidth]{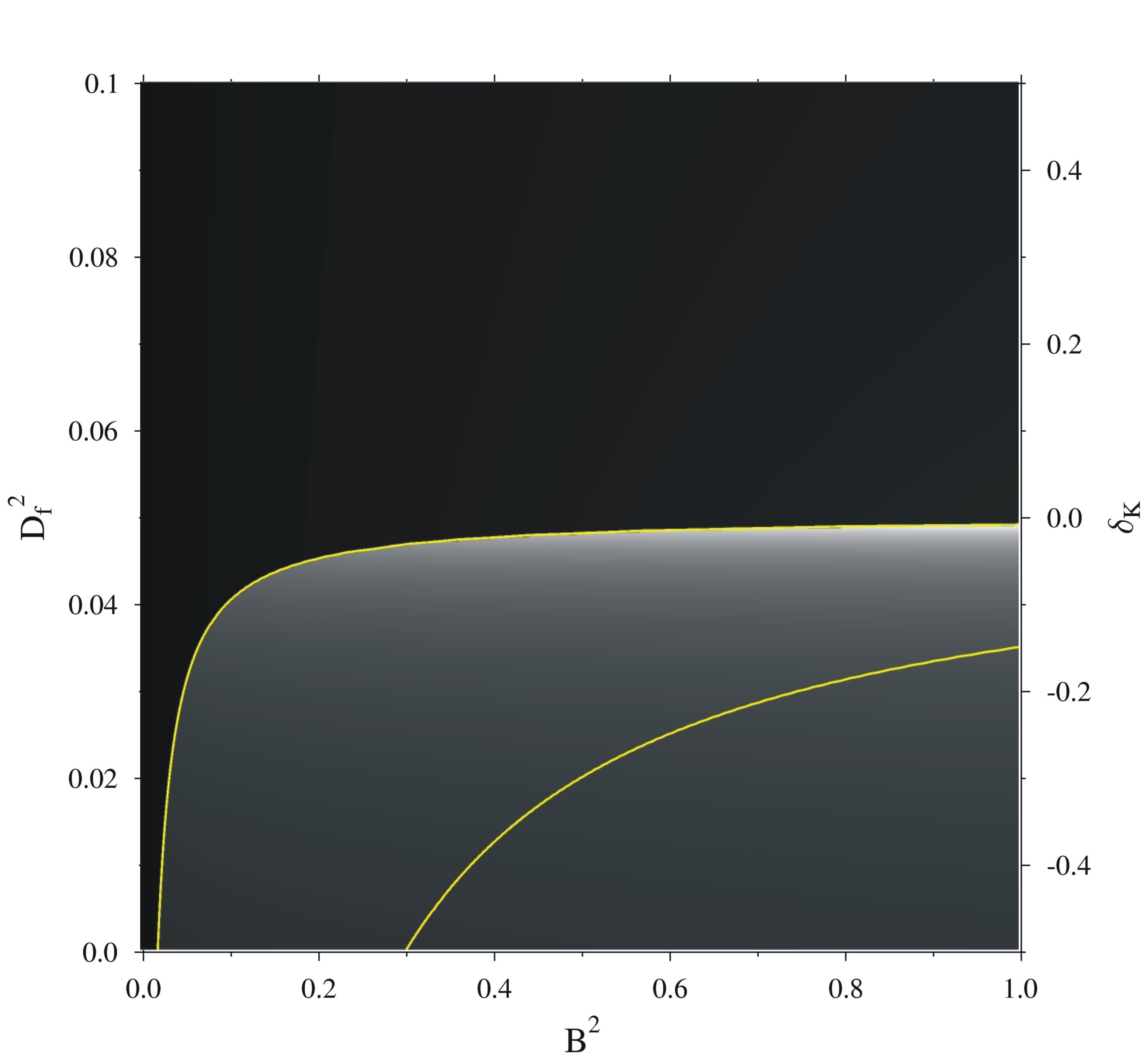}
        \label{fig6a}
    \end{minipage}
    \hfill
    \begin{minipage}{0.48\textwidth}
        \begin{flushleft}
            b)
        \end{flushleft}
        \vspace{-1em}
        \centering
        \includegraphics[width=\textwidth]{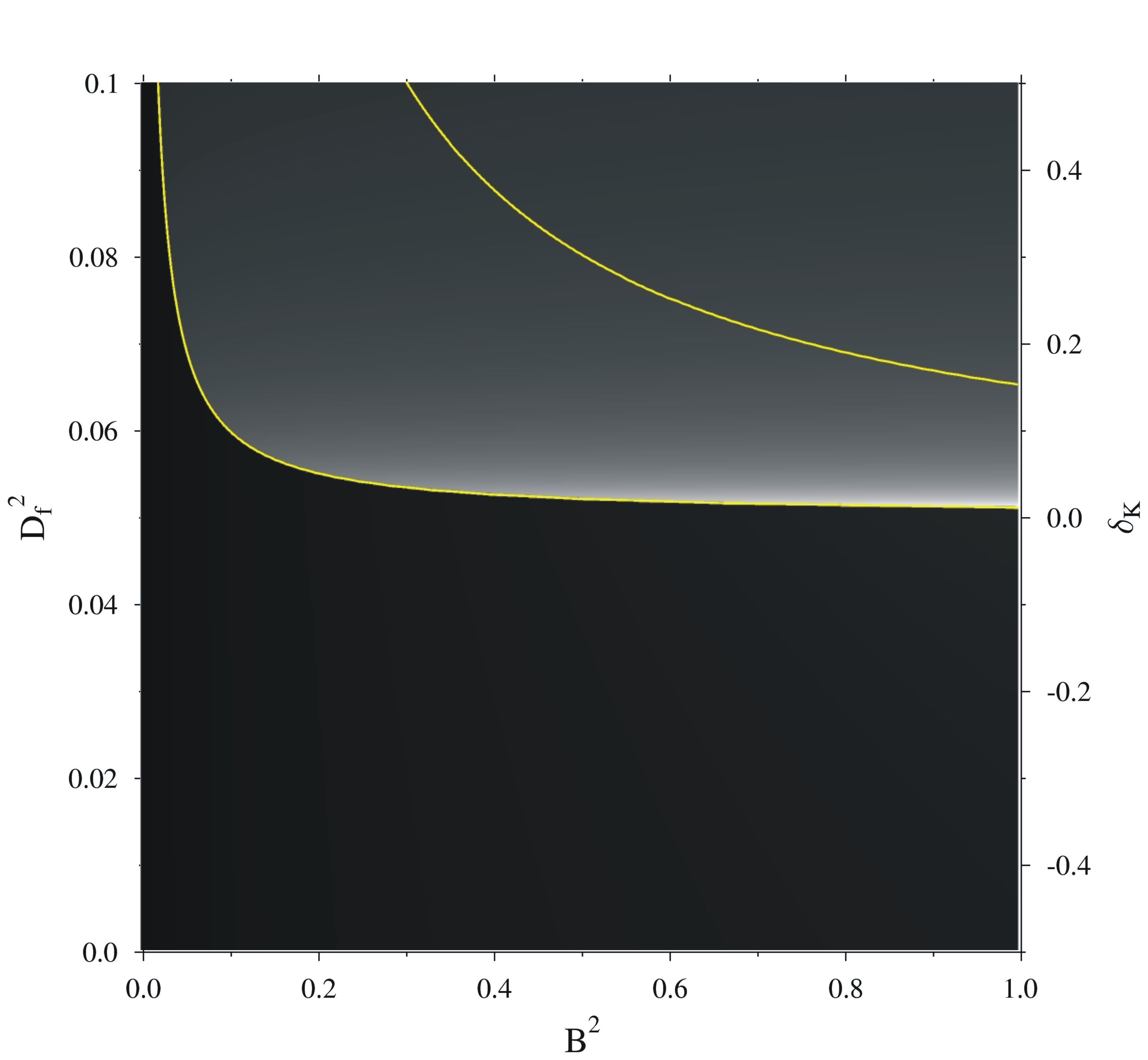}
        \label{fig6b}
    \end{minipage}
    \caption{The map of the steady-state mean magnon number $n_m=\beta^2$ in the parameter space of the external electric $E=D_f/g_{\text{ME}}$, $D_f=D/\sqrt{\omega_f^2+\gamma^2}$ and magnetic fields $B$. The grayscale defines the value of the mean magnon number. Black corresponds to $n_m=0$ and white corresponds to $n_m=$ max. Yellow lines define the boundaries of the bistability region for the detuning $\delta\omega$. The values of detuning are: a) $\delta\omega=1.0$, b) $\delta\omega=-1.0$.}
    \label{fig:6}
\end{figure*}
\begin{figure}[]
\centering
\includegraphics[width=0.95\columnwidth]{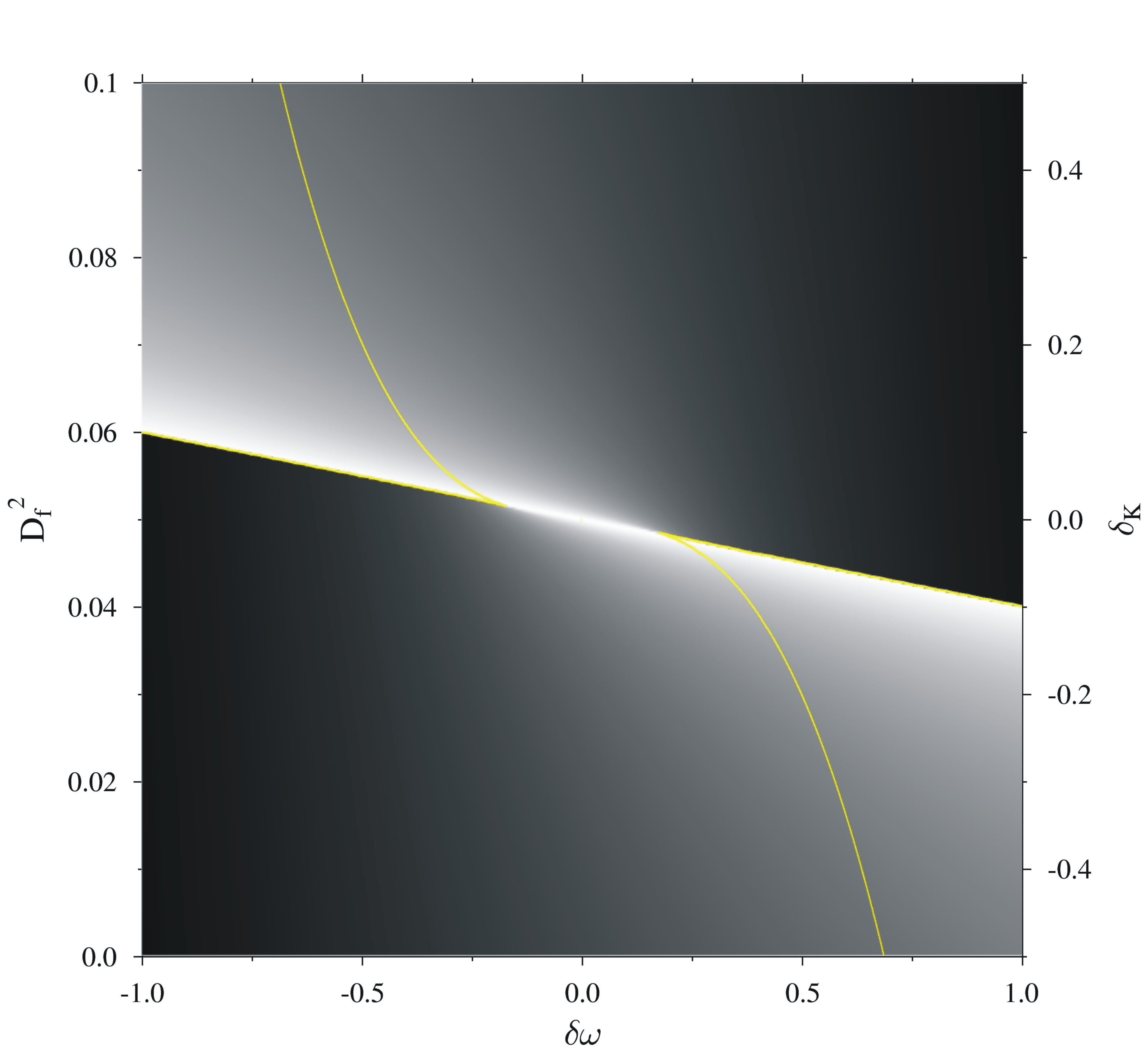}
\caption{The map of the steady-state mean magnon number $n_m=\beta^2$ in the parameter space of the external electric field $E=D_f/g_{\text{ME}}$, $D_f=D/\sqrt{\omega_f^2+\gamma^2}$ and the detuning $\delta\omega=\omega_0-\omega$ between magnon frequency $\omega$ and the frequency of the time-dependent magnetic field $\omega_0$. The grayscale defines the value of the mean magnon number. Black corresponds to $n_m=0$ and white corresponds to $n_m=$ max. Yellow lines define the boundaries of the bistability region for the value of magnetic field $B^2=0.2$.}
\label{fig7}
\end{figure}

In essence, results shown in Fig.\ref{fig3}, Fig.\ref{fig4} and Fig.\ref{fig5} define the borderlines between stability and instability regions. Maps of those regions in the parametric space of applied external fields $(D_f, B)$, $E=D_f/g_{\text{ME}}$, $D_f=D/\sqrt{\omega_f^2+\gamma^2}$ are plotted in Fig.\ref{fig:6}. Same regions in the parametric space ($\delta\omega$, $D_f$) of the detuning $\delta\omega=\omega-\omega_0$ and the electric field  are plotted in Fig.\ref{fig7}.   The change of the sign of the detuning leads to the mirror reflection with respect to the line $\delta_k=0$. 

\begin{figure*}[]
    \centering
    \begin{minipage}{\textwidth}
        \centering
        \includegraphics[width=\linewidth]{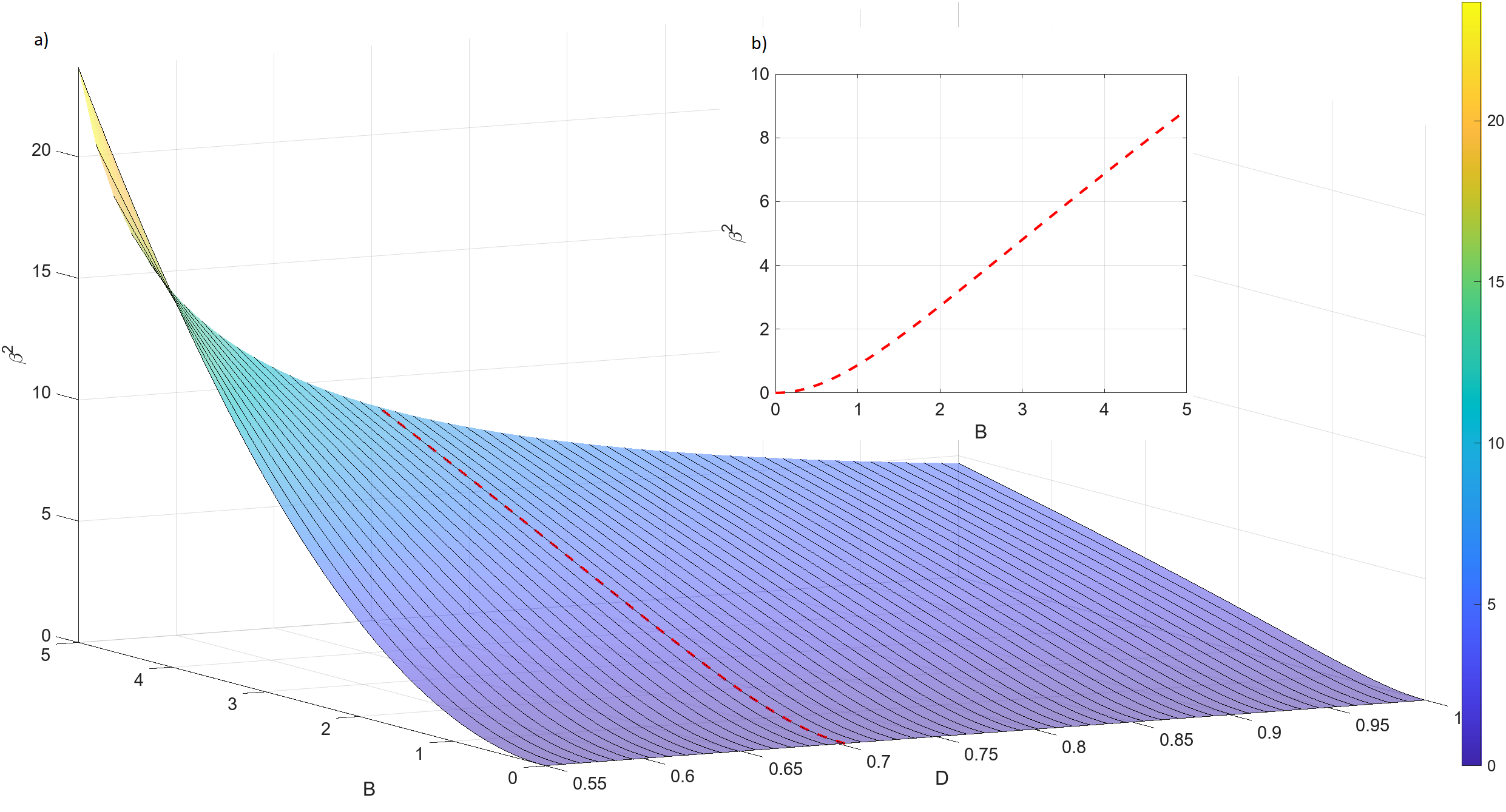}
        \caption{ a) The mean magnon number $\beta^2$ in the steady-state, as a function of applied external magnetic and electric fields $B$, $D$. b) The inset line shows the dependence of $\beta^2$ on $B$ for the fixed value of $D=0.7$. As we can see, dependence is monotonic, corresponding to the linear regime. The values of the parameters read: $K = 0.06, \delta\omega = 1.0, \omega_f = 5.0, \gamma_f = \gamma_m = 0.1$.}
        \label{fig:8}
    \end{minipage}
    \begin{minipage}{\textwidth}
        \centering
        \includegraphics[width=\linewidth]{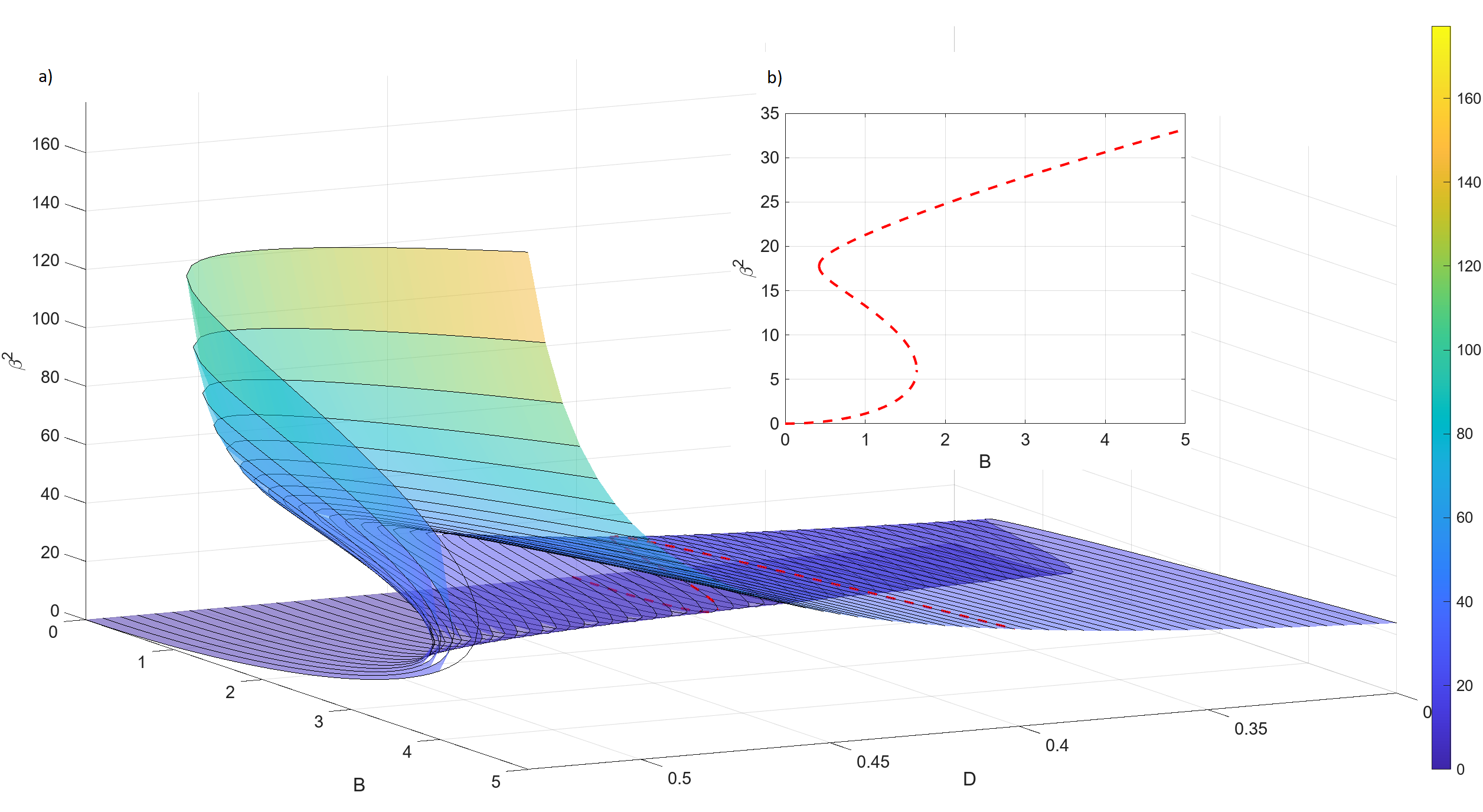}
        \caption{ a) The mean magnon number $\beta^2$ in the steady-state, as a function of applied external magnetic and electric fields $B$, $D$. The inset line shows the dependence of $\beta^2$ on $B$ for the fixed value of $D=0.4$. We can see that dependence is not monotonic, corresponding to the linear regime. b) The bistability region appears in the interval $0.3<B<1.6$. The values of the parameters read: $K = 0.06, \delta\omega = 1.0, \omega_f = 5.0, \gamma_f = \gamma_m = 0.1$.}
        \label{fig:9}
    \end{minipage}
\end{figure*}

For visual control and better clarity finally we plot instability regions in 3D plots
Fig.\ref{fig:8} and Fig.\ref{fig:9}. The inset lines correspond to the instability and stability regions respectively. \vspace{0.2cm}\\
The fixed point of the nonlinear system defines the system's steady state. However, one can pump an arbitrary number of magnons into the system through the nanoantenna and prepare a non-equilibrium state. Non-equilibrium effects are also in the scope of our interest, and therefore, when analyzing nonlinear effects, we exploit more general formalism and study the system's dynamics not only in the vicinity of the steady state fixed point but also in the vicinity of non-equilibrium states. 
\vspace{0.2cm}\\
We proceed with the analysis of the phase space of the dynamical system Eq.(\ref{semi-classical phase}). We linearize Eq.(\ref{semi-classical phase}) in the vicinity of the equilibrium point 
and write down the system of linearized equations:
\begin{eqnarray}\label{linearized}
&&\dot{\textbf{x}}=\textbf{F},\nonumber\\
&&\textbf{x}=\left\lbrace \alpha, \phi, \beta, \varphi \right\rbrace  ,\,\,\,\textbf{F}=\left\lbrace f_1(\textbf{x})...f_4(\textbf{x})\right\rbrace.
\end{eqnarray}
While the dimension of set Eq.(\ref{linearized}) is $n=4$, it is characterized by $n+2$ topologically different types of trajectories. One of them corresponds to the trajectory with a regular fixed point, and the rest $n+1$ trajectories are characterized by
exceptional fixed points (e.g. Saddles, Nodes, Stable spiral, and Unstable spiral) which following Petrovsky \cite{katok1995introduction} we denote by $\mathcal{O}^{p,q}$, where $p=0,1,...n$ and $p+q=n$. The points $\mathcal{O}^{p,q}$ with $p\neq n$ are saddle fixed points. The fixed points are found from the condition $\dot{\textbf{x}}=0$, i.e., $\textbf{F}(\textbf{x}^*)=0$. The corresponding characteristic equation in the vicinity of $\textbf{x}^*$ point reads:
\begin{eqnarray}\label{polinom}
\chi(\lambda)=\text{Det}\big\vert\big\vert\textbf{A}-\textbf{I}\lambda\big\vert\big\vert=\text{Det}\bigg\vert\bigg\vert\frac{\partial f_i}{\partial x_j}-\delta_{ij}\lambda\bigg\vert\bigg\vert=0.
\end{eqnarray}
Here $\textbf{I}$ is the $(4\otimes4)$ identity matrix. The equilibrium state $\textbf{x}^*[B, D, K]$ is the root of the equation
\begin{eqnarray}\label{bifurcation1}
\textbf{\textbf{F}}(\textbf{x}^*[B, D, K])=0. 
\end{eqnarray}
The type of the exceptional point is defined from the equation
\begin{eqnarray}\label{the roots}
\chi(\textbf{x}^*[B, D, K])=0.
\end{eqnarray}
The border of the equilibrium domain is given by
\begin{eqnarray}\label{bifurcation3}
&&\text{Im}(\lambda=i\Omega)=0.
\end{eqnarray}
In essence Eq.(\ref{bifurcation3}) defines a hypersurface $\mathcal{N}_\Omega$. We also define the hypersurface $\mathcal{N}_{\Omega=0}=\mathcal{N}_0$ and analyze what happens with the system when phase trajectory crosses hypersurface. Crossing of $\mathcal{N}_0$ when steering the parameters of the system demolishes the saddle fixed point $\mathcal{O}^{p,q}$ due to the merging with the states $\mathcal{O}^{p+1,q-1}$ or 
$\mathcal{O}^{p-1,q+1}$. On the other hand, $\mathcal{O}^{p,q}$ is stable when crossing
$\mathcal{N}_\Omega$. We implement this theoretical scheme to our system.
After linearization of Eq.(\ref{semi-classical phase}) $\textbf{X}=(\delta\alpha, \delta\phi, \delta\beta, \delta\varphi)^T$, where $\delta\alpha=\alpha-\alpha_1$, $\delta\phi=\phi-\phi_1$, $\delta\beta=\beta-\beta_1$, $\delta\varphi=\varphi-\varphi_1$, in the vicinity of arbitrary fixed point $(\alpha_1, \phi_1, \beta_1, \varphi_1)$ we obtain matrix equation:
\begin{eqnarray}\label{matrix equation}
\frac{d\textbf{X}}{dt}=\mathcal{A}_1\cdot\textbf{X},
\end{eqnarray}
where matrix $\mathcal{A}_1$ in the explicit form reads:
\begin{eqnarray}\label{MatrixMichal}
\left( \begin{matrix}
\delta\dot\alpha \\
\delta\dot\phi \\
\delta\dot\beta \\
\delta\dot\varphi
\end{matrix} \right)
\left(
\begin{matrix}
-\gamma_f & D\beta_1^2\sin\phi_1 & -2D\beta_1\cos\phi_1 & 0 \\
-\frac{D\beta_1^2}{\alpha_1^2}\sin\phi_1 &
\frac{D\beta_1^2}{\alpha_1}\cos\phi_1 &
2\frac{D\beta_1}{\alpha_1}\sin\phi_1 & 0 \\
0 & 0 & -\gamma_m & -B\sin\varphi_1  \\
2D\sin\phi_1 & 2D\alpha_1\cos\phi_1 &
\frac{B}{\beta_1^2}\sin\varphi_1-4K\beta_1 &
-\frac{B}{\beta_1}\cos\varphi_1
\end{matrix}
\right)
\left( \begin{matrix}
\delta\alpha \\
\delta\phi \\
\delta\beta \\
\delta\varphi
\end{matrix} \right).
\end{eqnarray}

\begin{figure*}
    \centering
    \begin{minipage}[t]{0.45\textwidth}
        \begin{flushleft}
            a)
        \end{flushleft}
        \vspace{-1em}
        \centering
        \includegraphics[width=\textwidth]{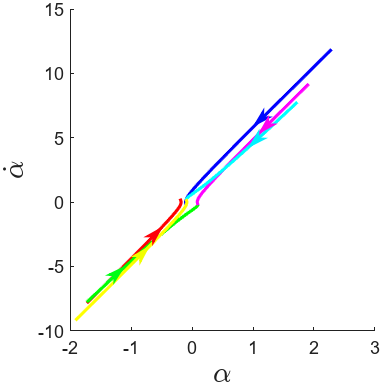}
        \label{fig15a}
    \end{minipage}
    \hfill
    \begin{minipage}[t]{0.45\textwidth}
        \begin{flushleft}
            b)
        \end{flushleft}
        \vspace{-1em}
        \centering
        \includegraphics[width=\textwidth]{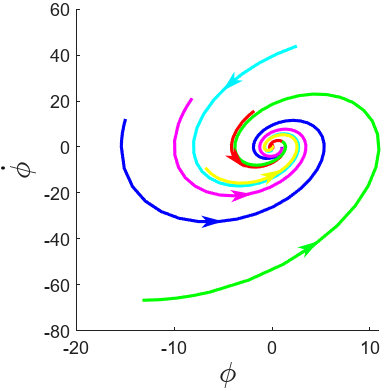}
        \label{fig15b}
    \end{minipage}
    \vspace{1em}
    \begin{minipage}[t]{0.45\textwidth}
        \begin{flushleft}
            c)
        \end{flushleft}
        \vspace{-1em}
        \centering
        \includegraphics[width=\textwidth]{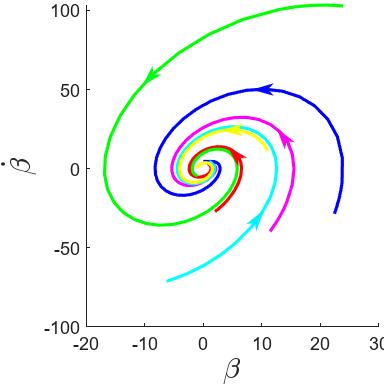}
        \label{fig15c}
    \end{minipage}
    \hfill
    \begin{minipage}[t]{0.45\textwidth}
        \begin{flushleft}
            d)
        \end{flushleft}
        \vspace{-1em}
        \centering
        \includegraphics[width=\textwidth]{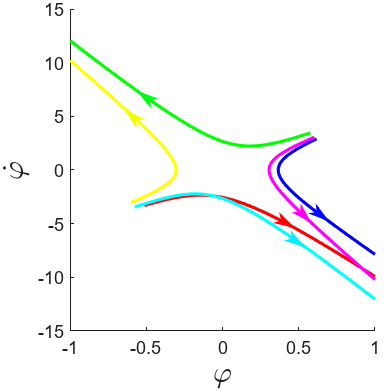}
        \label{fig15d}
    \end{minipage}
    \caption{Phase portraits for various types of phase trajectories. The arrows show the directions of the trajectories: a) The stable node for ($\dot{\alpha}$, $\alpha$) for the parameters: $D = 0.2$, $\beta^2 = 10$,  $K = 0.06$, $\omega = 35$,  $\omega_0 = 1.5$,  $\omega_f = 0.5$, $\gamma_f = \gamma_m = \gamma_0 = 0.1$, b) The stable spiral for ($\dot{\phi}$, $\phi$) and the parameters: $D = 0.4$,  $\beta^2 = 25$,  $K = 0.06$,  $\omega = 1.5$,  $\omega_0 = 1$,  $\omega_f = 5$,   $\gamma_f = \gamma_m = \gamma_0 = 0.1$, c) The stable spiral for ($\dot{\beta}$, $\beta$) obtained for the parameters: $D = 0.4$, $\beta^2 = 25$,  $K = 0.06$,  $\omega = 1.5$,  $\omega_0 = 1$, $\omega_f = 5$, $\gamma_f = \gamma_m = \gamma_0 = 0.1$, d) The saddle point for ($\dot{\varphi}$,  $\varphi$) obtained for the parameters: $D = 0.3$, $\beta^2 = 30$, $K = 0.06$, $\omega = 5$, $\omega_0 = 0.2$, $\omega_f = 1$, $\gamma_f = \gamma_m = \gamma_0 = 10$.}
    \label{phase_portraits}
\end{figure*}
In  Fig.(\ref{phase_portraits}), we plot phase portraits based on the Eq.(\ref{matrix equation}) and corresponding to the different regions of the phase space and the various sets of parameters given in Table 2. 
Below, magnon-photon entanglement is shown to be maximal in the region near the border between the stable node (a) and the stable spiral regions (b) and (c). The characteristic roots of the Eq.(\ref{MatrixMichal}) can be obtained analytically for the equal damping case $\gamma_m=\gamma_f=\gamma_0$ which is a physically reasonable assumption:
\begin{eqnarray}\label{new roots 1}
\lambda_{\pm,\pm}=(p\pm\sqrt{\Delta_{\pm}})/2,
\end{eqnarray}
where $\Delta_{\pm}=p^2-4q_{\pm}$, $p=-2\gamma_0$
and $q_{\pm}=\gamma_0^2-(\chi_1\pm\sqrt{\chi_1^2+\chi_2})/2$,  $\chi_1=-\omega_f^2-[\delta\omega+(\delta_K-4K)n_m](\delta\omega+\delta_Kn_m)$, $\chi_2=-4\omega_f[4D^2n_m+\omega_f(\delta\omega+(\delta_K-4K)n_m)](\delta\omega+\delta_Kn_m)$, with $\delta_k=\left(\frac{2D^2\omega_f}{\omega_f^2+\gamma_f^2}-2K\right)$ and equilibrium mean magnon number $n_m=\beta_1^2$.
Taking into account explicit expressions of the roots we identify the following three cases relevant to our problem:\vspace{0.3cm}\\
\textbf{I. Saddle point}. $p<0$, $\Delta_\pm>0$, $q_\pm<0$, and $\chi_1\pm\sqrt{\chi_1^2+\chi_2}>\gamma_0^2$. 
\vspace{0.3cm}\\
\textbf{II. Stable node}. $p<0$, $\Delta_\pm>0$, $q_\pm>0$, and $\gamma_0^2>\chi_1\pm\sqrt{\chi_1^2+\chi_2}>0$. 
\vspace{0.3cm}\\
\textbf{III. Stable spiral}. $p<0$, $\Delta_\pm<0$, $q_\pm>0$, and $\chi_1\pm\sqrt{\chi_1^2+\chi_2}<0$. 

Taking into account definitions below Eq. \ref{new roots 1} the cases above can be presented in their reduced form as shown in Table I.
\begin{table}[H]
    \caption{The phase space of the system. As we see, depending on the values of the parameters the system is characterized by Saddle point, Stable node or Stable spiral fixed points. }
    \centering
    \begin{tabular}{c|c|c|c}
        \multicolumn{3}{c|}{Conditions} & Phase type  \\ \hline
        $p<0$ & $\Delta_\pm>4\gamma_0^2$ & $q_\pm<0$ & Saddle point  \\
        $p<0$ & $2\gamma_0^2 < \Delta_{\pm} < 4\gamma_0^2$ & $0 < q_\pm < \gamma_0^2/2$ & Stable node  \\
        $p<0$ & $0 < \Delta_{\pm} < 2\gamma_0^2$ & $\gamma_0^2/2 < q_\pm < \gamma_0^2$ & Stable node  \\
        $p<0$ & $\Delta_\pm<0$ & $q_\pm > \gamma_0^2$ & Stable spiral 
    \end{tabular}
    \label{tab:my_label}
\end{table}

\begin{figure*}[ht]
    \centering
    \begin{minipage}{0.45\textwidth}
        \begin{flushleft}
            a)
        \end{flushleft}
        \vspace{-1em}
        \centering
        \includegraphics[width=\textwidth]{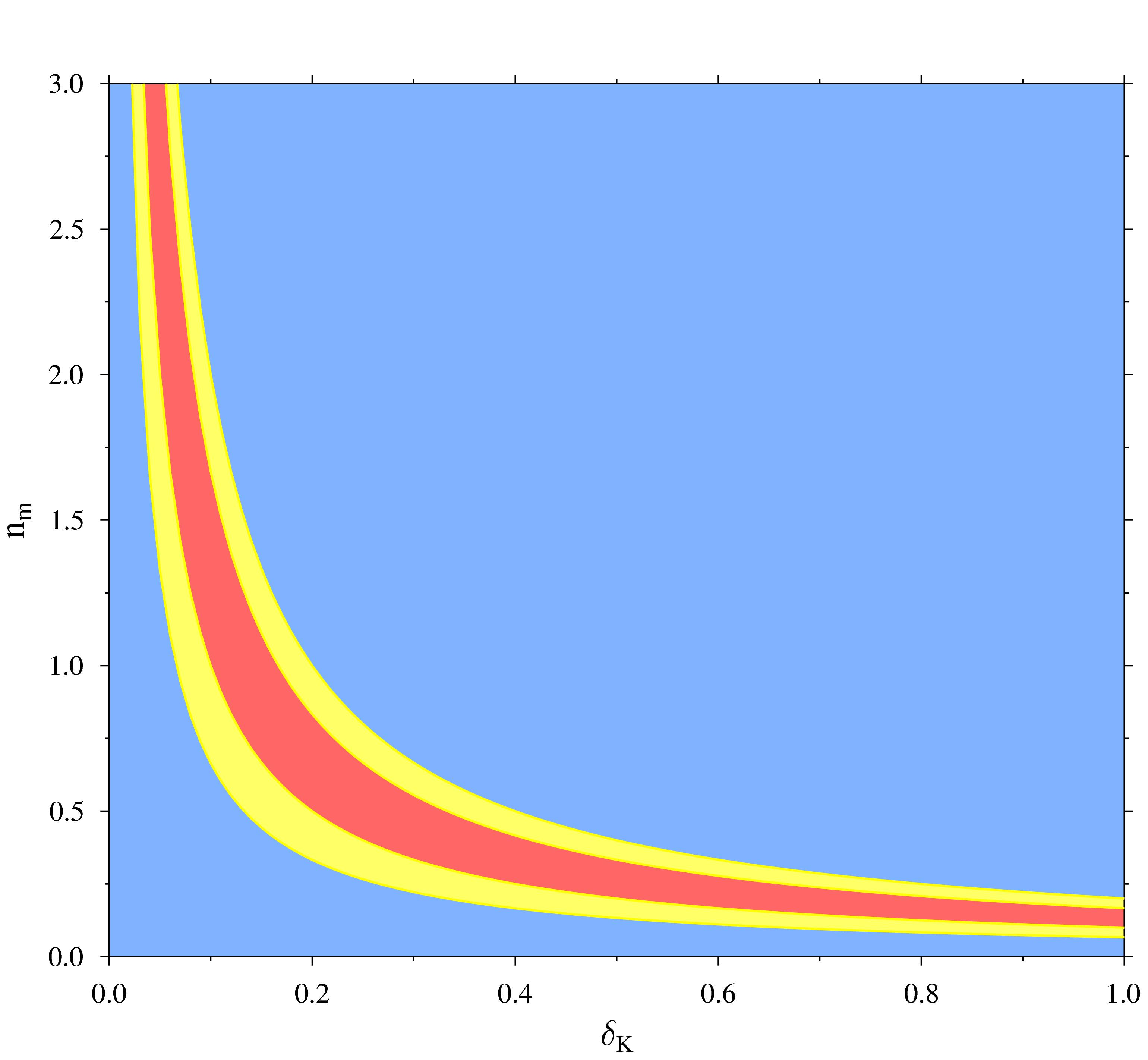}
        \label{fig10a}
    \end{minipage}
    \hfill
    \begin{minipage}{0.45\textwidth}
        \begin{flushleft}
            b)
        \end{flushleft}
        \vspace{-1em}
        \centering
        \includegraphics[width=\textwidth]{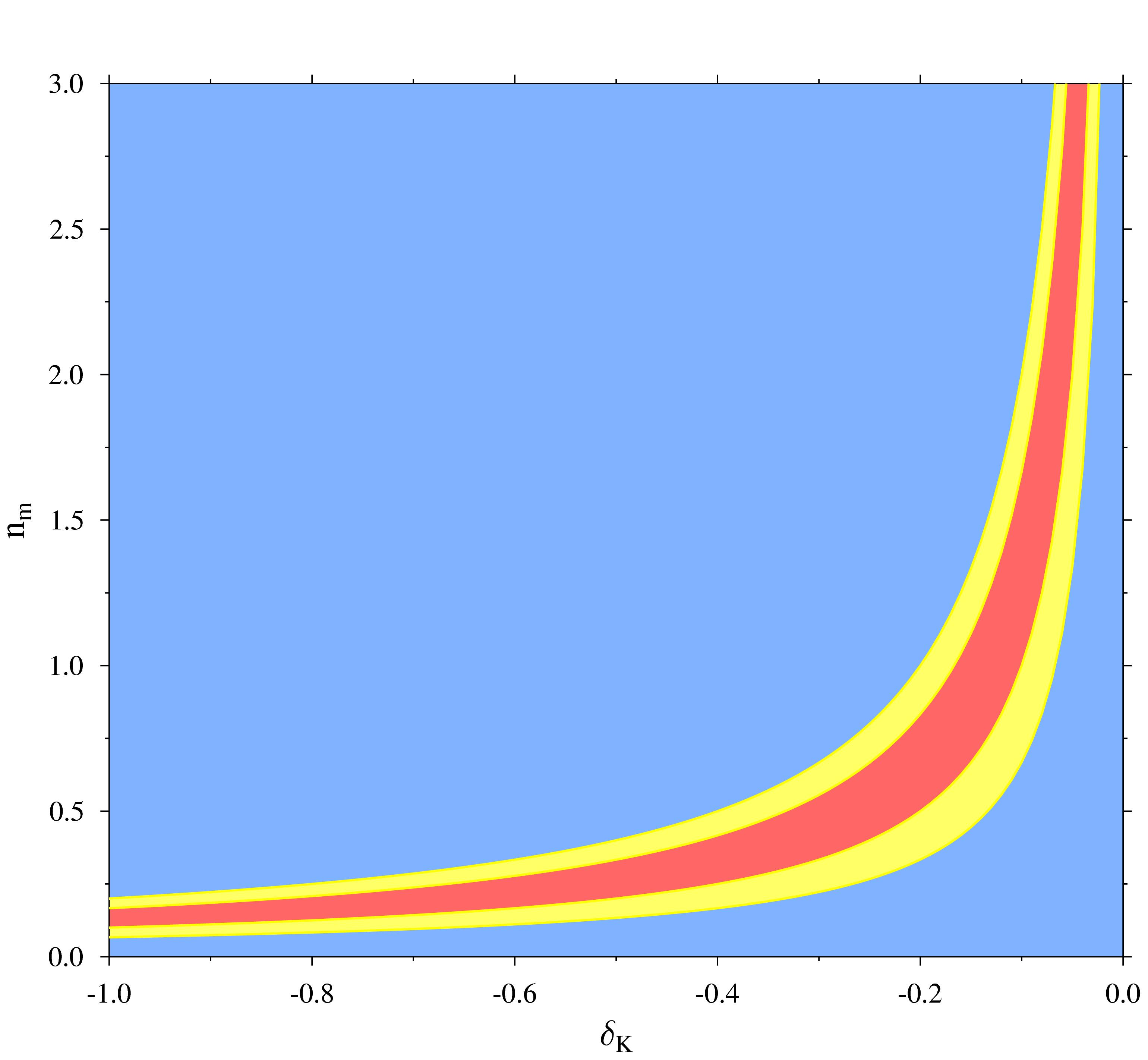}
        \label{fig10b}
    \end{minipage}
    \caption{The map of exceptional fixed points plotted for the value of detuning $\delta\omega$. The region of saddle points is highlighted in red. The map of the region of stable node points is highlighted in yellow. The map of the region of stable spiral points is highlighted in blue. On the \textbf{x} axis $\delta_k=\left(\frac{2D^2\omega_f}{\omega_f^2+\gamma_f^2}-2K\right)$ and on the \textbf{y} axis mean magnon number $n_m=\beta^2$. The values of detuning are: a) $\delta\omega=-0.2$, b) $\delta\omega=0.2$.}
    \label{fig:10}
\end{figure*}

\begin{figure*}
    \centering
    \begin{minipage}[t]{0.45\textwidth}
        \begin{flushleft}
            a)
        \end{flushleft}
        \vspace{-1em}
        \centering
        \includegraphics[width=\textwidth]{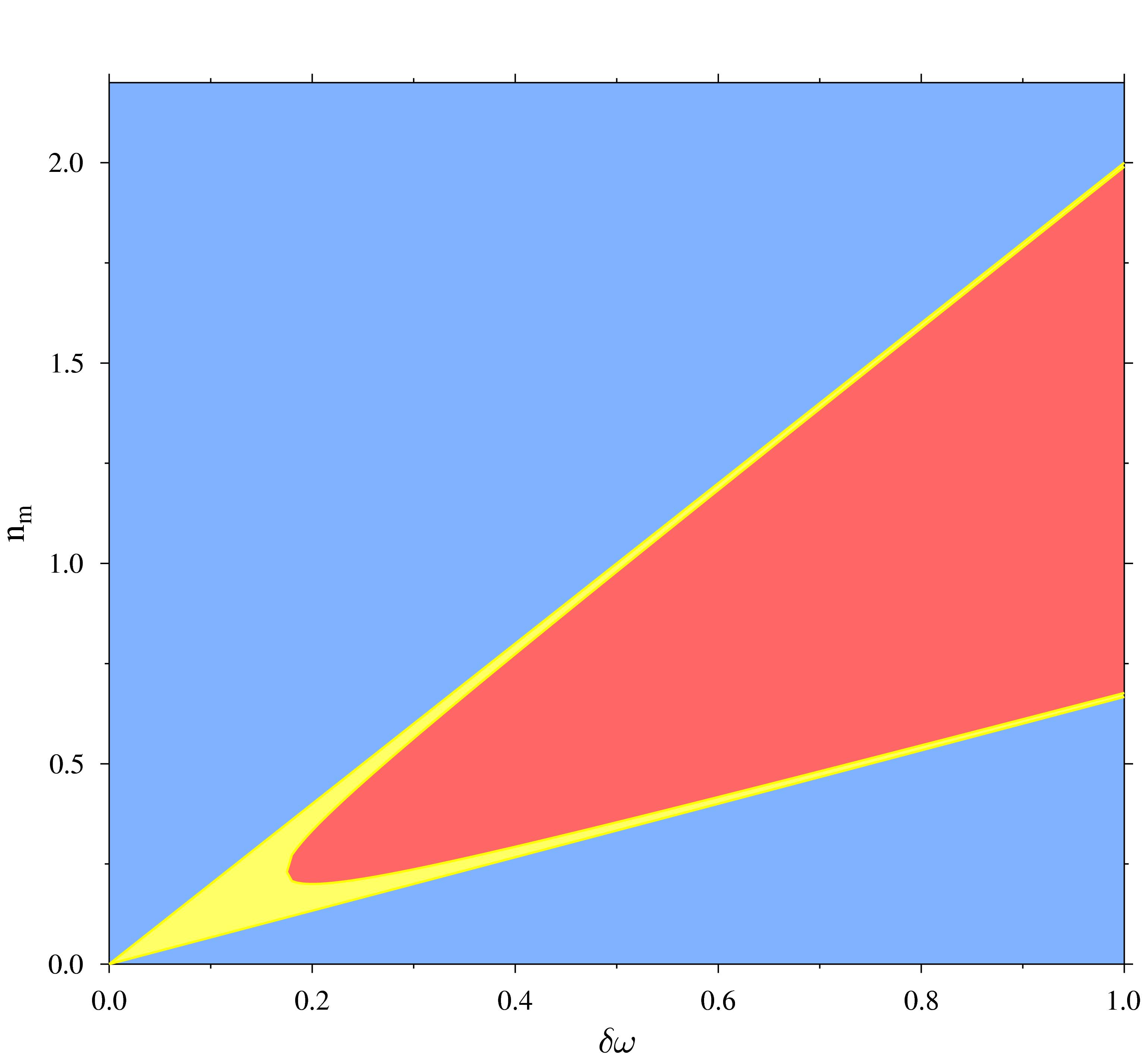}
        \label{fig11a}
    \end{minipage}
    \hfill
    \begin{minipage}[t]{0.45\textwidth}
        \begin{flushleft}
            b)
        \end{flushleft}
        \vspace{-1em}
        \centering
        \includegraphics[width=\textwidth]{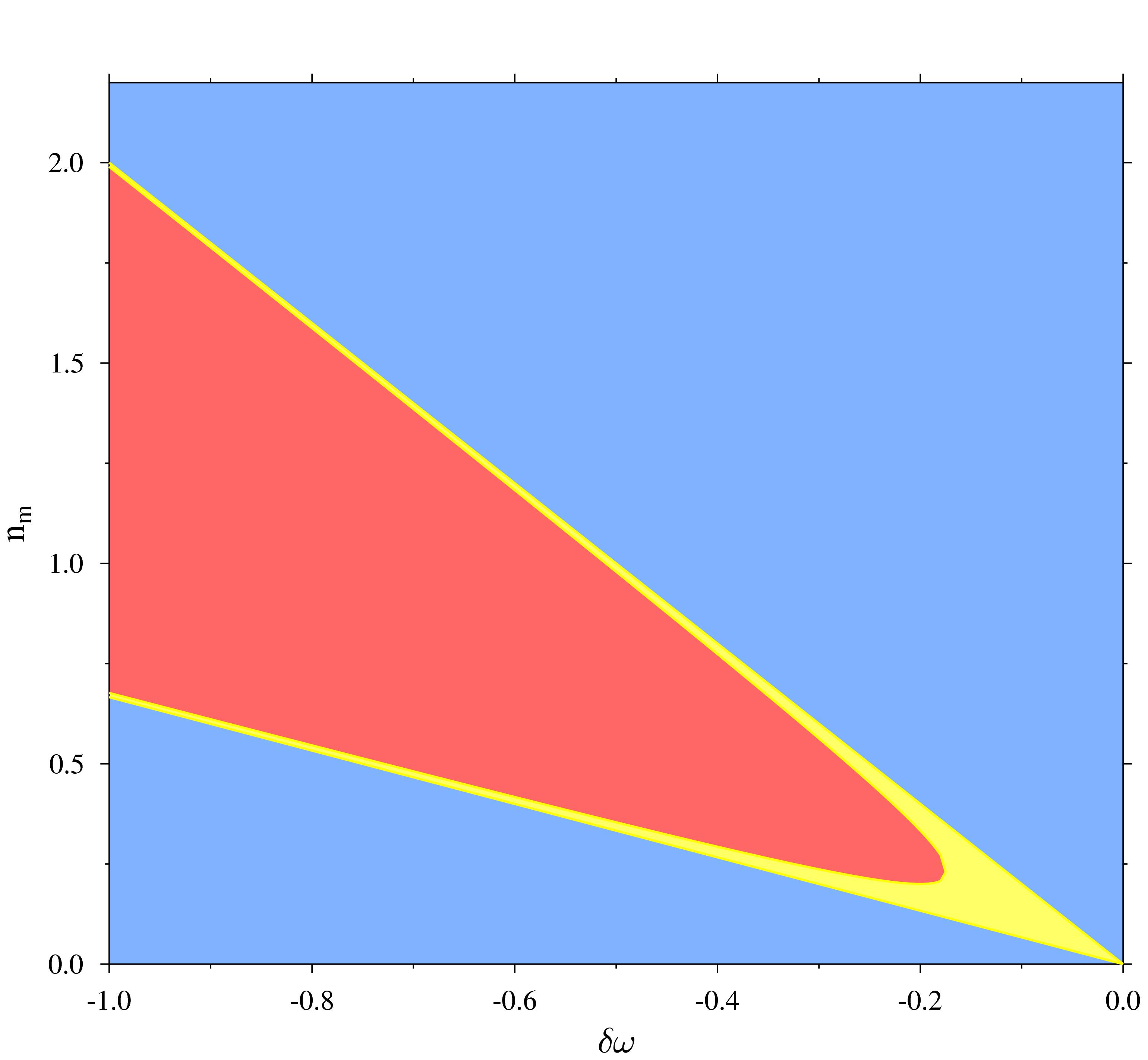}
        \label{fig11b}
    \end{minipage}
    \caption{The map of exceptional fixed points. The region of saddle points is highlighted in red. The map of the region of stable node points is highlighted in yellow. The map of the region of stable spiral points is highlighted in blue. On the \textbf{x} axis is shown detuning $\delta\omega=\omega_0-\omega$ between the frequency of time dependent magnetic field $\omega_0$ and magnon frequency $\omega$. On the \textbf{y} axis, mean magnon number $n_m=\beta^2$. The values of $\delta_k=\left(\frac{2D^2\omega_f}{\omega_f^2+\gamma_f^2}-2K\right)$ are: a) $\delta_k=-0.5$, b) $\delta_k=0.5$.}
    \label{fig:11}
\end{figure*}

Thus, depending on the values of the parameters, we can have different types of dynamics specified in Table \ref{tab:my_label}.  For illustrative purposes, the phase spaces of the system for the different sets of parameters are plotted in Fig.\ref{fig:10} and Fig.\ref{fig:11}. We clearly see that the most significant part of the parameter space belongs to the stable spiral points. The change of the sign of detuning from negative $\delta\omega=\omega_0-\omega=-0.2$ to positive  $\delta\omega=0.2$ shifts the region of stable node and stable spiral points from the small positive $\delta_k=\left(\frac{2D^2\omega_f}{\omega_f^2+\gamma_f^2}-2K\right)$ to the slight negative $\delta_k$. The effect observed in Fig.\ref{fig:10} means that the system's dynamic can be controlled through the frequency of the time-dependent magnetic field $\omega_0$ and amplitude of the external electric field $E=D/g_{\text{ME}}$. With the increase of the detuning $\delta\omega$, Fig.\ref{fig:11}, the saddle point's region becomes larger. Nevertheless, the symmetry upon simultaneous flips of signs 
$\delta\omega\rightarrow-\delta\omega$ and $\delta_K\rightarrow-\delta_K$ still holds. 

Further interest concerns the case $\chi_1^2+\chi_2<0$. Then
\begin{eqnarray}\label{new roots 2}
&&\lambda_{\pm,\pm}=\text{Re}(\lambda_{\pm,\pm})+\ii\text{Im}(\lambda_{\pm,\pm}),\nonumber\\
&&\text{Re}(\lambda_{\pm,\pm})=\frac{1}{2}\left(p^2\pm\sqrt{|\Delta'_\pm|^2+|\Delta''_\pm|}\cos\xi\right),\nonumber\\
&&\text{Im}(\lambda_{\pm,\pm})=\frac{1}{2}\sqrt{|\Delta'_\pm|^2+|\Delta''_\pm|}\sin\xi,
\end{eqnarray}
where 
\begin{eqnarray}\label{new roots 2 plus}
&&\Delta'_\pm=\chi_1/2,\,\,\,\Delta''_\pm=\sqrt{|\chi_1^2+\chi_2|}/2,\nonumber\\
&&\xi=\arctan\left(\Delta''_\pm/\Delta'_\pm\right).
\end{eqnarray}
\vspace{0.2cm}\\
\textbf{IV. Hopf Bifurcation}. Hopf bifurcation occurs when $\text{Re}(\lambda_{\pm,\pm})$ changes the sign from negative to positive and characteristic root crosses the imaginary axis. 
The values of the parameters $D$, $K$, $\omega$, $\omega_0$, $\omega_f$, $B$, $\gamma_0$, $n_m=\beta^2_1$ for each four cases is shown in the Table \ref{negativeroots}.
Considering the case $D=\sqrt{3K(\omega_f^2+\gamma^2_f)/\omega_f}$ from $\chi_1^2+\chi_2<0$ we deduce:
\begin{eqnarray}\label{2negativeroots}
&&D^2>\frac{\omega_f^4+\delta\omega^2(\delta\omega+\delta_Kn_m)^2}{16n_m\omega_f(\delta\omega+\delta_Kn_m)}-\frac{\omega_f^2\delta\omega}{8n_m}.
\end{eqnarray}
Or in the resonant case $\delta\omega=0$:
\begin{eqnarray}\label{3negativeroots}
&&D>\frac{\omega_f\sqrt{\omega_f}}{4n_m\sqrt{\delta_K}}.
\end{eqnarray}
Together with $D=\sqrt{3K(\omega_f^2+\gamma^2_f)/\omega_f}$, Eq.(\ref{2negativeroots}) and
Eq.(\ref{3negativeroots}) define bifurcation values of DMI constant. 
The particular values of the parameters for Hopf bifurcation for the general case are shown in the table \ref{negativeroots}. 
The analytic solution of the system Eq.(\ref{matrix equation}) reads:
\begin{eqnarray}\label{The general solution of the system 17}
&&\delta\alpha(t)=\sum\limits_{m=\alpha,\phi,\beta,\varphi}C_m\gamma_{m\alpha}\exp[\lambda_mt],\nonumber\\
&&\delta\phi(t)=\sum\limits_{m=\alpha,\phi,\beta,\varphi}C_m\gamma_{m\phi}\exp[\lambda_mt],\nonumber\\
&&\delta\beta(t)=\sum\limits_{m=\alpha,\phi,\beta,\varphi}C_m\gamma_{m\beta}\exp[\lambda_mt],\nonumber\\
&&\delta\varphi(t)=\sum\limits_{m=\alpha,\phi,\beta,\varphi}C_m\gamma_{m\varphi}\exp[\lambda_mt]. 
\end{eqnarray}
Here $\gamma_{nm}=(-1)^{1+m}\mathcal{M}_{1,m}\left\lbrace ||\mathcal{A}-\textbf{I}\lambda_n||\right\rbrace  $ is the co-factor of the minor of matrix Eq.(\ref{MatrixMichal}) and 
coefficients $C_m$ are defined from the initial conditions.
\begin{table*}[!ht]
    \caption{Parameters for which specific states of phase space exist.}
    \centering
    \begin{tabular}{p{3.5cm} p{1cm} p{1cm} p{1cm} p{1cm} p{1cm} p{1cm} p{2cm} }\label{negativeroots}
        ~ & $\beta^2$ & $D$ & $K$ & $\omega_f$ & $\omega_0$ & $\omega$ & $\gamma_0 = \gamma_m = \gamma_f$ \\ \hline
        \textbf{I. Saddle point} & 30 & 0.32 & 0.06 & 1.0 & 0.2 & 5.0 & 10 \\ 
        ~ & 30 & 0.33 & 0.06 & 1.0 & 0.2 & 5.0 & 10 \\ 
        ~ & 30 & 0.34 & 0.06 & 1.0 & 0.2 & 5.0 & 10 \\ 
        ~ & 25 & 0.30 & 0.06 & 1.0 & 0.2 & 5.0 & 10 \\ 
        ~ & 25 & 0.31 & 0.06 & 1.0 & 0.2 & 5.0 & 10 \\ \hline
        \textbf{II. Stable node} & 30 & 0.16 & 0.06 & 0.5 & 2.5 & 5.5 & 0.1 \\ 
        ~ & 30 & 0.20 & 0.06 & 0.5 & 2.5 & 5.5 & 0.1 \\ 
        ~ & 30 & 0.23 & 0.06 & 0.5 & 2.5 & 5.5 & 0.1 \\ 
        ~ & 30 & 0.18 & 0.06 & 0.5 & 2.0 & 2.5 & 0.1 \\ 
        ~ & 30 & 0.20 & 0.06 & 0.5 & 2.0 & 2.5 & 0.1 \\ 
        ~ & 30 & 0.25 & 0.06 & 0.5 & 2.0 & 2.5 & 0.1 \\ 
        ~ & 25 & 0.18 & 0.06 & 0.5 & 2.0 & 2.5 & 0.1 \\ 
        ~ & 25 & 0.20 & 0.06 & 0.5 & 2.0 & 2.5 & 0.1 \\ 
        ~ & 25 & 0.25 & 0.06 & 0.5 & 2.0 & 2.5 & 0.1 \\ 
        ~ & 10 & 0.20 & 0.06 & 0.5 & 1.5 & 5.0 & 0.1 \\ \hline 
        \textbf{III. Stable spiral} & 25 & 0.60 & 0.06 & 5.0 & 1.0 & 1.5 & 0.1 \\ 
        ~ & 25 & 0.55 & 0.06 & 5.0 & 1.0 & 1.5 & 0.1 \\ 
        ~ & 25 & 0.20 & 0.06 & 5.0 & 1.0 & 1.5 & 0.1 \\ 
        ~ & 30 & 0.20 & 0.06 & 5.0 & 1.0 & 1.5 & 0.1 \\ 
        ~ & 30 & 0.40 & 0.06 & 5.0 & 1.0 & 1.5 & 0.1 \\ 
        ~ & 30 & 0.55 & 0.06 & 5.0 & 1.0 & 1.5 & 0.1 \\ 
        ~ & 30 & 0.20 & 0.06 & 5.0 & 0.5 & 1.5 & 0.1 \\ 
        ~ & 30 & 0.40 & 0.06 & 5.0 & 0.5 & 1.5 & 0.1 \\ 
        ~ & 30 & 0.55 & 0.06 & 5.0 & 0.5 & 1.5 & 0.1 \\ \hline
        \textbf{IV. Hopf bifurcation} & 20 & 0.9489 & 0.06 & 5.0 & 1.0 & 2.0 & 0.1 \\
        ~ & 25 & 0.9489 & 0.06 & 5.0 & 1.0 & 2.0 & 0.1 \\
        ~ & 30 & 0.9489 & 0.06 & 5.0 & 1.0 & 2.0 & 0.1 \\
        ~ & 30 & 0.8488 & 0.06 & 4.0 & 1.0 & 2.0 & 0.1 \\
        ~ & 25 & 0.8488 & 0.06 & 4.0 & 1.0 & 2.0 & 0.1 \\
        ~ & 20 & 0.8488 & 0.06 & 4.0 & 1.0 & 2.0 & 0.1 \\ 
    \end{tabular}
\end{table*}
\section{Entanglement}
\label{sec:Entanglement}
To find steady state robust magnon-photon entanglement in the system we follow the recent work \cite{PhysRevB.107.115126}. Entanglement is quantified through the logarithmic negativity \cite{adesso2007entanglement,PhysRevLett.84.2726}:
\begin{eqnarray}\label{1entanglement helical weak}
&& E_N=\text{max}\big[0,-\ln[2\eta^-]\big],
\end{eqnarray}
where
\begin{eqnarray}\label{2entanglement helical weak}
&& \eta^-=2^{-1/2}\lbrace\Sigma(V_1)-[\Sigma(V_1)^2-4\det V_1]^{1/2}\rbrace^{1/2},
\end{eqnarray}
\begin{eqnarray}\label{3entanglement helical weak}
&& \Sigma(V_1)=\det G_1+\det B_1-2\det C_1,
\end{eqnarray}
and the matrix $\textbf{V}_1$ is defined as follows:
\begin{eqnarray}\label{4entanglement helical weak}
&& \mathcal{\boldmath{A}}_1 \textbf{V}_1+\textbf{V}_1 \mathcal{\boldmath{A}}_1^T=\textbf{W},\nonumber\\
&&\textbf{V}_1=\begin{pmatrix}
\textbf{G}_1 & \textbf{C}_1\\
\textbf{C}_1^T & \textbf{B}_1
\end{pmatrix}.
\end{eqnarray}
The explicit form of the matrix $\mathcal{\boldmath{A}}_1$ is given in Eq.(\ref{MatrixMichal}), and $\textbf{W}=-\text{diag}[\gamma_f(2N_f(\omega_f)+1),  \gamma_f(2N_f(\omega_f)+1), \gamma_m(2N_m(\omega)+1), \gamma_m(2N_m(\omega)+1)]$,
where $N(\omega)=[\exp(\hbar\omega/k_BT)-1]^{-1}$ is the boson distribution function.
We note that the equation for the steady covariance matrix Eq.(\ref{4entanglement helical weak}) is linear, and one should not expect nonlinear effects governed by the covariance matrix. On the other hand, $\mathcal{\boldmath{A}}_1$ contains information about different regions of the phase space of the nonlinear system and the corresponding fixed points. The value of $\mathcal{\boldmath{A}}_1$ is different in the different regions of the phase space. Thus, nonlinearity influences magnon-photon entanglement directly.
\begin{figure*}[ht!]
    \centering
    \begin{minipage}[t]{0.45\textwidth}
        \begin{flushleft}
            a)
        \end{flushleft}
        \vspace{-1em}
        \centering
        \includegraphics[width=\textwidth]{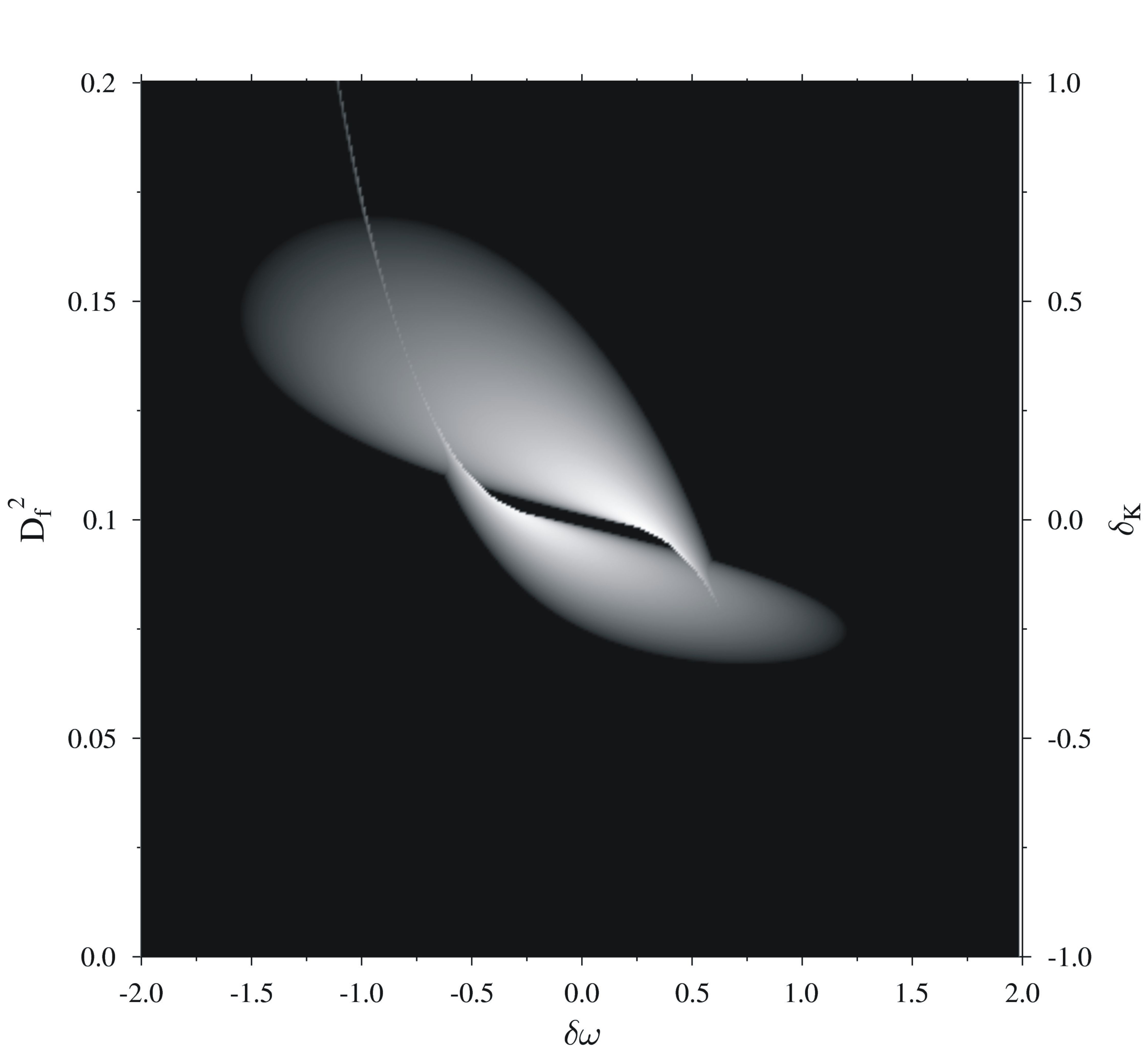}
        \label{fig12a}
    \end{minipage}
    \hfill
    \begin{minipage}[t]{0.45\textwidth}
        \begin{flushleft}
            b)
        \end{flushleft}
        \vspace{-1em}
        \centering
        \includegraphics[width=\textwidth]{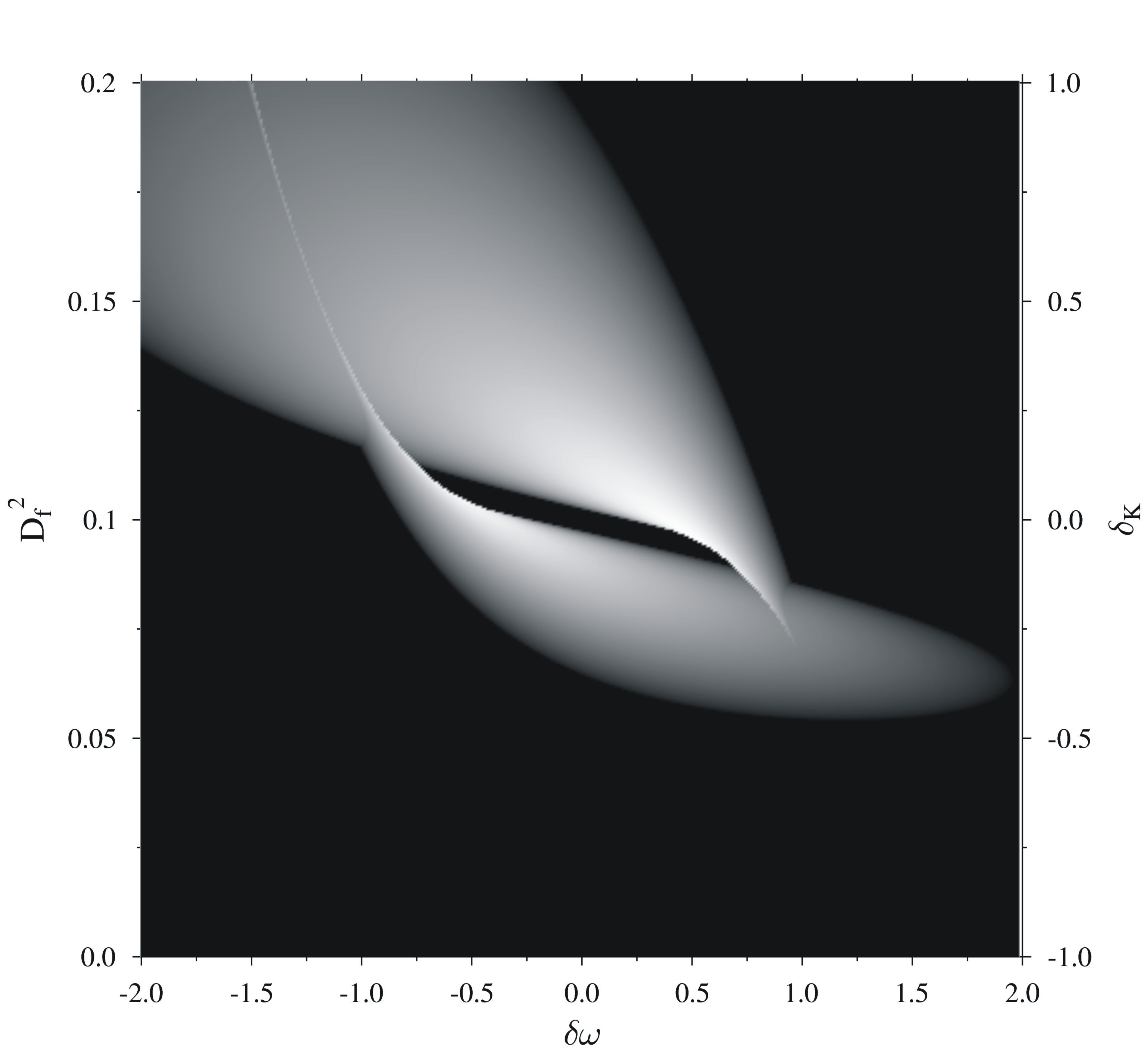}
        \label{fig12b}
    \end{minipage}
    \caption{Dependence of entanglement on the detuning $\delta\omega=\omega_0-\omega$ and parameter $\delta_k[D_f]=2D_f^2\omega_f-2K$, where $D_f=D/\sqrt{\omega_f^2+\gamma_f^2}$. The white color corresponds to the nonzero entanglement region. The temperature in the system $T=0.1$ Kelvin, meaning that $k_BT=\hbar\omega/2$. The values of the dimensionless magnetic field $B\equiv\gamma_e B/\omega$ are: a) $B^2=0.2$, b) $B^2=0.5$.}
    \label{fig:12}
\end{figure*}
Results of numerical calculations for the entanglement are plotted in Fig.\ref{fig:12}. In Fig.\ref{fig:12} we see two drop-shaped regions of nonzero entanglement. Surprisingly, we see the symmetry breaking concerning the mutual flip of the signs $(\delta\omega, \delta_k)\rightarrow(-\delta\omega, -\delta_k)$. The sizes of two nonzero entanglement regions are different. The reason for the observed asymmetry is the fact that not only the difference between magnon frequency $\omega$ and the frequency of the external magnetic field $\omega_0$ is essential but the value of the magnon frequency $\omega$ as well. Hence, the determination of the entanglement cannot be reduced solely to the value of detuning $\delta\omega=\omega_0-\omega$. As a consequence, the quantum properties of the system are characterized by symmetry lower than in the classical case Eq.(\ref{closed equation for the equilibrium magnon number}). 
\begin{figure*}[]
    \centering
    \begin{minipage}[t]{0.40\textwidth}
         \begin{flushleft}
            a)
        \end{flushleft}
        \vspace{-1em}
        \centering
        \includegraphics[width=\textwidth]{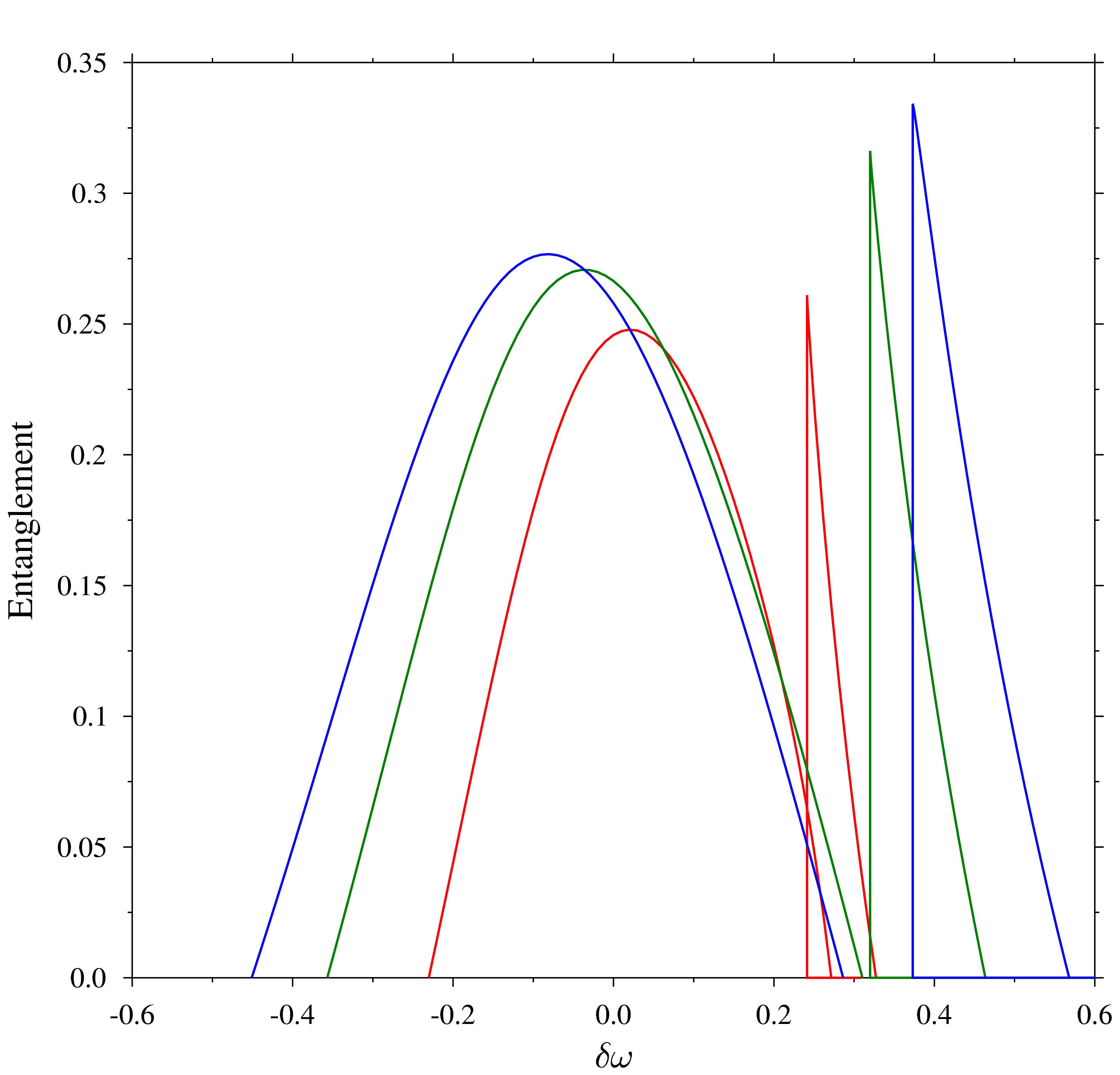}
        \label{fig13a}
    \end{minipage}
    \hfill
    \begin{minipage}[t]{0.40\textwidth}
        \begin{flushleft}
            b)
        \end{flushleft}
        \vspace{-1em}
        \centering
        \includegraphics[width=\textwidth]{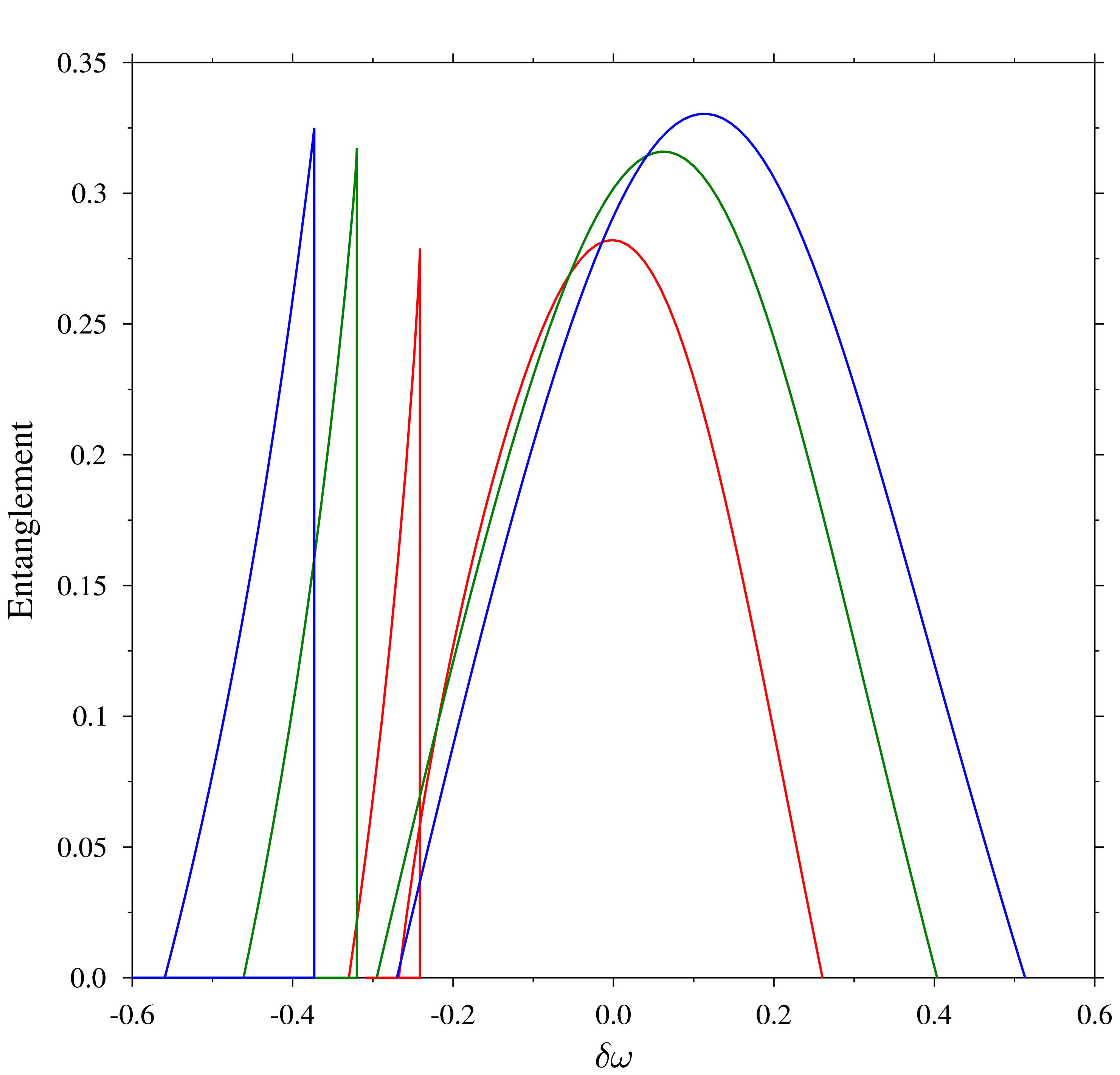}
        \label{fig13b}
    \end{minipage}
    \caption{The dependence of the entanglement on the detuning between the frequency of external time-dependent magnetic field and the magnon frequency $\delta\omega=\omega_0-\omega$ for the different values of magnetic fields: $B^2=0.1$ the red line, $B^2=0.2$ the green line, $B^2=0.3$ the blue line. The first maximum in the entanglement is associated with the small detuning case $|\delta\omega|<0.2$, while the second maximum occurs in the vicinity of a) $\delta\omega=0.4$ and b) $\delta\omega=-0.4$. With the increase of the amplitude of the time dependent magnetic field, the maximum of the entanglement moves towards the higher values of the detuning. This result is straightforward because the weak field in the sizeable detuning limit does not compensate for the thermal decay of magnons. The value of the temperature $T=0.1$ Kelvin meaning that $k_BT=\hbar\omega/2$. The values of parameter $\delta_k=\left(\frac{2D^2\omega_f}{\omega_f^2+\gamma_f^2}-2K\right)$ are: a) $\delta_K=-0.03$ b) $\delta_K=0.03$.}
    \label{fig:13}
\end{figure*}
The green lines plotted in Fig. \ref{fig:13} correspond to the $\delta_k=-0.03$ section (Fig. \ref{fig:13}a) and the $\delta_k=0.03$ section (Fig. \ref{fig:13}b) of central part of Fig. \ref{fig:12}a.
The central broad maximum of the entanglement is associated with the small detuning case $|\delta\omega|<0,2$, while the side sharp maximum in the vicinity of $\delta\omega=\pm0.4$ is counter intuitive.
Upon flipping the sign of the detuning the maximal value of the entanglement associated with the sizeable detuning case is almost the same.
Comparing the results for the entanglement plotted in Fig.\ref{fig:12} and Fig.\ref{fig:13} with the phase portrait of the system Fig.\ref{fig:10} and Fig.\ref{fig:11}, we conclude that the most significant values of the entanglement are linked with the border between Stable spiral and Stable node fixed point regions. 
In the red region (the Saddle fixed point region), the logarithmic negativity becomes imaginary, meaning that the entanglement is not defined in the Saddle fixed point region. 
\begin{figure*}
    \centering
    \begin{minipage}[]{0.45\textwidth}
        \begin{flushleft}
            a)
        \end{flushleft}
        \vspace{-1em}
        \centering
        \includegraphics[width=\textwidth]{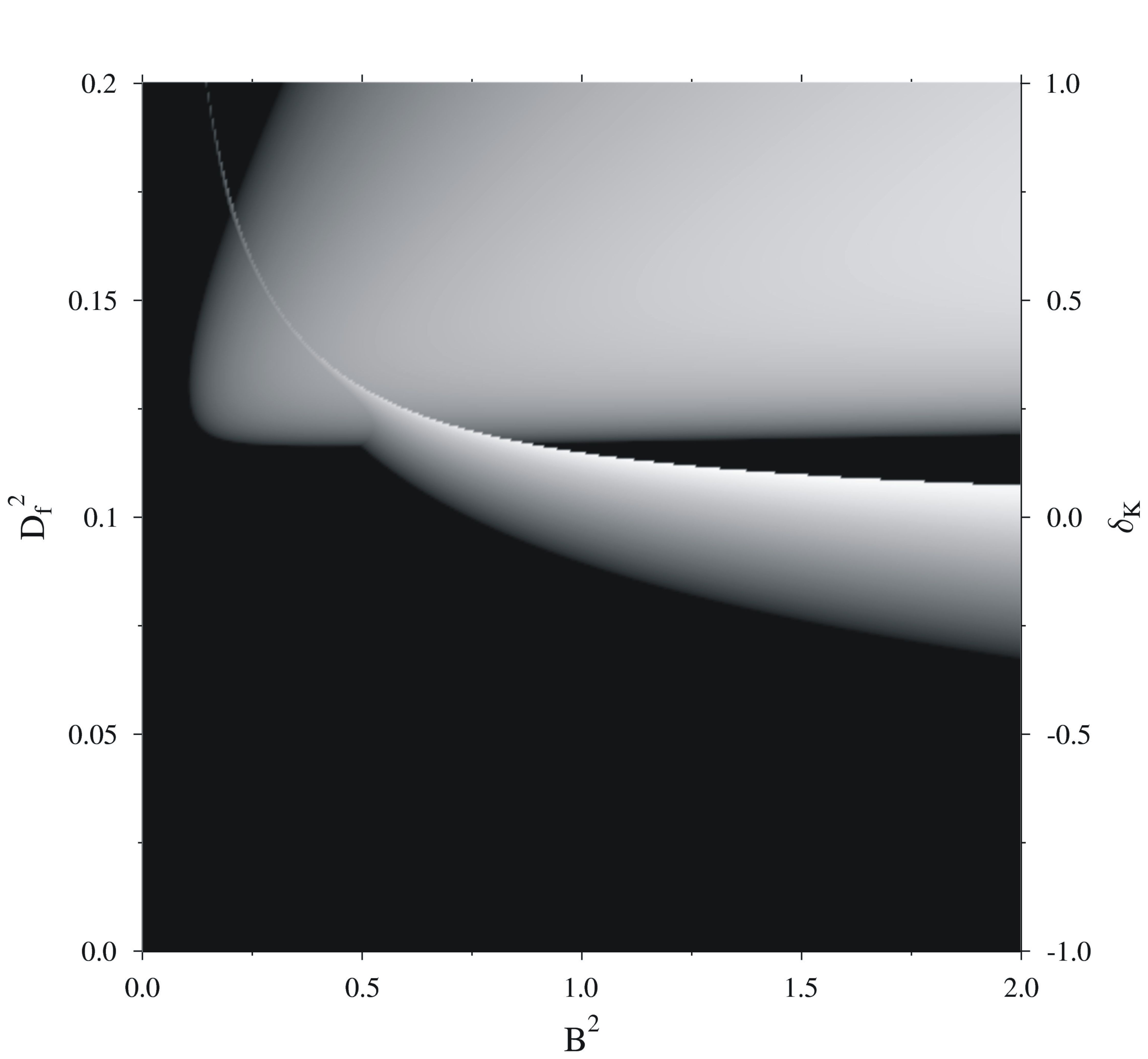}
        \label{fig14a}
    \end{minipage}
    \hfill
    \begin{minipage}[]{0.45\textwidth}
        \begin{flushleft}
            b)
        \end{flushleft}
        \vspace{-1em}
        \centering
        \includegraphics[width=\textwidth]{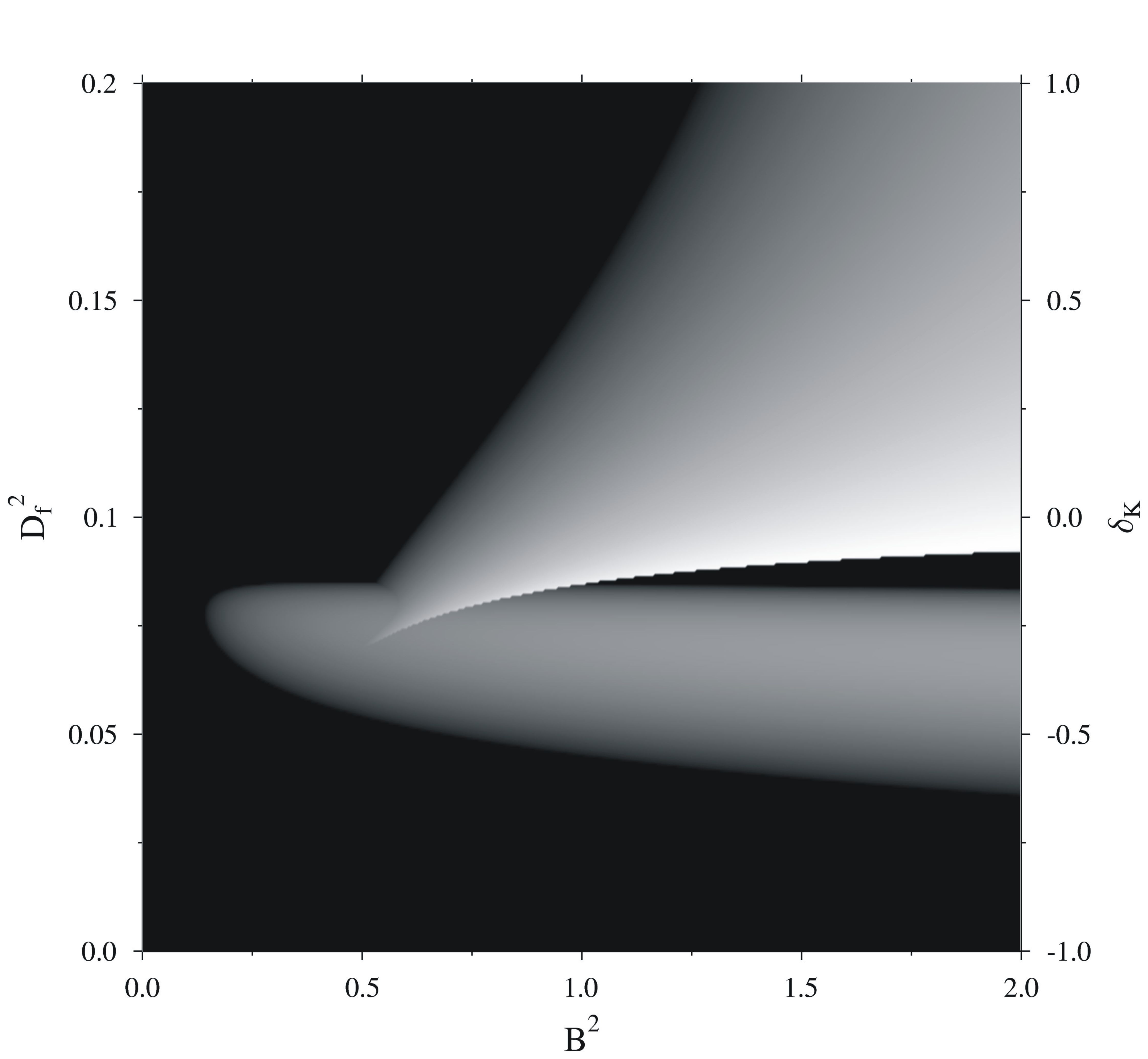}
        \label{fig14b}
    \end{minipage}
    \caption{Dependence of entanglement on the magnetic and electric fields $B$, $D_f$ and parameter $\delta_k=2D_f^2\omega_f-2K$, where $D_f=D/\sqrt{\omega_f^2+\gamma_f^2}$. The white color corresponds to the nonzero entanglement region. The temperature in the system $T=0.1$ Kelvin, meaning that $k_BT=\hbar\omega/2$. Detuning between magnon frequency $\omega$ and the frequency of the time-dependent magnetic field $\omega_0$ is equal to $\delta\omega=\omega_0-\omega$. The values of $\delta\omega$ are: a) $\delta\omega=-1$GHz, b) $\delta\omega=1$GHz.}
    \label{fig:14}
\end{figure*}

\begin{figure*}
    \centering
    \begin{minipage}[t]{0.45\textwidth}
        \begin{flushleft}
            a)
        \end{flushleft}
        \vspace{-1em}
        \centering
        \includegraphics[width=\textwidth]{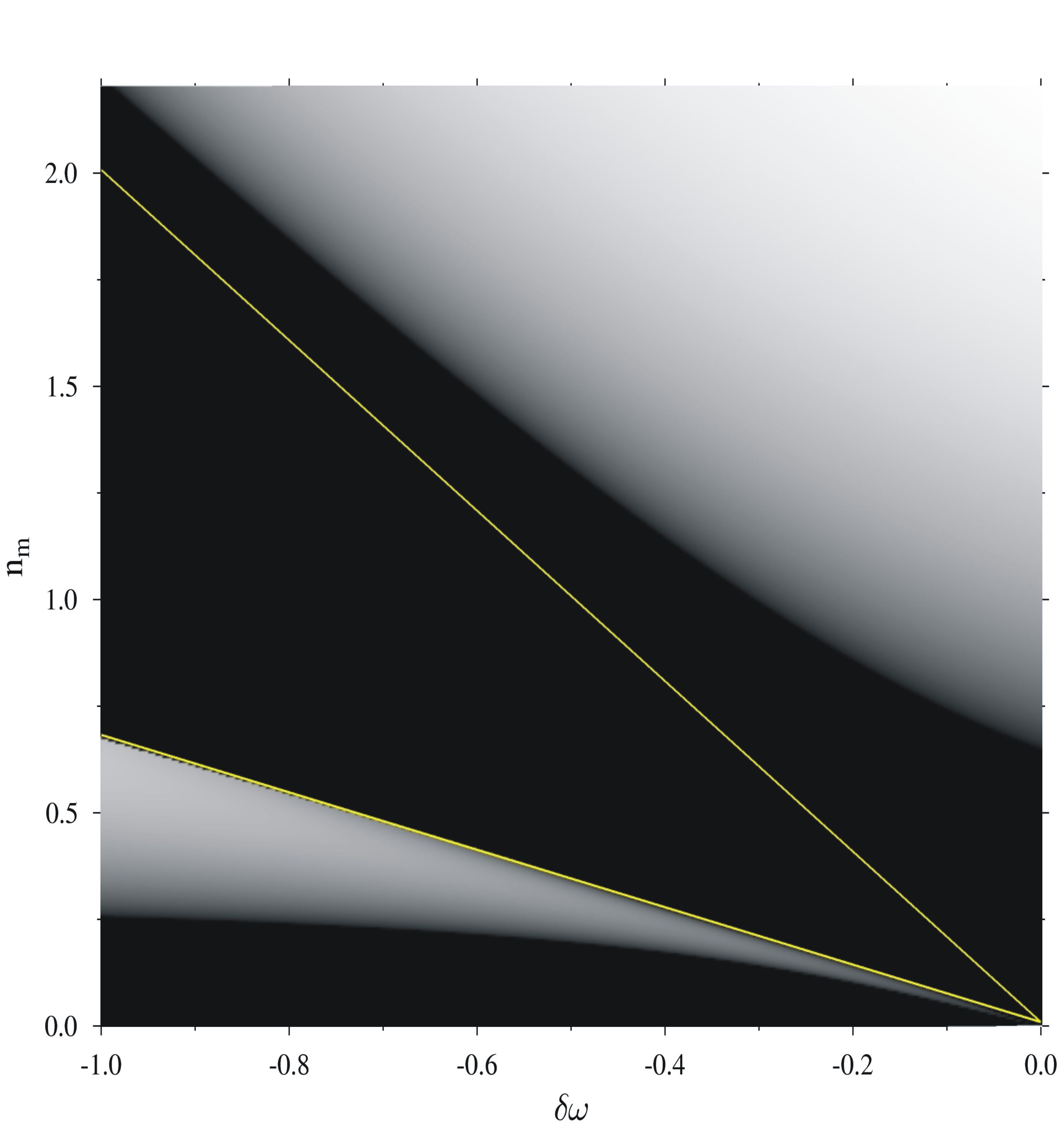}
        \label{fig15a}
    \end{minipage}
    \hfill
    \begin{minipage}[t]{0.45\textwidth}
        \begin{flushleft}
            b)
        \end{flushleft}
        \vspace{-1em}
        \centering
        \includegraphics[width=\textwidth]{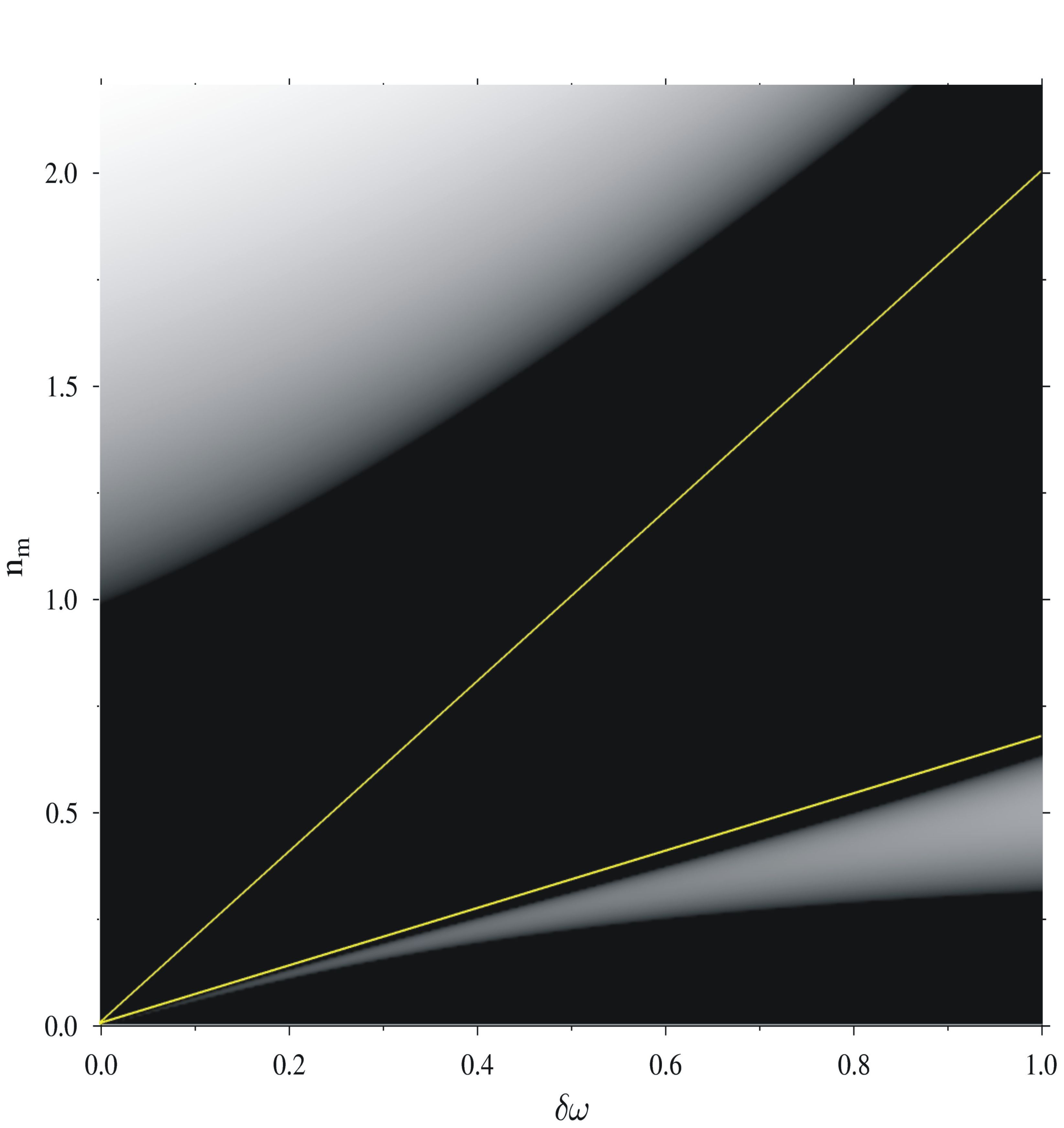}
        \label{fig15b}
    \end{minipage}
    \vspace{1em}
    \begin{minipage}[t]{0.45\textwidth}
        \begin{flushleft}
            c)
        \end{flushleft}
        \vspace{-1em}
        \centering
        \includegraphics[width=\textwidth]{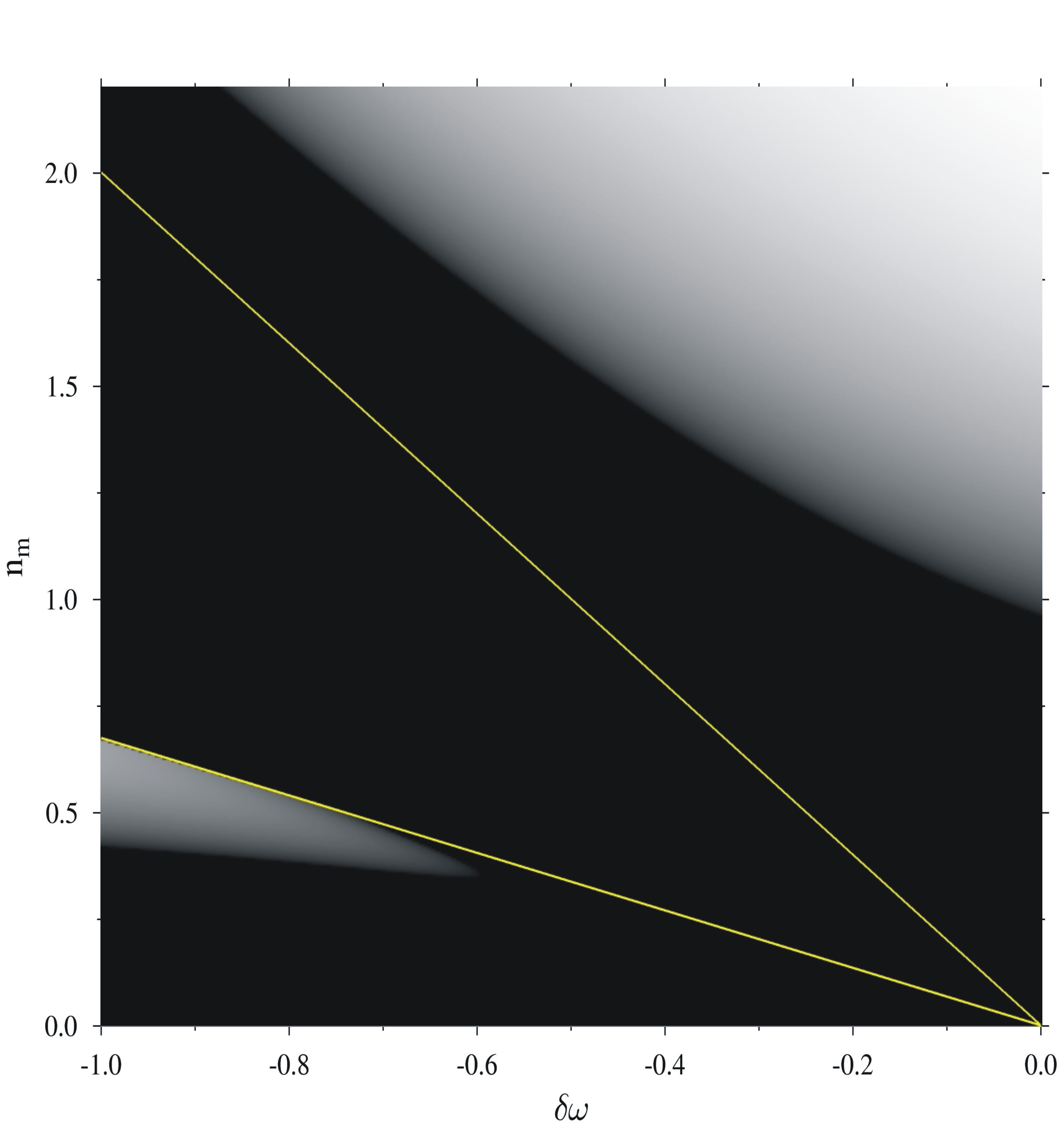}
        \label{fig15c}
    \end{minipage}
    \hfill
    \begin{minipage}[t]{0.45\textwidth}
        \begin{flushleft}
            d)
        \end{flushleft}
        \vspace{-1em}
        \centering
        \includegraphics[width=\textwidth]{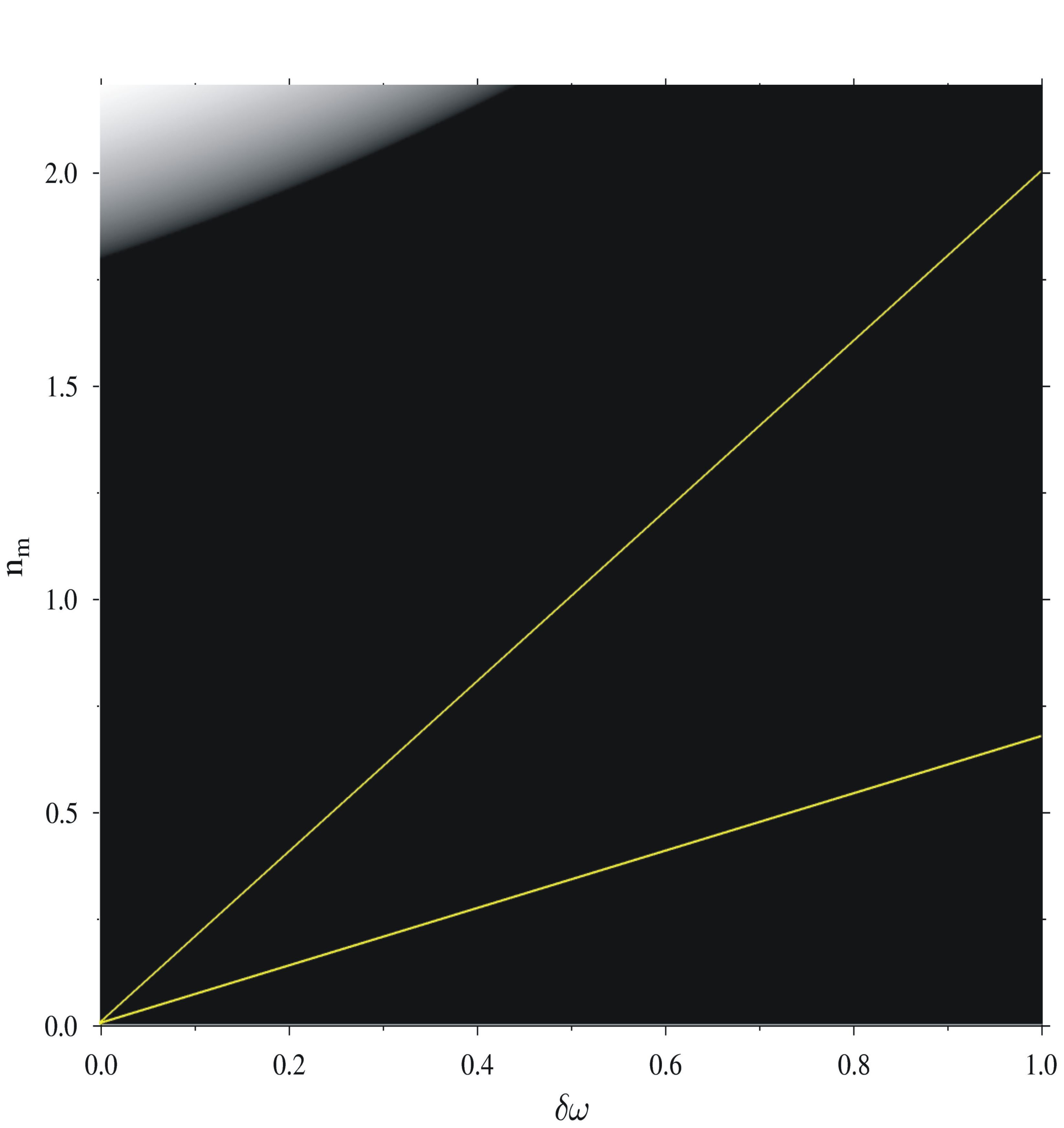}
        \label{fig15d}
    \end{minipage}
    \caption{Dependence of entanglement on the detuning $\delta\omega=\omega_0-\omega$ and magnon mean number $n_m=\beta^2$. The white color corresponds to the nonzero entanglement region. The yellow lines correspond to the border between Stable spiral and Stable node regions in Fig.\ref{fig:11} (the blue and yellow regions). The values of the magnetic field $B^2$, parameter $\delta_k=\left(\frac{2D^2\omega_f}{\omega_f^2+\gamma_f^2}-2K\right)$ and temperature $T$ in the system  are: a) $B^2=0.5$, $\delta_k=0.5$, $T=0$ Kelvin meaning that $k_BT=0$, b) $B^2=0.5$, $\delta_k=-0.5$, $T=0$ Kelvin meaning that $k_BT=0$, c) $B^2=0.5$, $\delta_k=0.5$, $T=0.1$ Kelvin meaning that $k_BT=\hbar\omega/2$, d) $B^2=0.5$, $\delta_k=-0.5$, $T=0.1$ Kelvin meaning that $k_BT=\hbar\omega/2$.}
    \label{fig:15}
\end{figure*}
In Fig.\ref{fig:14}, we plot the magnon-photon entanglement as a function of the applied external magnetic and electric fields $B, D=g_{\text{ME}}E$. The white region corresponds to the nonzero entanglement. As we see, entanglement in the system is zero for a weak electric field case. Comparing Fig.\ref{fig:14} with Fig.\ref{fig:6}, we see that the sign of the detuning between magnon frequency and the frequency of time-dependent magnetic field $\delta\omega=\omega_0-\omega$ influences the region of nonzero entanglement. Moreover, the sharp maxima of the entanglement are aligned along one of the boundary lines of the bistability region. Also in Fig.\ref{fig:12} one can see the trace of bistability border line presented in Fig.\ref{fig7}.
The yellow lines in Fig.\ref{fig:15} correspond to the border between Stable spiral and Stable node regions in Fig.\ref{fig:11} (the blue and yellow regions). As we see in Fig.\ref{fig:15}, the entanglement disappears after crossing the bottom border of the Stable node region and becomes nonzero only after crossing the upper yellow line. 
A very interesting fact is that the black region between yellow lines with zero entanglement is characterized by singularities of the covariance matrix Eq.(\ref{4entanglement helical weak}), and on the other hand, it corresponds to the Saddle point region of the dynamical system Fig.\ref{fig:11}. 

The temperature dependence of the entanglement we infer from Fig.\ref{fig:15}. Namely, the zero temperature case  $k_BT=0$ is plotted in Fig.\ref{fig:15}a and Fig.\ref{fig:15}b correspond for $\delta_K=-0.5$ and  $\delta_K=0.5$ respectively. Fig.\ref{fig:15}c and Fig.\ref{fig:15}d correspond to the nonzero temperature case for $\delta_K=-0.5$ (Fig. \ref{fig:15}c) and  $\delta_K=0.5$ (Fig. \ref{fig:15}d)

As we can see, the nonzero magnon-photon entanglement region shrinks with the increase of temperature. 

\section{Conclusions}

\begin{figure}[]
\centerline{\includegraphics[width=\columnwidth]{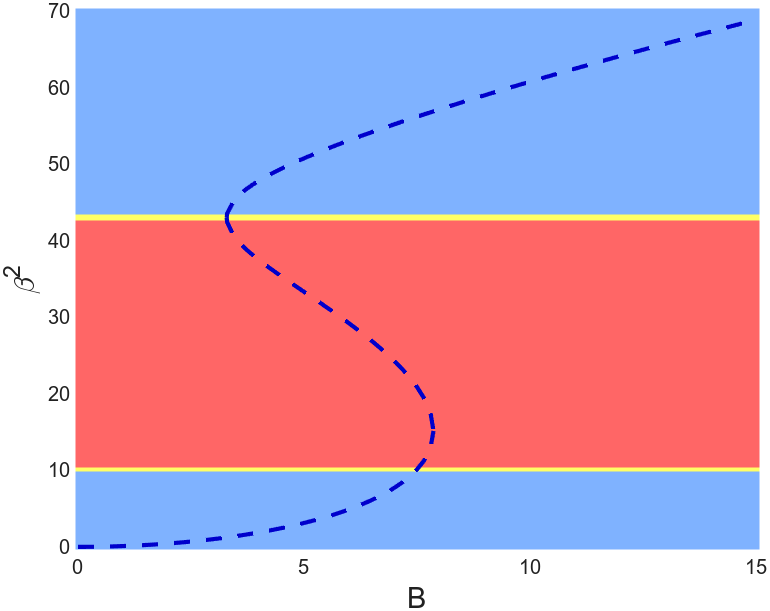}}
\caption{Schematic dependence of phase spaces on the magnetic field $B$ and the mean magnon number $\beta^2$. The dashed line indicates the dependence of $\beta^2$ on $B$. The region of saddle points is highlighted in red. The map of the region of stable node points is highlighted in yellow. The map of the region of stable spiral points is highlighted in blue. The values of the parameters read: $D = 0.16$, $K = 0.06$, $\omega_f = 0.5$, $\omega_0 = 2.5$, $\omega = 5.5$, $\gamma_f = \gamma_m = 0.5$.}
\label{fig16}
\end{figure}

In the present project, we studied the photon-magnon crystal system based on the YIG film with the periodic air holes drilled in the film.
The periodic air hols lead to the optical crystal properties, and for the confinement of the magnonic excitations, we proposed to exploit the magnon condensation effect and intraband magnon-magnon interaction originated from the perpendicular magnetocrystalline anisotropy (PMA) and bismuth doping on the YIG film. Then, the magnon-magnon interband interaction term in the magnon condensate mimics the effective Kerr effect. The magnonic spectrum has a profound minimum that grants a magnon confinement effect. Overall, confined magnons and photons in the photon-magnon crystal interact through the magneto-electric effect observed for YIG earlier. To quantify the entanglement between two continuous bosonic modes, such as magnons and photons, we followed the state-of-the-art method for today. Using the quantum Langevin equations subjected to thermal noise, we calculated the logarithmic negativity. When using this method in previous studies, the standard recipe was a linearization procedure, i.e., replacing quantum operators by their semiclassical expectation values calculated near the steady state. The cost of this approximation is the loss of information about the character of the fixed points. Namely, in the standard semi-classical approximation, a general quantum operator $\hat Q$ is replaced by $\hat Q=\langle\hat Q\rangle+\delta\hat Q$, where $\delta\hat Q$ is the deviation of operators from its steady state expectation value $\langle\hat Q\rangle$. In particular, magnon creation and annihilation operators are replaced by
$\langle\hat m^\dag\rangle=\beta^*e^{-i\varphi}$, $\langle\hat m\rangle=\beta e^{i\varphi}$.  The present work showed that phase plays a crucial role in nonlinear cases, leading to the instability and transitions between the different dynamical regimes. Before analyzing the quantum entanglement, we explored the nonlinear semiclassical dynamics in detail and precisely defined the phase space. It is well known that the typical nonlinear dynamical system holds bifurcation points and fixed points of different characters in its phase space: Saddle points, Stable and unstable spiral, and Stable node points. We showed that methods of the qualitative theory of nonlinear differential equations are also relevant for photon-magnon entanglement problems. We found that the entanglement is not defined in the  Saddle fixed point region (see red region in Fig. \ref{fig16}). The reason is the singularity in the covariance matrix that occurs in this region. On the other hand, we proved that the maximum of the entanglement corresponds to the region near the border between the Stable node and Stable spiral fixed point regions (the border between yellow and blue regions in Fig. \ref{fig16})). Besides, we calculated photon modes for a particular geometry of the photon-magnon crystal. We have analyzed the amplitude-frequency characteristics of the photon-magnon crystal and showed that due to the instability region, one could efficiently switch the mean magnon numbers in the system and control entanglement. We also studied the dependence of magnon-photon entanglement on the temperature and showed that the region of the nonzero entanglement shrinks at higher temperatures. We found that the entanglement is nonzero in the Stable spiral fixed point region as it is shown schematically in Fig. \ref{fig16}.

\clearpage

\appendix

\section{Derivation of dynamical equations}
\label{sec:Derivation of dynamical equations}

Let us consider the total Hamiltonian Eq.(\ref{total Hamiltonian})

\begin{eqnarray}
&&\hat H_{tot}=\omega_f\hat a^\dag\hat a+\omega \hat m^\dag\hat m+ K\hat m^\dag\hat m\hat m^\dag\hat m+\nonumber\\
&&\ii B\left(\hat m^\dag e^{-\ii\omega_0t}-\hat m e^{\ii\omega_0t}\right)+ \ii D\hat m^+\hat m\left(\hat a-\hat a^\dag\right).
\end{eqnarray}
Corresponding equations of motion read:
\begin{equation}
    \frac{d\hat{a}(t)}{dt} = -i \left[\hat{a}(t),\hat{H}_{eff} \right]-\gamma_f \hat{a}(t) + \sqrt{2\gamma_f}\hat{a}_{in}(t),
\end{equation}
and
\begin{equation}
    \frac{d\hat{m}(t)}{dt} = -i\left[\hat{m}(t),\hat{H}\right]-\gamma_m \hat{m}(t) + \sqrt{2\gamma_m}\hat{m}_{in}(t).
\end{equation}
Taking into account bosonic commutation relations:
\begin{eqnarray}
&&[\hat{m},\hat{m}^+] = 1,\,\,\,\,\, \hat{n} = \hat{m}^{+}\hat{m},\nonumber\\
&&[\hat{n}, \hat{m}] = [\hat{m}^{+}\hat{m}, \hat{m}] = \hat{m}[\hat{m}^+,\hat{m}] = -\hat{m},
\end{eqnarray}
and 
\begin{equation}
   [\hat{n}, \hat{m}^+] = [\hat{m}^{+}\hat{m}, \hat{m}^+] = \hat{m}^+[\hat{m},\hat{m}^+] = \hat{m}^+, 
\end{equation}
and commutation relations for the photon number operators:
\begin{eqnarray}
&&[\hat{a},\hat{a}^+]=1, \,\,\,\,\,\,\hat{n}_f=\hat{a}^+\hat{a},\\
&&[\hat{n}_f,\hat{a}]=[\hat{a}^+\hat{a},\hat{a}] =\hat{a}^+\hat{a}\hat{a} - \hat{a}\hat{a}^+\hat{a}=\nonumber\\
&&[\hat{a}^+,\hat{a}]\hat{a}=-\hat{a},\\
&& [\hat{n}_f,\hat{a}^+]=[\hat{a}^+\hat{a},\hat{a}^+] =\hat{a}^+\hat{a}\hat{a}^+ - \hat{a}^+\hat{a}^+\hat{a} = \nonumber\\
&& \hat{a}^+[\hat{a},\hat{a}^+]=\hat{a}^+,
\end{eqnarray}
we obtain equations
\begin{eqnarray}
&&[\hat{a}(t),\hat{H}] = \omega_f[\hat{a},\hat{a}^+\hat{a}]+iD[\hat{a},(\hat{a}-\hat{a}^+)]\hat{m}^+\hat{m} = \nonumber \\
&&\omega_f\hat{a} - iD\hat{m}^+\hat{m},\\
&&[\hat{m}(t),\hat{H}] = \omega \left[ \hat{m},\hat{m}^+\hat{m} \right]+iD(\hat{a}-\hat{a}^+)[\hat{m},\hat{m}^+\hat{m}] + \nonumber \\
&& iB[\hat{m},\hat{m}^+]e^{-i\omega_0t} + K[\hat{m},\hat{m}^{+}\hat{m}\hat{m}^{+}\hat{m}] = \nonumber\\
&&\omega\hat{m} + iD(\hat{a}-\hat{a}^+)\hat{m} + iBe^{-i\omega_0t}+K(\hat{m}^+\hat{m}\hat{m} + \hat{m}\hat{m}^+\hat{m}),\\
&&[\hat{m},\hat{m}^{+}\hat{m}\hat{m}^{+}\hat{m}] = \hat{m}^{+}\hat{m}[\hat{m},\hat{m}^{+}\hat{m}] + [\hat{m},\hat{m}^{+}\hat{m}]\hat{m}^{+}\hat{m}= \nonumber \\
&& \hat{m}^{+}\hat{m}\hat{m} + \hat{m}\hat{m}^+\hat{m}.
\end{eqnarray}
Taking into account the identity
\begin{eqnarray}
&&[A,BC] = ABC - BCA = ABC - BAC + BAC - BCA = \nonumber \\
&& [A,B]C + B[A,C],
\end{eqnarray}
we obtain the set of equations:
\begin{eqnarray}
&&\frac{d\hat{a}(t)}{dt} = -i\omega_f\hat{a} - D\hat{m}^{+}\hat{m} - \gamma_f\hat{a}(t) + \sqrt{2\gamma_f}\hat{a}_{in}(t),\\
&&\frac{d\hat{m}(t)}{dt} = -i\omega\hat{m} + D(\hat{a}-\hat{a}^+)\hat{m} + Be^{-i\omega_0t} - \nonumber \\
&& iK(\hat{m}^+\hat{m}\hat{m}+ \hat{m}\hat{m}^+\hat{m}) - \gamma_m\hat{m}(t) + \sqrt{2\gamma_m}\hat{m}_{in}(t),\\
&&\frac{d\hat{m}(t)}{dt} = -i\omega\hat{m} + D(\hat{a}-\hat{a}^+)\hat{m} + Be^{-i\omega_0t} - \nonumber \\ 
&& iK(2\hat{m}^+\hat{m}-1)\hat{m} - \gamma_m\hat{m}(t) + \sqrt{2\gamma_m}\hat{m}_{in}(t).
\end{eqnarray}
After the standard transformation:
\begin{eqnarray}
\hat{m}\to \hat{m}e^{-i\omega_0t},\,\,\,\,\hat{m}_{in}\to \hat{m}_{in}e^{-i\omega_0t},
\end{eqnarray}
the set of equations takes the form:
\begin{eqnarray}
&& \frac{d\hat{a}(t)}{dt} = -i\omega_f\hat{a} - D\hat{m}^{+}\hat{m} - \gamma_f\hat{a}(t) + \sqrt{2\gamma_f}\hat{a}_{in}(t),\nonumber \\
&& \Delta = \omega_0 - \omega,\\
&& \frac{d\hat{m}(t)}{dt} = -i\Delta\hat{m}+ D(\hat{a}-\hat{a}^+)\hat{m} - \nonumber \\
&& iK(\hat{m}^+\hat{m}\hat{m} + \hat{m}\hat{m}^+\hat{m}) - \gamma_m\hat{m}(t) + \sqrt{2\gamma_m}\hat{m}_{in}(t) + B.
\end{eqnarray}
When performing the averaging procedure we take into account the standard approach \cite{PhysRevLett.121.203601}:
\begin{eqnarray}
&& \left\langle \hat{a}_{in}(t) \right\rangle = \left\langle \hat{a}_{in}^+(t) \right\rangle = \left\langle \hat{m}_{in}(t) \right\rangle = \langle \hat{m}_{in}^+(t) \rangle = 0,\nonumber\\
&&\frac{d\left\langle \hat{a} \right\rangle}{dt} = -i\omega_f\left\langle \hat{a} \right\rangle - D \left\langle \hat{m}^+\hat{m} \right\rangle - \gamma_f\left\langle \hat{a} \right\rangle,
\end{eqnarray}
and obtain:
\begin{eqnarray}
&& \frac{d\left\langle \hat{m} \right\rangle}{dt} = -i\Delta\left\langle \hat{m} \right\rangle + D \left\langle (\hat{a}-\hat{a}^+)\hat{m} \right\rangle - \nonumber \\
&&iK\left( \left\langle \hat{m}^+\hat{m}\hat{m} \right\rangle +  \left\langle \hat{m}\hat{m}^+\hat{m} \right\rangle \right) - 
 \gamma_m\left\langle \hat{m} \right\rangle + B\left\langle \hat{Q} \right\rangle.
\end{eqnarray}
Here $Q=\left\langle \hat{Q} \right\rangle$  is the quantum-mechanical average of operator and $Q$ is the expectation value. We adopt the semi-classical averaging procedure \cite{PhysRevLett.121.203601}:
\begin{eqnarray}
&&\left\langle (\hat{a}-\hat{a}^+)\hat{m} \right\rangle = \left\langle (\hat{a}-\hat{a}^+) \right\rangle \left\langle \hat{m} \right\rangle = (a - a^+)m,\nonumber\\
&&\left\langle \hat{m}^+\hat{m} \right\rangle = \left| m \right|^2,\\
&&\left\langle \hat{m}^+\hat{m}\hat{m} \right\rangle = \left\langle \hat{m}\hat{m}^+\hat{m}\right\rangle = \left| m \right|^2 m.
\end{eqnarray}
Then the dynamical equations take the form:
\begin{eqnarray}
\label{A.23}
&&\frac{da}{dt}= -i\omega_fa - D \left| m \right|^2 -\gamma_f a,\\
&&\frac{dm}{dt}= \left( -i\Delta + D(a-a^+) - 2iK \left| m \right|^2 - \gamma_m \right) m + B.\nonumber
\end{eqnarray}
Introducing phases and amplitudes of the averaged quantities $a=\left| a \right| e^{i\phi} = \alpha e^{i\phi}, m = \left| m\right| e^{i\varphi} = \beta e^{i\varphi}$ we deduce:
\begin{eqnarray}
&&\frac{da}{dt} = \dot{\alpha} e^{i\phi} + i\dot{\phi}\alpha e^{i\phi} = -i\omega_f \alpha e^{i\phi} - D \beta^2 - \gamma_f a e^{i\phi},\nonumber\\
&&\dot{\alpha} + i \dot{\phi} \alpha = -i \omega_f \alpha - D \beta^2 e^{-i\phi} - \gamma_f a,\\
&&\dot{\alpha} = - D \beta^2 \cos{\phi} - \gamma_f a,\nonumber\\
&&\alpha \dot{\phi} = -\omega_f \alpha - D \beta^2 \sin{\phi}.\nonumber
\end{eqnarray}
Taking into account that:
\begin{eqnarray}
&&\frac{dm}{dt} = \dot\beta e^{i\varphi} + i\dot{\varphi}\beta e^{i\varphi}  \nonumber \\ 
&& = \left( -i\Delta + 2iD \alpha \sin{\phi} - 2iK\beta^2 - \gamma_m \right)\beta e^{i\varphi} + B,\\
&&\dot{\beta}+ i\dot{\varphi}\beta = \left( -i\Delta + 2iD \alpha \sin{\phi} - 2iK\beta^2 - \gamma_m \right)\beta + Be^{-i\varphi},\\
&&\beta \dot{\varphi} =  \left( -\Delta + 2D \alpha \sin{\phi} - 2K\beta^2\right)\beta - B\sin{\varphi},\\
&&\dot{\beta} = -\gamma_m \beta + B \cos{\varphi},
\end{eqnarray}
we obtain the set of equations:
\begin{eqnarray}
&&\dot{\alpha} = -D \beta^2 \cos{\phi} - \gamma_f a,\\
&&\alpha \dot{\phi} = -\omega_f \alpha + D \beta^2 \sin{\phi},\\
&&\dot{\beta} = -\gamma_m \beta + B \cos{\varphi}, \\
&&\beta \dot{\varphi} = \left( -\Delta + 2D \alpha \sin{\phi} - 2K\beta^2\right)\beta-\nonumber\\
&& B\sin{\varphi},
\end{eqnarray}
which we rewrite in the final form:
\begin{eqnarray}
&&\dot{\alpha}=-D\beta^2\cos\phi-\gamma_f\alpha,\nonumber\\
&&\dot{\phi}=-\omega_f+\frac{D\beta^2}{\alpha}\sin\phi,\nonumber\\
&&\dot{\beta}=-\gamma_m\beta+B\cos\varphi,\nonumber\\
&&\dot{\varphi}=\omega-\omega_0+2D\alpha\sin\phi-2K\beta^2-\frac{B\sin\varphi}{\beta}. 
\end{eqnarray}

\bibliographystyle{elsarticle-num} 
\bibliography{TEXT_21.10.24}

\begin{thebibliography}{10}
\expandafter\ifx\csname url\endcsname\relax
  \def\url#1{\texttt{#1}}\fi
\expandafter\ifx\csname urlprefix\endcsname\relax\def\urlprefix{URL }\fi
\expandafter\ifx\csname href\endcsname\relax
  \def\href#1#2{#2} \def\path#1{#1}\fi

\bibitem{PhysRevLett.111.127003}
H.~Huebl, C.~W. Zollitsch, J.~Lotze, F.~Hocke, M.~Greifenstein, A.~Marx, R.~Gross, S.~T.~B. Goennenwein, \href{https://link.aps.org/doi/10.1103/PhysRevLett.111.127003}{High cooperativity in coupled microwave resonator ferrimagnetic insulator hybrids}, Phys. Rev. Lett. 111 (2013) 127003.
\newblock \href {https://doi.org/10.1103/PhysRevLett.111.127003} {\path{doi:10.1103/PhysRevLett.111.127003}}.
\newline\urlprefix\url{https://link.aps.org/doi/10.1103/PhysRevLett.111.127003}

\bibitem{PhysRevLett.113.083603}
Y.~Tabuchi, S.~Ishino, T.~Ishikawa, R.~Yamazaki, K.~Usami, Y.~Nakamura, \href{https://link.aps.org/doi/10.1103/PhysRevLett.113.083603}{Hybridizing ferromagnetic magnons and microwave photons in the quantum limit}, Phys. Rev. Lett. 113 (2014) 083603.
\newblock \href {https://doi.org/10.1103/PhysRevLett.113.083603} {\path{doi:10.1103/PhysRevLett.113.083603}}.
\newline\urlprefix\url{https://link.aps.org/doi/10.1103/PhysRevLett.113.083603}

\bibitem{PhysRevLett.113.156401}
X.~Zhang, C.-L. Zou, L.~Jiang, H.~X. Tang, \href{https://link.aps.org/doi/10.1103/PhysRevLett.113.156401}{Strongly coupled magnons and cavity microwave photons}, Phys. Rev. Lett. 113 (2014) 156401.
\newblock \href {https://doi.org/10.1103/PhysRevLett.113.156401} {\path{doi:10.1103/PhysRevLett.113.156401}}.
\newline\urlprefix\url{https://link.aps.org/doi/10.1103/PhysRevLett.113.156401}

\bibitem{PhysRevB.91.104410}
J.~A. Haigh, N.~J. Lambert, A.~C. Doherty, A.~J. Ferguson, \href{https://link.aps.org/doi/10.1103/PhysRevB.91.104410}{Dispersive readout of ferromagnetic resonance for strongly coupled magnons and microwave photons}, Phys. Rev. B 91 (2015) 104410.
\newblock \href {https://doi.org/10.1103/PhysRevB.91.104410} {\path{doi:10.1103/PhysRevB.91.104410}}.
\newline\urlprefix\url{https://link.aps.org/doi/10.1103/PhysRevB.91.104410}

\bibitem{PhysRevResearch.3.013277}
J.~Graf, S.~Sharma, H.~Huebl, S.~V. Kusminskiy, \href{https://link.aps.org/doi/10.1103/PhysRevResearch.3.013277}{Design of an optomagnonic crystal: Towards optimal magnon-photon mode matching at the microscale}, Phys. Rev. Res. 3 (2021) 013277.
\newblock \href {https://doi.org/10.1103/PhysRevResearch.3.013277} {\path{doi:10.1103/PhysRevResearch.3.013277}}.
\newline\urlprefix\url{https://link.aps.org/doi/10.1103/PhysRevResearch.3.013277}

\bibitem{pantazopoulos2019high}
P.~A. Pantazopoulos, K.~L. Tsakmakidis, E.~Almpanis, G.~P. Zouros, N.~Stefanou, High-efficiency triple-resonant inelastic light scattering in planar optomagnonic cavities, New Journal of Physics 21~(9) (2019) 095001.

\bibitem{PhysRevB.96.104425}
P.~A. Pantazopoulos, N.~Stefanou, E.~Almpanis, N.~Papanikolaou, \href{https://link.aps.org/doi/10.1103/PhysRevB.96.104425}{Photomagnonic nanocavities for strong light--spin-wave interaction}, Phys. Rev. B 96 (2017) 104425.
\newblock \href {https://doi.org/10.1103/PhysRevB.96.104425} {\path{doi:10.1103/PhysRevB.96.104425}}.
\newline\urlprefix\url{https://link.aps.org/doi/10.1103/PhysRevB.96.104425}

\bibitem{PhysRevB.111.075415}
C.~Jasiukiewicz, A.~Sinner, I.~Weymann, T.~Doma\ifmmode~\acute{n}\else \'{n}\fi{}ski, L.~Chotorlishvili, \href{https://link.aps.org/doi/10.1103/PhysRevB.111.075415}{Entanglement between quantum dots transmitted via a majorana wire: Insights from the fermionic negativity, concurrence, and quantum mutual information}, Phys. Rev. B 111 (2025) 075415.
\newblock \href {https://doi.org/10.1103/PhysRevB.111.075415} {\path{doi:10.1103/PhysRevB.111.075415}}.
\newline\urlprefix\url{https://link.aps.org/doi/10.1103/PhysRevB.111.075415}

\bibitem{bachtold2022mesoscopic}
A.~Bachtold, J.~Moser, M.~Dykman, Mesoscopic physics of nanomechanical systems, Reviews of Modern Physics 94~(4) (2022) 045005.

\bibitem{aspelmeyer2014cavity}
M.~Aspelmeyer, T.~J. Kippenberg, F.~Marquardt, Cavity optomechanics, Reviews of Modern Physics 86~(4) (2014) 1391.

\bibitem{singh2022hybrid}
A.~Singh, L.~Chotorlishvili, Z.~Toklikishvili, I.~Tralle, S.~Mishra, Hybrid quantum--classical chaotic nems, Physica D: Nonlinear Phenomena 439 (2022) 133418.

\bibitem{PhysRevA.102.022816}
Y.~Chougale, J.~Talukdar, T.~Ramos, R.~Nath, \href{https://link.aps.org/doi/10.1103/PhysRevA.102.022816}{Dynamics of {Rydberg} excitations and quantum correlations in an atomic array coupled to a photonic crystal waveguide}, Phys. Rev. A 102 (2020) 022816.
\newblock \href {https://doi.org/10.1103/PhysRevA.102.022816} {\path{doi:10.1103/PhysRevA.102.022816}}.
\newline\urlprefix\url{https://link.aps.org/doi/10.1103/PhysRevA.102.022816}

\bibitem{manzoni2017designing}
M.~T. Manzoni, L.~Mathey, D.~E. Chang, Designing exotic many-body states of atomic spin and motion in photonic crystals, Nat. Commun. 8~(1) (2017) 1--9.

\bibitem{goban2014atom}
A.~Goban, C.-L. Hung, S.-P. Yu, J.~Hood, J.~Muniz, J.~Lee, M.~Martin, A.~McClung, K.~Choi, D.~E. Chang, et~al., Atom--light interactions in photonic crystals, Nat. Commun. 5~(1) (2014) 1--9.

\bibitem{hung2013trapped}
C.~Hung, S.~Meenehan, D.~Chang, O.~Painter, H.~Kimble, Trapped atoms in one-dimensional photonic crystals, New Journal of Physics 15~(8) (2013) 083026.

\bibitem{douglas2015quantum}
J.~S. Douglas, H.~Habibian, C.-L. Hung, A.~V. Gorshkov, H.~J. Kimble, D.~E. Chang, Quantum many-body models with cold atoms coupled to photonic crystals, Nature Photonics 9~(5) (2015) 326--331.

\bibitem{PhysRevLett.121.203601}
J.~Li, S.-Y. Zhu, G.~S. Agarwal, \href{https://link.aps.org/doi/10.1103/PhysRevLett.121.203601}{Magnon-photon-phonon entanglement in cavity magnomechanics}, Phys. Rev. Lett. 121 (2018) 203601.
\newblock \href {https://doi.org/10.1103/PhysRevLett.121.203601} {\path{doi:10.1103/PhysRevLett.121.203601}}.
\newline\urlprefix\url{https://link.aps.org/doi/10.1103/PhysRevLett.121.203601}

\bibitem{PhysRevB.107.115126}
Z.~Toklikishvili, L.~Chotorlishvili, R.~Khomeriki, V.~Jandieri, J.~Berakdar, \href{https://link.aps.org/doi/10.1103/PhysRevB.107.115126}{Electrically controlled entanglement of cavity photons with electromagnons}, Phys. Rev. B 107 (2023) 115126.
\newblock \href {https://doi.org/10.1103/PhysRevB.107.115126} {\path{doi:10.1103/PhysRevB.107.115126}}.
\newline\urlprefix\url{https://link.aps.org/doi/10.1103/PhysRevB.107.115126}

\bibitem{PhysRevB.106.054425}
H.~Pan, Y.~Yang, Z.~H. An, C.-M. Hu, \href{https://link.aps.org/doi/10.1103/PhysRevB.106.054425}{Bistability in dissipatively coupled cavity magnonics}, Phys. Rev. B 106 (2022) 054425.
\newblock \href {https://doi.org/10.1103/PhysRevB.106.054425} {\path{doi:10.1103/PhysRevB.106.054425}}.
\newline\urlprefix\url{https://link.aps.org/doi/10.1103/PhysRevB.106.054425}

\bibitem{PhysRevA.104.033708}
M.~Wang, C.~Kong, Z.-Y. Sun, D.~Zhang, Y.-Y. Wu, L.-L. Zheng, \href{https://link.aps.org/doi/10.1103/PhysRevA.104.033708}{Nonreciprocal high-order sidebands induced by magnon kerr nonlinearity}, Phys. Rev. A 104 (2021) 033708.
\newblock \href {https://doi.org/10.1103/PhysRevA.104.033708} {\path{doi:10.1103/PhysRevA.104.033708}}.
\newline\urlprefix\url{https://link.aps.org/doi/10.1103/PhysRevA.104.033708}

\bibitem{PhysRevA.106.012419}
Z.-B. Yang, W.-J. Wu, J.~Li, Y.-P. Wang, J.~Q. You, \href{https://link.aps.org/doi/10.1103/PhysRevA.106.012419}{Steady-entangled-state generation via the cross-kerr effect in a ferrimagnetic crystal}, Phys. Rev. A 106 (2022) 012419.
\newblock \href {https://doi.org/10.1103/PhysRevA.106.012419} {\path{doi:10.1103/PhysRevA.106.012419}}.
\newline\urlprefix\url{https://link.aps.org/doi/10.1103/PhysRevA.106.012419}

\bibitem{PhysRevApplied.12.034001}
C.~Kong, H.~Xiong, Y.~Wu, \href{https://link.aps.org/doi/10.1103/PhysRevApplied.12.034001}{Magnon-induced nonreciprocity based on the magnon kerr effect}, Phys. Rev. Appl. 12 (2019) 034001.
\newblock \href {https://doi.org/10.1103/PhysRevApplied.12.034001} {\path{doi:10.1103/PhysRevApplied.12.034001}}.
\newline\urlprefix\url{https://link.aps.org/doi/10.1103/PhysRevApplied.12.034001}

\bibitem{PhysRevB.107.064417}
G.-Q. Zhang, Y.~Wang, W.~Xiong, \href{https://link.aps.org/doi/10.1103/PhysRevB.107.064417}{Detection sensitivity enhancement of magnon kerr nonlinearity in cavity magnonics induced by coherent perfect absorption}, Phys. Rev. B 107 (2023) 064417.
\newblock \href {https://doi.org/10.1103/PhysRevB.107.064417} {\path{doi:10.1103/PhysRevB.107.064417}}.
\newline\urlprefix\url{https://link.aps.org/doi/10.1103/PhysRevB.107.064417}

\bibitem{PhysRevLett.129.123601}
R.-C. Shen, J.~Li, Z.-Y. Fan, Y.-P. Wang, J.~Q. You, \href{https://link.aps.org/doi/10.1103/PhysRevLett.129.123601}{Mechanical bistability in kerr-modified cavity magnomechanics}, Phys. Rev. Lett. 129 (2022) 123601.
\newblock \href {https://doi.org/10.1103/PhysRevLett.129.123601} {\path{doi:10.1103/PhysRevLett.129.123601}}.
\newline\urlprefix\url{https://link.aps.org/doi/10.1103/PhysRevLett.129.123601}

\bibitem{amazioug2023enhancement}
M.~Amazioug, B.~Teklu, M.~Asjad, Enhancement of magnon--photon--phonon entanglement in a cavity magnomechanics with coherent feedback loop, Scientific Reports 13~(1) (2023) 3833.

\bibitem{amazioug2023feedback}
M.~Amazioug, S.~Singh, B.~Teklu, M.~Asjad, Feedback control of quantum correlations in a cavity magnomechanical system with magnon squeezing, Entropy 25~(10) (2023) 1462.

\bibitem{adani2024critical}
M.~Adani, S.~Cavazzoni, B.~Teklu, P.~Bordone, M.~G. Paris, Critical metrology of minimally accessible anisotropic spin chains, Scientific Reports 14~(1) (2024) 19933.

\bibitem{PhysRevB.111.024315}
R.~K. Shukla, L.~Chotorlishvili, S.~K. Mishra, F.~Iemini, \href{https://link.aps.org/doi/10.1103/PhysRevB.111.024315}{Prethermal floquet time crystals in chiral multiferroic chains and applications as quantum sensors of ac fields}, Phys. Rev. B 111 (2025) 024315.
\newblock \href {https://doi.org/10.1103/PhysRevB.111.024315} {\path{doi:10.1103/PhysRevB.111.024315}}.
\newline\urlprefix\url{https://link.aps.org/doi/10.1103/PhysRevB.111.024315}

\bibitem{PhysRevLett.84.2726}
R.~Simon, \href{https://link.aps.org/doi/10.1103/PhysRevLett.84.2726}{Peres-horodecki separability criterion for continuous variable systems}, Phys. Rev. Lett. 84 (2000) 2726--2729.
\newblock \href {https://doi.org/10.1103/PhysRevLett.84.2726} {\path{doi:10.1103/PhysRevLett.84.2726}}.
\newline\urlprefix\url{https://link.aps.org/doi/10.1103/PhysRevLett.84.2726}

\bibitem{PhysRevLett.84.2722}
L.-M. Duan, G.~Giedke, J.~I. Cirac, P.~Zoller, \href{https://link.aps.org/doi/10.1103/PhysRevLett.84.2722}{Inseparability criterion for continuous variable systems}, Phys. Rev. Lett. 84 (2000) 2722--2725.
\newblock \href {https://doi.org/10.1103/PhysRevLett.84.2722} {\path{doi:10.1103/PhysRevLett.84.2722}}.
\newline\urlprefix\url{https://link.aps.org/doi/10.1103/PhysRevLett.84.2722}

\bibitem{PhysRevLett.96.050503}
M.~Hillery, M.~S. Zubairy, \href{https://link.aps.org/doi/10.1103/PhysRevLett.96.050503}{Entanglement conditions for two-mode states}, Phys. Rev. Lett. 96 (2006) 050503.
\newblock \href {https://doi.org/10.1103/PhysRevLett.96.050503} {\path{doi:10.1103/PhysRevLett.96.050503}}.
\newline\urlprefix\url{https://link.aps.org/doi/10.1103/PhysRevLett.96.050503}

\bibitem{adesso2007entanglement}
G.~Adesso, F.~Illuminati, Entanglement in continuous-variable systems: recent advances and current perspectives, Journal of Physics A: Mathematical and Theoretical 40~(28) (2007) 7821.

\bibitem{PhysRevLett.95.090503}
M.~B. Plenio, \href{https://link.aps.org/doi/10.1103/PhysRevLett.95.090503}{Logarithmic negativity: A full entanglement monotone that is not convex}, Phys. Rev. Lett. 95 (2005) 090503.
\newblock \href {https://doi.org/10.1103/PhysRevLett.95.090503} {\path{doi:10.1103/PhysRevLett.95.090503}}.
\newline\urlprefix\url{https://link.aps.org/doi/10.1103/PhysRevLett.95.090503}

\bibitem{PhysRevLett.106.247203}
T.~Liu, G.~Vignale, \href{https://link.aps.org/doi/10.1103/PhysRevLett.106.247203}{Electric control of spin currents and spin-wave logic}, Phys. Rev. Lett. 106 (2011) 247203.
\newblock \href {https://doi.org/10.1103/PhysRevLett.106.247203} {\path{doi:10.1103/PhysRevLett.106.247203}}.
\newline\urlprefix\url{https://link.aps.org/doi/10.1103/PhysRevLett.106.247203}

\bibitem{PhysRevB.80.140412}
Y.~Yamasaki, Y.~Kohara, Y.~Tokura, \href{https://link.aps.org/doi/10.1103/PhysRevB.80.140412}{Quantum magnetoelectric effect in iron garnet}, Phys. Rev. B 80 (2009) 140412.
\newblock \href {https://doi.org/10.1103/PhysRevB.80.140412} {\path{doi:10.1103/PhysRevB.80.140412}}.
\newline\urlprefix\url{https://link.aps.org/doi/10.1103/PhysRevB.80.140412}

\bibitem{trybus2024dielectric}
M.~Trybus, L.~Chotorlishvili, E.~Jartych, Dielectric and magnetoelectric properties of tgs--magnetite composite, Molecules 29~(6) (2024) 1378.

\bibitem{kolesnikov2024energy}
S.~Kolesnikov, E.~Sapronova, Energy barriers for the spontaneous magnetization reversal of atomic co chains on the surface pt (664) in the model of dzyaloshinskii--moriya interaction, Journal of Surface Investigation: X-ray, Synchrotron and Neutron Techniques 18~(1) (2024) 150--155.

\bibitem{kolesnikov2023influence}
S.~Kolesnikov, E.~S. Sapronova, I.~N. Kolesnikova, An influence of the dzyaloshinskii-moriya interaction on the magnetization reversal process of the finite-size co chains on pt (664) surface, Journal of Magnetism and Magnetic Materials 579 (2023) 170869.

\bibitem{kolesnikov2022influence}
S.~V. Kolesnikov, E.~S. Sapronova, Influence of dzyaloshinskii--moriya and dipole--dipole interactions on spontaneous magnetization reversal time of finite-length co chains on pt (664) surfaces, IEEE Magnetics Letters 13 (2022) 1--5.

\bibitem{wang2020optical}
X.-G. Wang, L.~Chotorlishvili, V.~K. Dugaev, A.~Ernst, I.~V. Maznichenko, N.~Arnold, C.~Jia, J.~Berakdar, I.~Mertig, J.~Barna{\'s}, The optical tweezer of skyrmions, npj Computational Materials 6~(1) (2020) 140.

\bibitem{demokritov2006bose}
S.~O. Demokritov, V.~E. Demidov, O.~Dzyapko, G.~A. Melkov, A.~A. Serga, B.~Hillebrands, A.~N. Slavin, Bose--einstein condensation of quasi-equilibrium magnons at room temperature under pumping, Nature 443~(7110) (2006) 430--433.

\bibitem{schneider2020bose}
M.~Schneider, T.~Br{\"a}cher, D.~Breitbach, V.~Lauer, P.~Pirro, D.~A. Bozhko, H.~Y. Musiienko-Shmarova, B.~Heinz, Q.~Wang, T.~Meyer, et~al., Bose--einstein condensation of quasiparticles by rapid cooling, Nature Nanotechnology 15~(6) (2020) 457--461.

\bibitem{mohseni2020bose}
M.~Mohseni, A.~Qaiumzadeh, A.~A. Serga, A.~Brataas, B.~Hillebrands, P.~Pirro, Bose--einstein condensation of nonequilibrium magnons in confined systems, New Journal of Physics 22~(8) (2020) 083080.

\bibitem{PhysRevResearch.6.L012011}
T.~Frostad, P.~Pirro, A.~A. Serga, B.~Hillebrands, A.~Brataas, A.~Qaiumzadeh, \href{https://link.aps.org/doi/10.1103/PhysRevResearch.6.L012011}{Anisotropy-assisted magnon condensation in ferromagnetic thin films}, Phys. Rev. Res. 6 (2024) L012011.
\newblock \href {https://doi.org/10.1103/PhysRevResearch.6.L012011} {\path{doi:10.1103/PhysRevResearch.6.L012011}}.
\newline\urlprefix\url{https://link.aps.org/doi/10.1103/PhysRevResearch.6.L012011}

\bibitem{bukharaev2018straintronics}
A.~A. Bukharaev, A.~K. Zvezdin, A.~P. Pyatakov, Y.~K. Fetisov, Straintronics: a new trend in micro-and nanoelectronics and materials science, Physics-Uspekhi 61~(12) (2018) 1175.

\bibitem{seki2012observation}
S.~Seki, X.~Yu, S.~Ishiwata, Y.~Tokura, Observation of skyrmions in a multiferroic material, Science 336~(6078) (2012) 198--201.

\bibitem{PhysRevB.96.054440}
S.~Stagraczy\ifmmode~\acute{n}\else \'{n}\fi{}ski, L.~Chotorlishvili, M.~Sch\"uler, M.~Mierzejewski, J.~Berakdar, \href{https://link.aps.org/doi/10.1103/PhysRevB.96.054440}{Many-body localization phase in a spin-driven chiral multiferroic chain}, Phys. Rev. B 96 (2017) 054440.
\newblock \href {https://doi.org/10.1103/PhysRevB.96.054440} {\path{doi:10.1103/PhysRevB.96.054440}}.
\newline\urlprefix\url{https://link.aps.org/doi/10.1103/PhysRevB.96.054440}

\bibitem{khomeriki2016positive}
R.~Khomeriki, L.~Chotorlishvili, I.~Tralle, J.~Berakdar, Positive--negative birefringence in multiferroic layered metasurfaces, Nano Letters 16~(11) (2016) 7290--7294.

\bibitem{PhysRevB.91.041408}
R.~Khomeriki, L.~Chotorlishvili, B.~A. Malomed, J.~Berakdar, \href{https://link.aps.org/doi/10.1103/PhysRevB.91.041408}{Creation and amplification of electromagnon solitons by electric field in nanostructured multiferroics}, Phys. Rev. B 91 (2015) 041408.
\newblock \href {https://doi.org/10.1103/PhysRevB.91.041408} {\path{doi:10.1103/PhysRevB.91.041408}}.
\newline\urlprefix\url{https://link.aps.org/doi/10.1103/PhysRevB.91.041408}

\bibitem{PhysRevLett.125.227201}
X.-G. Wang, L.~Chotorlishvili, N.~Arnold, V.~K. Dugaev, I.~Maznichenko, J.~Barna\ifmmode~\acute{s}\else \'{s}\fi{}, P.~A. Buczek, S.~S.~P. Parkin, A.~Ernst, \href{https://link.aps.org/doi/10.1103/PhysRevLett.125.227201}{Plasmonic skyrmion lattice based on the magnetoelectric effect}, Phys. Rev. Lett. 125 (2020) 227201.
\newblock \href {https://doi.org/10.1103/PhysRevLett.125.227201} {\path{doi:10.1103/PhysRevLett.125.227201}}.
\newline\urlprefix\url{https://link.aps.org/doi/10.1103/PhysRevLett.125.227201}

\bibitem{PhysRevB.88.184404}
J.-H. Moon, S.-M. Seo, K.-J. Lee, K.-W. Kim, J.~Ryu, H.-W. Lee, R.~D. McMichael, M.~D. Stiles, \href{https://link.aps.org/doi/10.1103/PhysRevB.88.184404}{Spin-wave propagation in the presence of interfacial dzyaloshinskii-moriya interaction}, Phys. Rev. B 88 (2013) 184404.
\newblock \href {https://doi.org/10.1103/PhysRevB.88.184404} {\path{doi:10.1103/PhysRevB.88.184404}}.
\newline\urlprefix\url{https://link.aps.org/doi/10.1103/PhysRevB.88.184404}

\bibitem{tiablikov2013methods}
S.~V. Tiablikov, Methods in the quantum theory of magnetism, Springer, 2013.

\bibitem{jandieri20191}
V.~Jandieri, P.~Baccarelli, G.~Valerio, G.~Schettini, 1-d periodic lattice sums for complex and leaky waves in 2-d structures using higher order ewald formulation, IEEE Transactions on Antennas and Propagation 67~(4) (2019) 2364--2378.

\bibitem{jandieri2020modal}
V.~Jandieri, P.~Baccarelli, G.~Valerio, K.~Yasumoto, G.~Schettini, Modal propagation in periodic chains of circular rods: Real and complex solutions, IEEE Photonics Technology Letters 32~(17) (2020) 1053--1056.

\bibitem{yasumoto1999efficient}
K.~Yasumoto, K.~Yoshitomi, Efficient calculation of lattice sums for free-space periodic green's function, IEEE Transactions on Antennas and Propagation 47~(6) (1999) 1050--1055.

\bibitem{chotorlishvili2010quantum}
L.~Chotorlishvili, A.~Ugulava, Quantum chaos and its kinetic stage of evolution, Physica D: Nonlinear Phenomena 239~(3-4) (2010) 103--122.

\bibitem{PhysRevE.70.026219}
A.~Ugulava, L.~Chotorlishvili, K.~Nickoladze, \href{https://link.aps.org/doi/10.1103/PhysRevE.70.026219}{Quantum-mechanical research on nonlinear resonance and the problem of quantum chaos}, Phys. Rev. E 70 (2004) 026219.
\newblock \href {https://doi.org/10.1103/PhysRevE.70.026219} {\path{doi:10.1103/PhysRevE.70.026219}}.
\newline\urlprefix\url{https://link.aps.org/doi/10.1103/PhysRevE.70.026219}

\bibitem{PhysRevE.71.056211}
A.~Ugulava, L.~Chotorlishvili, K.~Nickoladze, \href{https://link.aps.org/doi/10.1103/PhysRevE.71.056211}{Irreversible evolution of quantum chaos}, Phys. Rev. E 71 (2005) 056211.
\newblock \href {https://doi.org/10.1103/PhysRevE.71.056211} {\path{doi:10.1103/PhysRevE.71.056211}}.
\newline\urlprefix\url{https://link.aps.org/doi/10.1103/PhysRevE.71.056211}

\bibitem{katok1995introduction}
A.~Katok, A.~Katok, B.~Hasselblatt, Introduction to the modern theory of dynamical systems, no.~54, Cambridge university press, 1995.

\bibitem{rabinovich2012oscillations}
M.~I. Rabinovich, D.~I. Trubetskov, Oscillations and waves: in linear and nonlinear systems, Vol.~50, Springer Science \& Business Media, 2012.

\end{thebibliography}

\end{document}